\documentclass[10pt]{iopart}
\usepackage{url} 
\usepackage{pdflscape} 
\usepackage{amsmath}
\usepackage{amssymb} 
\usepackage{epsfig} 
\usepackage{amsfonts}
\usepackage{graphicx}
\usepackage{subfigure}
\usepackage{graphicx}
\usepackage{times,fancyhdr}

\usepackage{enumitem}
\setlist[enumerate,2]{label=\roman*)}

\def\case#1/#2{\textstyle\frac{#1}{#2}}

\newcommand{\be}{\begin{equation}}
\newcommand{\ee}{\end{equation}}
\newcommand{\ben}{\begin{eqnarray}}
\newcommand{\een}{\end{eqnarray}}
\usepackage{mdframed}
\usepackage{grffile}
\usepackage{amssymb}
\usepackage{ulem}
\usepackage[T1]{fontenc}
\usepackage[colorlinks=true,
            linkcolor=blue,
           urlcolor=blue,
           citecolor=blue]{hyperref}
\usepackage{subfigure}

\newtheorem{defn}{Definition}

\providecommand{\U}[1]{\protect\rule{.1in}{.1in}}
\newcommand{\mincir}{\raise
-3.truept\hbox{\rlap{\hbox{$\sim$}}\raise4.truept\hbox{$<$}\ }}
\newcommand{\magcir}{\raise
-3.truept\hbox{\rlap{\hbox{$\sim$}}\raise4.truept\hbox{$>$}\ }}

\def\S{{}^3\!S_+}

\begin{document}

\title{Generalized Scalar Field Cosmologies: A Global Dynamical Systems Formulation}

\author{Genly Leon}
\address{Departamento  de  Matem\'aticas,  Universidad  Cat\'olica  del  Norte, Avda. Angamos  0610,  Casilla  1280  Antofagasta,  Chile.}
\ead{genly.leon@ucn.cl}

\author{Felipe Orlando Franz Silva}
\address{Departamento  de  Matem\'aticas,  Universidad  Cat\'olica  del  Norte, Avda. Angamos  0610,  Casilla  1280  Antofagasta,  Chile.}
\ead{felipe.franz@alumnos.ucn.cl}

\begin{abstract}
Local and global phase-space descriptions and averaging methods are used to find qualitative features of solutions for the FLRW and the Bianchi I metrics in the context of scalar field cosmologies with arbitrary potentials and arbitrary couplings to matter. The stability of the equilibrium points in a phase-space as well as the dynamics in the regime where the scalar field diverges are studied. Equilibrium points that represent some solutions of cosmological interest such as: several types of scaling solutions, a kinetic dominated solution representing a stiff fluid, a solution dominated by an effective energy density of geometric origin, a quintessence scalar field dominated solution, the vacuum de Sitter solution associated to the minimum of the potential, and a non-interacting matter dominated solution are obtained. All reveal a very rich cosmological phenomenology. 
\end{abstract}


\maketitle

\section{Introduction}

Gravitational scalar field  theories of special interest are Jordan-Brans-Dicke Theories \cite{Jordan:1958zz,Brans:1961sx}, Horndeski Theories \cite{Horndeski:1974wa},  Teleparalell Analogue of Horndeski Theories \cite{Bahamonde:2019shr,Bahamonde:2019ipm,Bahamonde:2020cfv}, Inflationary Models \cite{Guth:1980zm}, Extended Quintessence, Modified Gravity, Ho\v{r}ava-Lifschitz and the Galileons, etc.,  \cite{Copeland:1993jj,Ibanez:1995zs,Chimento:1995da,Lidsey:1995np,Coley:1997nk,Copeland:1998fz,Coley:1999mj,Coley:1999uh,Coley:2000zw,Coley:2000yc,Coley:2003tf,Elizalde:2004mq,Capozziello:2005tf,Curbelo:2005dh,Gonzalez:2005ie,Gonzalez:2006cj,Lazkoz:2006pa,Lazkoz:2007mx,Elizalde:2008yf,Leon:2009dt,Leon:2009rc,Leon:2009ce,Leon:2010pu,Basilakos:2011rx,Xu:2012jf,Leon:2012mt,Leon:2013qh,Fadragas:2013ina,Kofinas:2014aka,Leon:2014yua,Paliathanasis:2014yfa,DeArcia:2015ztd,Paliathanasis:2015gga,Leon:2015via,Barrow:2016qkh,Barrow:2016wiy,Cruz:2017ecg,Paliathanasis:2017ocj,Alhulaimi:2017ocb,Dimakis:2017kwx,Giacomini:2017yuk,Karpathopoulos:2017arc,DeArcia:2018pjp,Tsamparlis:2018nyo,Paliathanasis:2018vru,Basilakos:2019dof,VanDenHoogen:2018anx,Leon:2018lnd,Leon:2018skk,Leon:2019mbo,Paliathanasis:2019qch,Leon:2019jnu,Paliathanasis:2019pcl,Barrow:2018zav,Quiros:2019ktw,Leon:2019iwj,Giacomini:2020zmv,Paliathanasis:2020abu,Giacomini:2020grc,Copeland:1997et}. There are several studies in literature that provide both global and local dynamical systems analysis for scalar field cosmologies with arbitrary potentials and arbitrary couplings, say \cite{Foster:1998sk,Miritzis:2003ym,Dania&Yunelsy,Leon:2008de,Giambo:2009byn,Nojiri:2019riz,Humieja:2019ywy,Matsumoto:2017gnx,Matsumoto:2015hua,Solomon:2015hja,Harko:2015pma,Minazzoli:2014xua,Skugoreva:2013ooa,Jamil:2012vb,Miritzis:2011zz,Hrycyna:2007gd,Giambo:2008ck,Leon:2010ai,Leon:2014bta,Leon:2014rra,Fadragas:2014mra,Tzanni:2014eja} (see paper \cite{PaperI} and discussions therein). 

In \cite{PaperI}, the action for a general class of Scalar Tensor Theory of Gravity written in the so-called Einstein's frame  was studied. This action is given by \cite{Kaloper:1997sh,Gonzalez:2007ht}:
\begin{align}&
\mathcal{L}=\int  d{ }^4 x \sqrt{|g|}\left\{\frac{1}{2} R-\frac{1}{2} g^{\mu
\nu}\nabla_\mu\phi\nabla_\nu\phi-V(\phi)-\Lambda+\chi(\phi)^{-2}
\mathcal{L}(\mu,\nabla\mu,\chi(\phi)^{-1}g_{\alpha\beta})\right\},\label{eq1}
\end{align} where a system of units in which $8\pi G=c=\hbar=1$ is used. 
In this equation $R$ is the curvature scalar, $\phi$ is the 
scalar field,  $\nabla_\alpha$
is the covariant derivative, $V(\phi)$ is the self-interaction potential, $\Lambda\geq 0$ is the Cosmological Constant, $\chi(\phi)^{-2}$ is the coupling function. $\mathcal{L}$
is the matter Lagrangian, and $\mu$ is the collective name for the
matter degrees of freedom. 

For the action \eqref{eq1}, the matter energy-momentum tensor is defined by:
\begin{equation}T_{\alpha
\beta}=-\frac{2}{\sqrt{|g|}}\frac{\delta}{\delta g^{\alpha
\beta}}\left\{\sqrt{|g|}
 \chi^{-2}\mathcal{L}(\mu,\nabla\mu,\chi^{-1}g_{\alpha
 \beta})\right\}.\label{Tab}\end{equation}
The ``energy exchange'' vector is defined by: 
\begin{equation}
\label{exchange}
    Q_\beta\equiv\nabla^\alpha T_{\alpha \beta}=-\frac{1}{2}T\frac{1}{\chi(\phi)}\frac{\mathrm{d}\chi(\phi)}{\mathrm{d}\phi}\nabla_{\beta}\phi,\;
 T=T^\alpha_\alpha,
\end{equation}
 where $T$ is the trace of the energy-momentum tensor. The matter is assumed in the form of  a perfect fluid   with energy density $\rho_m\geq 0$ and pressure  $p_m= (\gamma-1) \rho_m, \gamma \in [1,2]$. The expression \eqref{exchange} implies that the scalar field $\phi$ is non--minimally coupled to matter with coupling function $\chi(\phi)$. 
 Additionally, the geometric properties of the metric are incorporated in the form of the function:
\begin{align}
&G_0(a)=\left\{ \begin{array}{cc}
-3\frac{k}{a^2}, k=0, \pm 1, & \text{FLRW}\\
\frac{\sigma_0^2}{a^6}, & \text{Bianchi I}
\end{array} 
\right..
\end{align} 
The following field equations holds:
\begin{subequations}
	\label{Non_min2}
	\begin{align}
	& \dot{H}=-\frac{1}{2}\left(\gamma \rho_m+y^2\right)+\frac{1}{6}a G_0'(a), \label{Rachd}\\
	&\dot{\rho_m}=-3\gamma H\rho_m-\frac{1}{2}(4-3\gamma)\rho_m y\frac{d\ln\chi(\phi)}{d \phi}, \label{consb}\\
	& \dot a= a H,\\
	&\dot y=-3 H y -\frac{d V(\phi)}{d\phi}+\frac{1}{2}(4-3\gamma)\rho_m \frac{d\ln\chi(\phi)}{d \phi} \label{13EQ},\\
	& \dot\phi=y, \\
	& 3H^2=\rho_m+\frac{1}{2}\dot\phi^2+V(\phi)+\Lambda+G_0(a),\label{Fried2b}
	\end{align}
\end{subequations}
The phase-space can be defined using equation \eqref{Fried2b} as follows:
\begin{equation}
\label{Fried2bB}
   \left\{(H, \rho_m, a, y, \phi)\in \mathbb{R}^5: 3H^2=\rho_m+\frac{1}{2}y^2+V(\phi)+\Lambda+G_0(a)\right\}.
\end{equation}
In paper \cite{PaperI} some theorems related to the asymptotic behavior of a very general cosmological model given by system \eqref{Non_min2} were presented. In particular, the conditions of a scalar field potential $V\in  C^{2}(\mathbb{R})$  under which $\displaystyle{\lim_{t\rightarrow  \infty} \dot \phi =0}$ were discussed; they are: non-negativity of the potential which is zero only at the origin and the boundedness of both  $V^{\prime}(\phi)$ and  $V(\phi)$  (Theorem 2.1.3 in \cite{PaperI}). Furthermore, some extra conditions for having $\displaystyle{\lim_{t\rightarrow  \infty} \phi(t) \in \lbrace -\infty , 0 , + \infty \rbrace }$ were presented. They are the previous conditions with the addition of $V^{\prime}(\phi)>0$ for $\phi>0$ and $V^{\prime}(\phi)<0$ for $\phi<0$  (Theorem 2.2.3 in \cite{PaperI}).
Mild conditions under the potential for having  $\lim_{t \rightarrow  + \infty} \dot \phi =0$ and $\lim_{t \rightarrow  + \infty} \phi (t)= +\infty$ (Theorem 2.3.3 in \cite{PaperI}) were considered.
It was examined to which extent the hypotheses of the theorems proved in \cite{PaperI} can be relaxed in order to obtain the same conclusions or provide a counterexample, by means of generalized harmonic self-interacting potentials: $V_1(\phi)= \mu^3 \left[\frac{\phi^2}{\mu} + b f \cos\left(\delta + \frac{\phi}{f}\right)\right]$, $b\neq 0$ and 
$V_2(\phi)= \mu ^3 \left[b f \left(\cos (\delta )-\cos \left(\delta +\frac{\phi }{f}\right)\right)+\frac{\phi ^2}{\mu}\right]
$, $b\neq 0$. Harmonic potentials plus cosine corrections were introduced in the context of inflation in loop-quantum cosmology in \cite{Sharma:2018vnv}. 
Finally, the Hubble--normalized formulation for a scalar field cosmology with generalized harmonic potential $V(\phi)= \frac{\phi^2}{2} + f \left[1-\cos\left( \frac{\phi}{f}\right)\right]$, $f> 0$, nonminimally coupled to matter with coupling function  $\chi=\chi_0 e^{\frac{\lambda \phi}{4-3\gamma}}$, where  $\lambda$ is a constant and  $0\leq \gamma \leq 2, \quad \gamma \neq \frac{4}{3}$  was performed for FLRW metrics and for Bianchi I metric.  Using Theorem 2.4 in \cite{PaperI}, the late time attractors correspond to the non zero local minimums of the potential, i.e., $\phi=\phi_*$, satisfying $\sin(\phi^*/f)+\phi^*=0$, $0<|\phi^*|\leq 1, \cos \left(\frac{\phi^*}{f}\right)+f>0$.  Whenever  $\phi^*$ is a local non zero minimum of $V(\phi)$: $\lim_{t\rightarrow \infty} \phi(t) =\phi^*,\lim_{t\rightarrow \infty} \dot\phi(t)=0, \lim_{t\rightarrow \infty} H(t)= \frac{\sqrt{V({\phi^*})}}{\sqrt{3}}$, and  $ \lim_{t\rightarrow \infty} G_0(a)=0$. The global minimum of $V(\phi)$ at $\phi=0$  is unstable to curvature perturbations for $\gamma>\frac{2}{3}$ in case of negatively curved FLRW model. This confirms the result in \cite{Giambo:2019ymx} that for $\gamma> 2/3$, in a non-degenerated minima with zero critical value the curvature has a dominant effect on the late evolution of the universe and will eventually dominate both the perfect fluid and the scalar field. For Bianchi I model the global minimum $V(0)=0$ is unstable to shear perturbations.

In this companion paper  complementary formulations based on  \cite{Alho:2014fha} will be presented, including the analysis of  the oscillations entering the system via Klein--Gordon equation \cite{Fajman:2020yjb}.

It is useful to define the quantities
$\lambda=-\frac{V^{\prime }(\phi)}{V(\phi)}, \quad f\equiv\frac{V^{\prime \prime
}(\phi)}{V(\phi)}-\frac{{V^{\prime}(\phi)}^2}{V(\phi)^2}$ to obtain qualitative information about the past and future asymptotic structure of the equations's solutions in scalar field theories with potential $V(\phi)$. This is the basis of the so-called method of $f$- devisers. 
For the common scalar field potentials the function $f(\lambda)$ is found as
follows. The monomial potential $V(\phi)=\frac{1}{2n}(\mu
\phi)^{2n}$, $\mu >0, n=1,2,\ldots$ \cite{Alho:2015cza} has $f(\lambda)=-\frac{\lambda^2}{%
2 n}$. The so-called $E$--model studied from the dynamical systems point of
view in \cite{Alho:2017opd} has potential $V(\phi)=V_0\left(1-e^{-\sqrt{%
\frac{2}{3 \alpha}} \phi}\right)^{2 n}$. The corresponding $f$-deviser is $%
f(\lambda)=-\frac{\lambda \left(\lambda-\sqrt{6} \mu \right)}{2 n}$, where $\mu=\frac{n}{3 
\sqrt{\alpha}}$. The exponential potential plus a cosmological constant $V(\phi)=V_{0}e^{-l\phi}+V_1$ \cite%
{Yearsley:1996yg,Pavluchenko:2003ge,Cardenas:2002np} has $f(\lambda)=-\lambda(\lambda-l)$. The hyperbolic potentials:   $V(\phi)=V_{0}\left(\cosh\left( \xi \phi \right)-1\right)$ \cite%
{Pavluchenko:2003ge,Ratra:1987rm,Wetterich:1987fm,Matos:2009hf,Copeland:2009be,Leyva:2009zz,delCampo:2013vka,Sahni:1999qe,Sahni:1999gb,Lidsey:2001nj,Matos:2000ng} with $%
f(\lambda)=-\frac{1}{2}(\lambda^2-\xi^2)$ and $V(\phi)=V_{0}\sinh^{-\alpha}(\beta\phi)$ 
\cite%
{Pavluchenko:2003ge,Ratra:1987rm,Wetterich:1987fm,Copeland:2009be,Leyva:2009zz,Sahni:1999gb,UrenaLopez:2000aj}
with   $f(\lambda)=\frac{\lambda^2}{\alpha}-\alpha\beta^2$. The double exponential potential $%
V(\phi)=V_{0}\left(e^{\alpha\phi}+e^{\beta\phi}\right)$ \cite%
{Gonzalez:2006cj,Barreiro:1999zs,Gonzalez:2007hw} has $f(\lambda)=-(\lambda+\alpha)(\lambda+%
\beta)$. 
This paper's proposal, for a scalar field non-minimally coupled to matter, relies on defining $\lambda= -\frac{W'(\phi)}{W(\phi)}, \quad f=\frac{W''(\phi)}{W(\phi)}-\frac{W'(\phi)^2}{W(\phi)^2},  \quad  g= - \frac{\chi '(\phi)}{\chi (\phi)}$,
where $W(\phi)=\Lambda + V(\phi)$, $W(\phi)\geq 0$. Then, it is assumed that $f$ and $g$ can be explicitly written as functions of $\lambda$, as in the original $f$-deviser method.  This is a significant advantage due to the analysis for arbitrary potential can be performed as a first step; then, just substitute the desired forms instead of repeating the whole procedure for every distinct potential. Similar approaches were used in \cite{Leon:2018skk,Giacomini:2020grc} for two-field Jordan-Brans-Dicke cosmologies.

This paper is organized as follows.  A local dynamical systems analysis using Hubble normalized equations is performed in Section
\ref{SECT:6.1}. Then, a dynamical systems formulation  using global dynamical systems variables based on Alho \& Uggla's approach \cite{Alho:2014fha} is given in Section \ref{SECT:3.3}.  
Section \ref{SECT:3.4} is devoted to the asymptotic analysis as $\phi\rightarrow \infty$ for arbitrary $V(\phi)$ and $\chi(\phi)$. In Section \ref{sect3.2.2}, the physical interpretation of the solutions that were found is discussed. Section \ref{Section2.3} is devoted to equations analysis of a scalar field model with potential $V(\phi)= V_0 e^{-\lambda \phi}$ in a vacuum. 
{In Section \ref{Section2.3B}, the potential of the so-called $E$--model: $%
V(\phi )=V_{0}\left( 1-e^{-\sqrt{\frac{2}{3\alpha }}\phi }\right) ^{2n}$ is investigated. This potential was discussed in \cite{Alho:2017opd} for a conventional scalar
field cosmology and in \cite{Leon:2019mbo} for Hořava–Lifshitz cosmology.}  
In Sections \ref{Sect:2.4} and \ref{Sect:2.5}, scalar field cosmologies under the potentials $V_1(\phi)= \mu^3 \left[\frac{\phi^2}{\mu} + b f \cos\left(\delta + \frac{\phi}{f}\right)\right]$, $b\neq 0$ and 
$V_2(\phi)= \mu ^3 \left[b f \left(\cos (\delta )-\cos \left(\delta +\frac{\phi }{f}\right)\right)+\frac{\phi ^2}{\mu}\right]
$, $b\neq 0$  are respectively studied. Sections \ref{Sect:2.4.1} and \ref{Sect:2.5.1} are devoted to the analysis of the  dynamics as $\phi\rightarrow \infty$. In Sections \ref{Sect:2.4.2} and \ref{Sect:2.5.2}, the study of the oscillatory behavior is presented.  
In Section \ref{discussion}, the main results are summarized. Finally, Section \ref{Sect:7} is devoted to conclusions. 
\section{Local dynamical systems analysis for arbitrary $V(\phi)$ and $\chi(\phi)$}
\label{SECT:6.1}
For the analysis of the system \eqref{Non_min2} for an arbitrary $V(\phi)$ and arbitrary $\chi(\phi)$ are defined: 
\begin{align}
  & \lambda(\phi)= -\frac{W'(\phi)}{W(\phi)}, \quad f(\phi)=\frac{W''(\phi)}{W(\phi)}-\frac{W'(\phi)^2}{W(\phi)^2},  \quad  g(\phi)= - \frac{\chi '(\phi)}{\chi (\phi)}, \label{parametrizationA}
\end{align}
where $W(\phi)=\Lambda + V(\phi)$, $W(\phi)\geq 0$. The conditions \eqref{parametrizationA} can be  alternatively written as:
\begin{equation}
\label{eqA8}
\frac{d \lambda}{d \phi}= -f( \phi), \quad 
\frac{d W}{d \phi}= -\lambda(\phi) W(\phi), \quad
\frac{d \chi}{d \phi}=-g(\phi) \chi(\phi).
\end{equation}
The idea is to assume that $f$ and $g$ can be explicitly written as functions of $\lambda$. Hence,  \eqref{eqA8} is written as:
\begin{equation}
\frac{d \phi}{d \lambda}= -\frac{1}{f(\lambda)}, \quad 
\frac{d W}{d \lambda}= \frac{\lambda  W(\lambda)}{f(\lambda)}, \quad
\frac{d \chi}{d \lambda}=\frac{g(\lambda) \chi(\lambda)}{f(\lambda)}, 
\end{equation}
which can be integrated in quadrature as:
\begin{equation}
 \phi (\lambda )=\phi (1)-\int_1^{\lambda } \frac{1}{f(s)} \, ds, \quad
 W (\lambda )=W(1) e^{\int_1^{\lambda } \frac{s}{f(s)} \, ds}, \quad
 \chi (\lambda )=\chi(1) e^{\int_1^{\lambda } \frac{g(s)}{f(s)} \, ds}.
\end{equation}
The last equations can be used to generate the potentials and couplings by giving the functions $f(\lambda)$ and $g(\lambda)$ as input. 

Defining the new variables:
\begin{equation}
 x= \frac{\dot\phi}{\sqrt{6}H}, \quad \Omega_m= \frac{\rho_m}{3 H^2}, \quad \Omega_0= \frac{G_0(a)}{3 H^2}, \quad \lambda=-\frac{V'(\phi)}{\Lambda +V(\phi)},
\end{equation}
where $ G_0(a)=q a^{-p}, p\geq 0$, and the time variable $\tau=\ln a$, the following dynamical system is obtained:
\begin{small}
\begin{subequations}
\label{systH1}
\begin{align}
&\frac{d x}{d\tau}=\frac{1}{2} x \left(3 \gamma  \Omega_m+p \Omega_0+6
   x^2-6\right)-\sqrt{\frac{3}{2}} \lambda  \left(x^2+\Omega_0+\Omega_m-1\right)  +\frac{\sqrt{6}}{4} (3 \gamma -4) \Omega_m g(\lambda
   ),\\
&\frac{d\Omega_m}{d\tau}= \frac{1}{2} \Omega_m \left(\sqrt{6} (4-3
   \gamma ) x g(\lambda )+2 \left(3 \gamma  (\Omega_m-1)+p
   \Omega_0+6 x^2\right)\right),\\
&\frac{d\Omega_0}{d\tau}= \Omega_0 \left(3 \gamma\Omega_m+p (\Omega_0-1)+6 x^2\right),\\
&\frac{d\lambda}{d\tau}= -\sqrt{6} x f(\lambda ),   
\end{align}
\end{subequations}
\end{small}
with the restriction
\begin{equation}
    x^2+\Omega _{m}+\Omega_{0}=1-\frac{W(\phi)}{3H^2} \leq 1. 
\end{equation}

 The equilibrium points of the system \eqref{systH1} are the following: 
\begin{enumerate}
    \item[$A_1(\hat{\lambda})$:] $(x, \Omega_m, \Omega_0, \lambda)=\left(\frac{(4-3 \gamma ) g(\hat{\lambda})}{\sqrt{6} (\gamma -2)},
   1-\frac{(4-3 \gamma )^2 g(\hat{\lambda})^2}{6 (\gamma
   -2)^2}, 0, \hat{\lambda}\right)$, where  $\hat{\lambda}$ denotes any value of $\lambda$ satisfying $f(\hat{\lambda})=0$, represents a matter - kinetic scaling solution. 
For $1\leq \gamma< 2$, it follows that  $A_1(\hat{\lambda})$ is a sink under one of the conditions (i)- (x) in  \ref{AppA}. If it exists, it will never be a source. 
   
   \item[$A_2(\hat{\lambda})$:] $(x, \Omega_m, \Omega_0, \lambda)=\left(\frac{\sqrt{\frac{2}{3}} (p-3 \gamma )}{(3 \gamma -4) g\left(\hat{\lambda
   }\right)}, \frac{2 (6-p) (p-3 \gamma )}{3 (4-3 \gamma )^2
   g(\hat{\lambda})^2}, \frac{2 (p-3 \gamma ) (\gamma
   -2)}{(4-3 \gamma )^2 g(\hat{\lambda})^2}+1, \hat{\lambda}\right)$ represents a matter- scalar field - geometric ``fluid'' scaling solution. For $1\leq \gamma< 2$, it follows that 
$A_2(\hat{\lambda})$ is a sink under one of the conditions (i) - (viii) in  \ref{AppA}.
It is non--hyperbolic for $p=6$, and a saddle for $p=2$. If exists, it will never be  a source. 
   
   \item[$A_3(\hat{\lambda})$:] $(x, \Omega_m, \Omega_0, \lambda)=\left(\frac{\sqrt{6} \gamma }{(4-3 \gamma ) g(\hat{\lambda})+2
   \hat{\lambda }}, \frac{2 (4-3 \gamma ) g(\hat{\lambda})
   \hat{\lambda }+4 \left(\hat{\lambda }^2-3 \gamma \right)}{\left((3
   \gamma -4) g(\hat{\lambda})-2 \hat{\lambda }\right)^2}, 0
   , \hat{\lambda}\right)$ represents a matter - scalar field scaling solution. 
In this case, it can be proceeded semi-analytically, that is, for non-minimal coupling $g\equiv 0$  and assuming $1\leq \gamma<2$, $A_3(\hat{\lambda})$ is a sink for: 
 \begin{enumerate}
     \item $1\leq \gamma <2, p>3 \gamma, -\frac{2 \sqrt{6} \gamma }{\sqrt{9 \gamma -2}}\leq \hat{\lambda }<-\sqrt{3} \sqrt{\gamma }, f'(\hat{\lambda })<0$, or 
     \item $1\leq \gamma <2, p>3 \gamma,
   \sqrt{3} \sqrt{\gamma }<\hat{\lambda }\leq \frac{2 \sqrt{6} \gamma }{\sqrt{9 \gamma -2}}, f'(\hat{\lambda })>0$. 
 \end{enumerate}
  Otherwise, it is a saddle.  For non-minimal coupling the analysis has to be done numerically. 
  
   \item[$A_4(\hat{\lambda})$:] $(x, \Omega_m, \Omega_0, \lambda)=\left(-1, 0, 0,\hat{\lambda}\right)$ is a kinetic dominated solution representing a stiff fluid. 
   $A_4(\hat{\lambda})$ is a source for: 
   \begin{enumerate}
       \item $1\leq \gamma <\frac{4}{3},  0\leq p<6,  g(\hat{\lambda })<\frac{\sqrt{6} (\gamma -2)}{3 \gamma -4},  f'(\hat{\lambda })>0,  \hat{\lambda }>-\sqrt{6}$, or 
       \item $\gamma =\frac{4}{3},   0\leq p<6,  f'(\hat{\lambda })>0,  \hat{\lambda }>-\sqrt{6}$, or 
   \item $\frac{4}{3}<\gamma <2,  0\leq p<6,  g(\hat{\lambda })>\frac{\sqrt{6} (\gamma -2)}{3 \gamma -4}, 
   f'(\hat{\lambda })>0,  \hat{\lambda }>-\sqrt{6}$.
   \end{enumerate}

   $A_4(\hat{\lambda})$ is a sink for:
   \begin{enumerate}
       \item $1\leq \gamma <\frac{4}{3}, p>6, g(\hat{\lambda })>\frac{\sqrt{6} (\gamma -2)}{3 \gamma -4}, f'(\hat{\lambda })<0, \hat{\lambda }<-\sqrt{6}$, or 
       \item $\frac{4}{3}<\gamma <2, p>6, g(\hat{\lambda })<\frac{\sqrt{6} (\gamma -2)}{3 \gamma -4}, f'(\hat{\lambda })<0, \hat{\lambda }<-\sqrt{6}$.
   \end{enumerate}
   
   \item[$A_5(\hat{\lambda})$:] $(x, \Omega_m, \Omega_0, \lambda)=\left(0, 0, 1, \hat{\lambda}\right)$ is a solution dominated by the effective energy density of $G_0(a)$. It is non--hyperbolic with a 3D unstable manifold for $p>6$.
   
   \item[$A_6(\hat{\lambda})$:] $(x, \Omega_m, \Omega_0, \lambda)=\left(1, 0, 0, \hat{\lambda}\right)$  is a kinetic dominated solution representing a stiff fluid. 
   $A_6(\hat{\lambda})$ is a source for:
   \begin{enumerate}
       \item $1\leq \gamma <\frac{4}{3},  p<6,  g(\hat{\lambda })>-\frac{\sqrt{6} (\gamma -2)}{3 \gamma -4},  f'(\hat{\lambda })<0,  \hat{\lambda }<\sqrt{6}$, or 
       \item $\gamma =\frac{4}{3},  p<6,    f'(\hat{\lambda })<0,  \hat{\lambda }<\sqrt{6}$, or 
   \item $\frac{4}{3}<\gamma<2,  p<6,  g(\hat{\lambda })<-\frac{\sqrt{6} (\gamma -2)}{3 \gamma -4},  f'(\hat{\lambda })<0, 
   \hat{\lambda }<\sqrt{6}$.
   \end{enumerate}
      $A_6(\hat{\lambda})$ is a  sink for: 
    \begin{enumerate}
        \item $1\leq \gamma <\frac{4}{3},  p>6,  g(\hat{\lambda })<-\frac{\sqrt{6} (\gamma -2)}{3 \gamma -4},  f'(\hat{\lambda })>0,  \hat{\lambda }>\sqrt{6}$, or 
        \item $1\leq \gamma >\frac{4}{3},  p>6,   g(\hat{\lambda })>-\frac{\sqrt{6} (\gamma -2)}{3 \gamma -4},  f'(\hat{\lambda })>0,  \hat{\lambda }>\sqrt{6}$.
    \end{enumerate}
   
   \item[$A_7(\hat{\lambda})$:] $(x, \Omega_m, \Omega_0, \lambda)=\left(\frac{p}{\sqrt{6} \hat{\lambda }}, 0, 1-\frac{p}{\hat{\lambda }^2} , \hat{\lambda}\right)$ is a scaling solution where the energy density of the scalar field and the effective energy density from $G_0(a)$ scale with the same order of magnitude.    For $1\leq \gamma\leq 2$, $A_7(\hat{\lambda})$, it is a sink for one of the conditions (i) - (xliv) in  \ref{AppA}.  It will never be a source.
   
   \item[$A_8(\hat{\lambda})$:] $(x, \Omega_m, \Omega_0, \lambda)=\left(\frac{\hat{\lambda }}{\sqrt{6}}, 0, 0, \hat{\lambda}\right)$  represents the typical quintessence scalar field dominated solution. Assuming $1\leq \gamma \leq 2$,  $A_8(\hat{\lambda})$ is a sink for: 
   \begin{enumerate}
       \item $1\leq \gamma <\frac{4}{3},  p>\hat{\lambda }^2,  f'(\hat{\lambda })>0,  0<\hat{\lambda }<\sqrt{6},  g(\hat{\lambda })<\frac{6 \gamma -2 \hat{\lambda }^2}{4 \hat{\lambda }-3 \gamma  \hat{\lambda}}$, or 
   \item $1\leq \gamma <\frac{4}{3},  p>\hat{\lambda }^2,  g(\hat{\lambda })>\frac{6 \gamma -2 \hat{\lambda }^2}{4 \hat{\lambda }-3 \gamma  \hat{\lambda }},  -\sqrt{6}<\hat{\lambda }<0, 
   f'(\hat{\lambda })<0$, or 
   \item $\gamma=\frac{4}{3},  p>\hat{\lambda }^2,  -2<\hat{\lambda }<0,  f'(\hat{\lambda })<0$, or 
   \item $\gamma=\frac{4}{3},  p>\hat{\lambda }^2,  f'(\hat{\lambda})>0,  0<\hat{\lambda }<2$, or 
   \item $\frac{4}{3}<\gamma <2,  p>\hat{\lambda }^2,  g(\hat{\lambda })>\frac{6 \gamma -2 \hat{\lambda }^2}{4 \hat{\lambda }-3 \gamma  \hat{\lambda }}, 
   f'(\hat{\lambda })>0,  0<\hat{\lambda }<\sqrt{6}$, or 
   \item $\frac{4}{3}<\gamma <2,  p>\hat{\lambda }^2,  -\sqrt{6}<\hat{\lambda }<0,  g(\hat{\lambda })<\frac{6 \gamma -2 \hat{\lambda
   }^2}{4 \hat{\lambda }-3 \gamma  \hat{\lambda }},  f'(\hat{\lambda })<0$.
   \end{enumerate}
 It will never be a source.
   
   \item[$A_9$:] $(x, \Omega_m, \Omega_0, \lambda)=\left(0, 0, 0, 0\right)$ represents the vacuum de Sitter solution associated to the minimum of the potential. It is a sink for $p>0, \gamma>0, f(0)>0$. Otherwise, it is a saddle. 
   
   \item[$A_{10}(\tilde{\lambda})$:] $(x, \Omega_m, \Omega_0, \lambda)=\left(0, 1, 0, \tilde{\lambda}\right)$, where we denote by  $\tilde{\lambda}$, the values of $\lambda$ for which  $g(\lambda)=0$. It represents a non-interacting matter dominated solution. It is a saddle. 
\end{enumerate}

\section{Global dynamical analysis for arbitrary $V(\phi)$ and $\chi(\phi)$}
\label{SECT:3.3}
In Section \ref{SECT:6.1}, the stability of the equilibrium points using Hubble- normalized equations, which is essentially based on the Copeland, Liddle \& Wands's approach \cite{Copeland:1997et} was investigated. It is well-known that this procedure is well-suited to investigate local stability features of the equilibrium points. However, it does not provide a global description of the phase space when generically $\phi$ diverges or when $H\rightarrow 0$; in these cases the method fails. For this reason,  a new global systems analysis  for arbitrary $V(\phi)$ and $\chi(\phi)$ as $|\phi|\rightarrow \infty$ is presented based on Alho \& Uggla's approach \cite{Alho:2014fha}. Doing so, setting $\Lambda=0$ for simplicity. 

Now, it is assumed that the following limits exist:
\begin{equation}
N=  \lim_{\phi \rightarrow + \infty}  \frac{V'(\phi)}{V(\phi)}, \;\text{and} \quad M=  \lim_{\phi \rightarrow + \infty}  \frac{\chi'(\phi)}{\chi(\phi)}.
\end{equation} This permits the definition of the new functions  $W_{V}(\phi), W_{\chi}(\phi)$: 
\begin{equation}
  W_{V}(\phi)=  \frac{V'(\phi)}{V(\phi)} -N, \quad W_{\chi}(\phi)= \frac{\chi'(\phi)}{\chi(\phi)} - M, 
\end{equation}
where the constants $N$ and $M$ are such that
\begin{equation}
  \lim_{\phi \rightarrow + \infty}   W_{V}(\phi)=0, \quad  \lim_{\phi \rightarrow + \infty}   W_{\chi}(\phi)=0.
\end{equation}
After, the following variables are defined:
\begin{align}
&T=\frac{m}{m+H}, \quad \theta=  \tan ^{-1} \left(\frac{\dot \phi}{\sqrt{2 V(\phi)+ 2 \rho_m + 2 G_0(a)}}\right), \nonumber \\
& \Omega_m= \frac{\rho_m}{3 H^2}, \quad \Omega_0= \frac{G_0(a)}{3 H^2}, \quad m>0, 
\label{EQ26}
\end{align}
where  $ G_0(a)=q a^{-p}, p\geq 0$,  
such that:
\begin{align}
  & V(\phi)=  \frac{3 m^2 (1-T)^2 (\cos (2 \theta )-2 \Omega_{0}-2 \Omega_{m}+1)}{2 T^2}, \nonumber \\
  & H = m \left(\frac{1-T}{T}\right),\quad \dot{\phi}= \frac{\sqrt{6} m (1-T) \sin (\theta )}{T},\quad
\rho_{m}=\frac{3 m^2(1-T)^2 \Omega_{m}}{T^2}.
\end{align}
Finally, the time variable  $\tau= \ln a$ is defined; resulting the following unconstrained dynamical system:
\begin{subequations}
\label{23-syst}
\begin{align}
   & \frac{d T}{d\tau}= \frac{1}{2} (1-T) T \left(3 \gamma  \Omega_m+6 \sin ^2(\theta )+p \Omega_0\right),\\
   & \frac{d \theta}{d\tau}= -\frac{\sec (\theta )}{4 \sqrt{6}} \Bigg\{3 \sqrt{6} \sin (3 \theta )+6 \left(\cos (2 \theta ) - 2
   \Omega_0+1\right) (N+W_V(\phi)) \nonumber \\
   & +\Omega_m \left(-6 \sqrt{6} \gamma  \sin (\theta )-12 (2 M+N+W_V(\phi)+2 W_{\chi}(\phi))+18 \gamma  (M+W_{\chi}(\phi))\right)\nonumber \\
   & +\sqrt{6} \sin (\theta ) (3-2 p \Omega_0)\Bigg\},\\
   & \frac{d \Omega_m}{d\tau}= -\frac{1}{2}  \Omega_m \Bigg\{6 \cos (2 \theta )-\sqrt{6} (3 \gamma -4) \sin (\theta ) (M+W_{\chi}(\phi))\nonumber \\
   & -2 (3 \gamma 
   (\Omega_m-1)+p \Omega_0+3)\Bigg\},\\
   & \frac{d \Omega_0}{d\tau}= - \Omega_0 (-3 \gamma  \Omega_m+3 \cos (2 \theta )-p \Omega_0+p-3),\\
   & \frac{d \phi}{d\tau}=\sqrt{6} \sin (\theta ).
\end{align}
\end{subequations}
Rather to discuss all the equilibrium points of \eqref{23-syst} (which is a complementary formulation of the system \eqref{systH1}), the focus will be to study what happens in the region of $\phi\rightarrow +\infty$. Due to the symmetries of the model, it is not necessary to study the limit $\phi\rightarrow -\infty$, which can be accomplished by taking the reversal $\phi \rightarrow -\phi$.

\subsection{Asymptotic analysis as $\phi\rightarrow \infty$}
\label{SECT:3.4}
In this section the stability of the equilibrium points of \eqref{23-syst} as $\phi\rightarrow \infty$ (for functions $V$ and $\chi$  well-behaved at infinity of exponential orders of $N$ and $M$, respectively) is discussed.  This analysis at infinity  complements  the analysis of Section \ref{SECT:6.1}.

\begin{defn}[Definition 1 \cite{Foster:1998sk}] 
\label{kWBI1}
Let $V:\mathbb{R}\longrightarrow \mathbb{R}$ be a $C^2(\mathbb{R})$ nonnegative function. Let there exist
some $\phi_0> 0$ for which $V(\phi)>0$ for all $\phi>\phi_0$ and some number
$N<\infty$, such that the function:
\begin{equation} W_V:   [\phi_0, \infty) \longrightarrow \mathbb{R}, \quad
\phi \longrightarrow \frac{V^{\prime}(\phi)}{V(\phi)}-N 
\end{equation}
is well-defined and satisfies 
\begin{equation}
\lim_{\phi\rightarrow \infty} W_V(\phi)=0,
\end{equation}
then, $V$ is said to be well-behaved at infinity (WBI) of exponential order $N$.
\end{defn}

\begin{defn}[Definition 2 \cite{Foster:1998sk}]
\label{kWBI2}
A $C^k(\mathbb{R})$ function $V(\phi)$ is a class k WBI function, if it is WBI of exponential order $N$, and there are
 $\phi_0> 0$, and a coordinate transformation $\varphi=h(\phi)$ which maps the interval
$[\phi_0,\infty)$  onto $(0, \epsilon]$, where $\epsilon=h(\phi_0)$, satisfying $\lim_{\phi\rightarrow +\infty}h(\phi)=0$, and with the following additional
properties:
\begin{enumerate}
\item $h$ is $C^{k+1}$ and strictly decreasing.
\item The functions: 
\begin{equation}
 \bar{W}_V=\left\{\begin{array}{cc}
\frac{V'(h^{-1}(\varphi))}{V(h^{-1}(\varphi))}-N, & \varphi>0,\\
0, & \varphi=0 \end{array}\right.
\end{equation} and 
\begin{equation}
\bar{h'}(\varphi)=\left\{\begin{array}{cc}
h'(h^{-1}(\varphi)), & \varphi>0,\\
\lim_{\phi\rightarrow \infty} h'(\phi), & \varphi=0 \end{array}\right.
\end{equation} are $C^k$ on the closed interval $[0, \epsilon]$ and
\item \begin{equation}
\frac{d \bar{W}_V}{d \varphi}(0)=\frac{d \bar{h'}}{d \varphi}(0)=0.
\end{equation}
\end{enumerate}
\end{defn}
The last hypotheses are equivalent to:
$$\lim_{\varphi \to 0} \, \frac{W_{\chi}'\left(h^{-1}(\varphi )\right)}{h'\left(h^{-1}(\varphi )\right)} \; \text{and} \; \lim_{\varphi \to 0} \, \frac{h''\left(h^{-1}(\varphi )\right)}{h'\left(h^{-1}(\varphi )\right)}=0.$$
Assuming the $C^2$ functions $V(\phi)$ and  $\chi(\phi)$ are class 2 WBI,  the unconstrained dynamical system is obtained:
\begin{subequations}
\label{31-syst} 
\begin{align}
   & \frac{d T}{d{\tau}}= \frac{1}{2} (1-T)T \left(3 \gamma  \Omega_m+6 \sin ^2(\theta )+p \Omega_0\right),\\
   & \frac{d \theta}{d\tau}= -\frac{\sec (\theta )}{4 \sqrt{6}} \Bigg\{3 \sqrt{6} \sin (3 \theta )+6 \left(\cos (2 \theta ) - 2
   \Omega_0+1\right) (N+\bar{W}_V(\varphi)) \nonumber \\
   & +\Omega_m \left(-6 \sqrt{6} \gamma  \sin (\theta )-12 (2 M+N+\bar{W}_V(\varphi)+2 \bar{W}_{\chi}(\varphi))+18 \gamma  (M+\bar{W}_{\chi}(\varphi))\right)\nonumber \\
   & +\sqrt{6} \sin (\theta ) (3-2 p \Omega_0)\Bigg\},\end{align}
   \begin{align}
   & \frac{d \Omega_m}{d{\tau}}= -\frac{1}{2} \Omega_m \Bigg\{6 \cos (2 \theta )-\sqrt{6} (3 \gamma -4) \sin (\theta ) (M+\bar{W}_{\chi}(\varphi))\nonumber \\
   & -2 (3 \gamma (\Omega_m-1)+p \Omega_0+3)\Bigg\},\\
   & \frac{d \Omega_0}{d{\tau}}= - \Omega_0 (-3 \gamma  \Omega_m+3 \cos (2 \theta )-p \Omega_0+p-3),\\
   & \frac{d \varphi}{d{\tau}}=\sqrt{6} \bar{h'}(\varphi)\sin (\theta ),
\end{align}
\end{subequations}
defined on the phase space: 
\begin{align}
   & \Big\{(T, \theta, \Omega_m, \Omega_0, \varphi)\in\mathbb{R}^5: 0\leq T\leq 1, -\frac{\pi}{2}\leq \theta \leq \frac{\pi}{2}, 0\leq \varphi\leq h({\phi_0}), \nonumber\\
    & {2 T^2} V(h^{-1}(\varphi))= 3 m^2 (1-T)^2 (\cos (2 \theta )-2 \Omega_{0}-2 \Omega_{m}+1)\Big\}.
\end{align} $\theta$ is unique modulo $2\pi$. It has been chosen such that $\cos \theta\geq 0$. In the following list	$\tan^{-1}[x,y]$
gives the arc tangent of $y/x$, taking into account on which quadrant the point $(x,y)$ is in.  When $x^2+y^2=1$, $\tan^{-1}[x,y]$ gives the number $\theta$  such as $x=\cos\theta$ and $y=\sin\theta$.
\\
The equilibrium points of system \eqref{31-syst} with $\varphi=0$  (i.e., corresponding to $\phi\rightarrow \infty$) are the following:
\begin{enumerate}
    \item[$B_1$:] $\left(0,\tan ^{-1}\left[\sqrt{1-\frac{N^2}{6}},-\frac{N}{\sqrt{6}}\right]+2 \pi  c_1,0,0,0\right), c_1\in \mathbb{Z}$, represents a scalar field dominated solution satisfying $H\rightarrow \infty$. It is always a non--hyperbolic saddle. 
    
    \item[$B_2$:] $\left(1,\tan ^{-1}\left[\sqrt{1-\frac{N^2}{6}},-\frac{N}{\sqrt{6}}\right]+2 \pi  c_1,0,0,0\right), c_1\in \mathbb{Z}$, represents a scalar field dominated solution, satisfying $H\rightarrow 0$.  The case of physical interest is when $B_2$ is non--hyperbolic with a 4D stable manifold under one of the conditions (i)- (x) of \ref{AppB}. 
   
    \item[$B_3$:] $\left(0,2 \pi  c_1,0,1,0\right), c_1\in \mathbb{Z}$, with eigenvalues  $\left\{0,\frac{p-6}{2},\frac{p}{2},p,p-3 \gamma \right\}$. It represents a geometric ``fluid'' dominated solution with $H\rightarrow \infty$. The physical interesting situation is when $B_3$ is non--hyperbolic with a 4D unstable manifold for $p>6, 1\leq \gamma \leq 2$. It is a non--hyperbolic saddle otherwise. 
        
    \item[$B_4$:] $\left(1,2 \pi  c_1,0,1,0\right), c_1\in \mathbb{Z}$, represents a geometric ``fluid'' dominated solution with $H\rightarrow 0$. It is a non--hyperbolic saddle. 
    
    \item[$B_5$:] $\left(0,\tan ^{-1}\left[\sqrt{1-\frac{p^2}{6 N^2}},-\frac{p}{\sqrt{6} N}\right]+2 \pi  c_1,0,1-\frac{p}{N^2},0\right), c_1\in \mathbb{Z}$, represents a scaling solution where neither the energy density of the geometric ``fluid'', nor the energy density of the scalar field completely dominates, that satisfies $H\rightarrow \infty$. It is a non--hyperbolic saddle. 
    
    \item[$B_6$:] $\left(1,\tan ^{-1}\left[\sqrt{1-\frac{p^2}{6 N^2}},-\frac{p}{\sqrt{6} N}\right]+2 \pi  c_1,0,1-\frac{p}{N^2},0\right), c_1\in \mathbb{Z}$, represents a scaling solution where neither the energy density of geometric ``fluid'' nor the energy density of the scalar field completely dominates, that satisfies $H\rightarrow 0$. The situation of physical interest is when $B_6$ is non--hyperbolic with a 4D stable manifold under one of the conditions (i)- (xl) in \ref{AppB}. 
    
    \item[$B_7$:] $\scriptstyle \left(0,\tan ^{-1}\left[\sqrt{1-\frac{2 (p-3 \gamma )^2}{3 M^2(4-3 \gamma )^2}},\frac{2(p-3 \gamma)}{\sqrt{6}M(4 -3 \gamma)}\right]+2 \pi  c_1,\frac{2 (6-p) (p-3 \gamma )}{3 (4-3 \gamma )^2 M^2},\frac{(4-3 \gamma )^2
   M^2+2 (\gamma -2) (p-3 \gamma )}{(4-3 \gamma )^2 M^2},0\right), c_1\in \mathbb{Z}$, represents a matter- scalar field - geometric ``fluid'' scaling solution with $H\rightarrow \infty$. It is a non--hyperbolic saddle.

    \item[$B_8$:] $\scriptstyle\left(1,\tan ^{-1}\left[\sqrt{1-\frac{2 (p-3 \gamma )^2}{3 M^2(4-3 \gamma )^2}},\frac{2(p-3 \gamma)}{\sqrt{6}M(4 -3 \gamma)}\right]+2 \pi  c_1,\frac{2 (6-p) (p-3 \gamma )}{3 (4-3 \gamma )^2 M^2},\frac{(4-3 \gamma )^2
   M^2+2 (\gamma -2) (p-3 \gamma )}{(4-3 \gamma )^2 M^2},0\right), c_1\in \mathbb{Z}$, represents a matter- scalar field - geometric ``fluid'' scaling solution with $H\rightarrow 0$. The situation of physical interest is when $B_8$ is non--hyperbolic with a 4D stable manifold  under the conditions (i) - (xi) of \ref{AppB}. 
   
   \item[$B_9$:] $\left(0,\tan ^{-1}\left[ \frac{\sqrt{6 (2-\gamma )^2-(4-3 \gamma )^2 M^2}}{\sqrt{6}(2-\gamma )},\frac{(4-3 \gamma ) M}{\sqrt{6}(2-\gamma)}\right]+2 \pi  c_1,\frac{6 (2-\gamma )^2-(4-3 \gamma )^2 M^2}{6 (2-\gamma )^2},0,0\right), c_1\in \mathbb{Z}$, represents a matter- scalar field scaling solution with $H\rightarrow \infty$. It is a non--hyperbolic saddle. 
   
      \item[$B_{10}$:] $\left(1,\tan ^{-1}\left[ \frac{\sqrt{6 (2-\gamma )^2-(4-3 \gamma )^2 M^2}}{\sqrt{6}(2-\gamma )},\frac{(4-3 \gamma ) M}{\sqrt{6}(2-\gamma)}\right]+2 \pi  c_1,\frac{6 (2-\gamma )^2-(4-3 \gamma )^2 M^2}{6 (2-\gamma )^2},0,0\right), c_1\in \mathbb{Z}$, represents a matter- scalar field scaling solution with $H\rightarrow 0$.
  The situation of physical interest is when $B_{10}$ is non--hyperbolic with a 4D stable manifold for: 
   \begin{enumerate}
    \item $1\leq \gamma <\frac{4}{3}, \;  N<-\sqrt{6}, \;  \frac{N}{3 \gamma -4}-\sqrt{\frac{6 \gamma ^2-12 \gamma +N^2}{(3 \gamma -4)^2}}<M<\frac{\sqrt{6} \gamma -2 \sqrt{6}}{3 \gamma -4}, \;  p>\frac{6 \gamma ^2-12 \gamma -9 \gamma
   ^2 M^2+24 \gamma  M^2-16 M^2}{2 \gamma -4}$, or 
   \item $\frac{4}{3}<\gamma <2, \;  N<-\sqrt{6}, \;  \frac{\sqrt{6} \gamma -2 \sqrt{6}}{3 \gamma -4}<M<\sqrt{\frac{6 \gamma ^2-12 \gamma +N^2}{(3 \gamma -4)^2}}+\frac{N}{3
   \gamma -4}, \;  p>\frac{6 \gamma ^2-12 \gamma -9 \gamma ^2 M^2+24 \gamma  M^2-16 M^2}{2 \gamma -4}$, or 
   \item $1\leq \gamma <\frac{4}{3}, \;  N>\sqrt{6}, \;  \frac{2 \sqrt{6}-\sqrt{6} \gamma }{3 \gamma -4}<M<\sqrt{\frac{6
   \gamma ^2-12 \gamma +N^2}{(3 \gamma -4)^2}}+\frac{N}{3 \gamma -4}, \;  p>\frac{6 \gamma ^2-12 \gamma -9 \gamma ^2 M^2+24 \gamma  M^2-16 M^2}{2 \gamma -4}$, or 
   \item $\frac{4}{3}<\gamma <2, \;  N>\sqrt{6}, \;  \frac{N}{3
   \gamma -4}-\sqrt{\frac{6 \gamma ^2-12 \gamma +N^2}{(3 \gamma -4)^2}}<M<\frac{2 \sqrt{6}-\sqrt{6} \gamma }{3 \gamma -4}, \;  p>\frac{6 \gamma ^2-12 \gamma -9 \gamma ^2 M^2+24 \gamma  M^2-16 M^2}{2 \gamma -4}$.
   \end{enumerate}
     
      \item[$B_{11}$:] $\scriptstyle\left(0,\tan^{-1}\left[\sqrt{1-\frac{6 \gamma ^2}{\left(2 N+M (4-3 \gamma )\right)^2}},-\frac{\sqrt{6}
\gamma }{2 N+M (4-3 \gamma )}\right]+2 \pi  c_1, \frac{4 N (2 M+N)-6 (2+M N) \gamma }{(2 N+M (4-3 \gamma ))^2},0,0\right), c_1\in \mathbb{Z}$, represents a matter- scalar field scaling solution with $H\rightarrow \infty$.
It is a hyperbolic saddle. 
     \item[$B_{12}$:] $\scriptstyle\left(1,\tan^{-1}\left[\sqrt{1-\frac{6 \gamma ^2}{\left(2 N+M (4-3 \gamma )\right)^2}},-\frac{\sqrt{6}
\gamma }{2 N+M (4-3 \gamma )}\right]+2 \pi  c_1,
\frac{4 N (2 M+N)-6 (2+M N) \gamma }{(2 N+M (4-3 \gamma ))^2},0,0\right), c_1\in \mathbb{Z}$ represents a matter- scalar field scaling solution with $H\rightarrow 0$.
The situation of physical interest is when $B_{12}$ is non--hyperbolic with a 4D stable manifold under one of the conditions (i) - (lviii) in \ref{AppB}. 
\end{enumerate}

\subsection{Discussion}
\label{sect3.2.2}
Firstly, a dynamical system analysis of the system \eqref{Non_min2} for arbitrary $V(\phi)$ and arbitrary $\chi(\phi)$ formulated as system \eqref{systH1}, was provided. Secondly, to complement the analysis of system \eqref{systH1}, the dynamics of the system \eqref{Non_min2} at the limit $\phi\rightarrow +\infty$ formulated as system \eqref{31-syst}, was analyzed. In both cases,  the equilibrium points in the finite region of the phase space (local analysis) as well as in the limit $\phi\rightarrow +\infty$ (global analysis at infinity) were exhaustively examined; obtaining equilibrium points that represent some solutions of cosmological interest, such as several types of scaling solutions.  I.e., a kinetic dominated solution representing a stiff fluid, a solution dominated by an effective energy density of geometric origin, a quintessence scalar field dominated solution, the vacuum de Sitter solution associated to the minimum of the potential, and a non-interacting matter dominated solution. 

\section{Example: a scalar field model with potential  $V(\phi)= V_0 e^{-\lambda \phi}$ in vacuum}
\label{Section2.3}

In this section, a scalar field model with potential $V(\phi)= V_0 e^{-\lambda \phi}$ in vacuum for the flat FLRW metric is considered. That is, $N=-\lambda, M=0, W_{V}\equiv 0, W_{\chi}\equiv 0$, $\Omega_m\equiv 0, \Omega_0\equiv 0, \rho_m\equiv 0, G_0 \equiv 0$ and $\chi\equiv 1$. 

The  traditional formulation of the stability analysis of the equilibrium points in a cosmological setup uses Hubble- normalized equations, which is essentially based on the Copeland, Liddle \& Wands's approach \cite{Copeland:1997et}. It is well-known that this procedure is well-suited to investigate local stability features of the equilibrium points. 

Defining the variables: 
\begin{equation}\label{VARXY}
x= \frac{\dot \phi}{\sqrt{6}H}, \quad  y= \frac{\sqrt{V(\phi)}}{\sqrt{3}H}, 
\end{equation}
and $\tau= \ln a$, through $d \tau = H d t, H>0$, the following dynamical system is obtained:  
\begin{align}
&\frac{d x}{d \tau}= 3 x^3-3 x + \sqrt{\frac{3}{2}} \lambda  y^2,  \quad \frac{d y}{d \tau}=3 x y \left(x-\frac{\sqrt{6}}{6} \lambda \right).
\end{align}
This system can be reduced in one dimension using the relation $x^2+y^2=1$: 
\begin{equation}
\label{sistemauni}
\frac{d x}{d \tau}= f(x):=- 3 \left(1-x^2\right) \left(x-\frac{\sqrt{6}}{6} \lambda\right). 
\end{equation}
Equation \eqref{sistemauni} is integrable and leads to 
\begin{align}
& \tau:= \ln a= c_1+\ln\left[(1-x)^{\frac{1}{6-\sqrt{6} \lambda }} (x+1)^{\frac{1}{\sqrt{6} \lambda +6}} \left|6 x-\sqrt{6} \lambda \right|^{\frac{2}{\lambda ^2-6}}\right].
\end{align}
That is, the scaling factor $a$ can be expressed as a function of $x$ by 
\begin{align}\label{cuadratura}
a(x)=e^{c_1}(1-x)^{\frac{1}{6-\sqrt{6} \lambda }} (x+1)^{\frac{1}{\sqrt{6} \lambda +6}} \left|6 x-\sqrt{6} \lambda \right|^{\frac{2}{\lambda ^2-6}}, x\in[-1,1].
\end{align}
Additionally, 
\begin{equation}
\label{Phisistemauni}
\frac{d \phi}{d x}=\frac{\frac{d \phi}{d \tau}}{\frac{d x}{d \tau}}= -\frac{2\sqrt{6} x}{\left(1-x^2\right) \left(6 x -\sqrt{6} \lambda\right)}, 
\end{equation}
from which it follows by integration that: 
\begin{align}
\label{quadratura2}
 & \phi(x)=c_2+\ln \left[(1-x)^{\frac{1}{\sqrt{6}-\lambda }} (x+1)^{-\frac{1}{\lambda +\sqrt{6}}} \left|6 x-\sqrt{6} \lambda \right|^{\frac{2 \lambda
   }{\lambda ^2-6}}\right].
\end{align}
On the other hand, from \eqref{VARXY} with $V(\phi)= V_0 e^{-\lambda \phi}$ holds: 
\begin{equation}
\label{quadratura3}
H(x)=e^{-\frac{c_2 \lambda }{2}} \sqrt{\frac{{V_0}}{3-3 x^2}} (1-x)^{\frac{1}{\sqrt{6}-\lambda }} (x+1)^{-\frac{1}{\lambda +\sqrt{6}}} \left|6
   x-\sqrt{6} \lambda \right|^{\frac{2 \lambda }{\lambda ^2-6}}. 
\end{equation}
Finally, it follows that:
\begin{equation}
  \frac{dt}{dx}= \frac{2 \sqrt{3} e^{\frac{c_2 \lambda }{2}} (1-x)^{\frac{1}{\lambda -\sqrt{6}}} (x+1)^{\frac{1}{\lambda +\sqrt{6}}} \left|6 x-\sqrt{6} \lambda\right|^{-\frac{2 \lambda }{\lambda ^2-6}}}{\sqrt{V_0(1-x^2)} \left(6 x-\sqrt{6} \lambda \right)},
\end{equation}
\begin{landscape}
\begin{figure}[t]
\begin{center}
\subfigure[\label{T11}]{\includegraphics[scale=0.4]{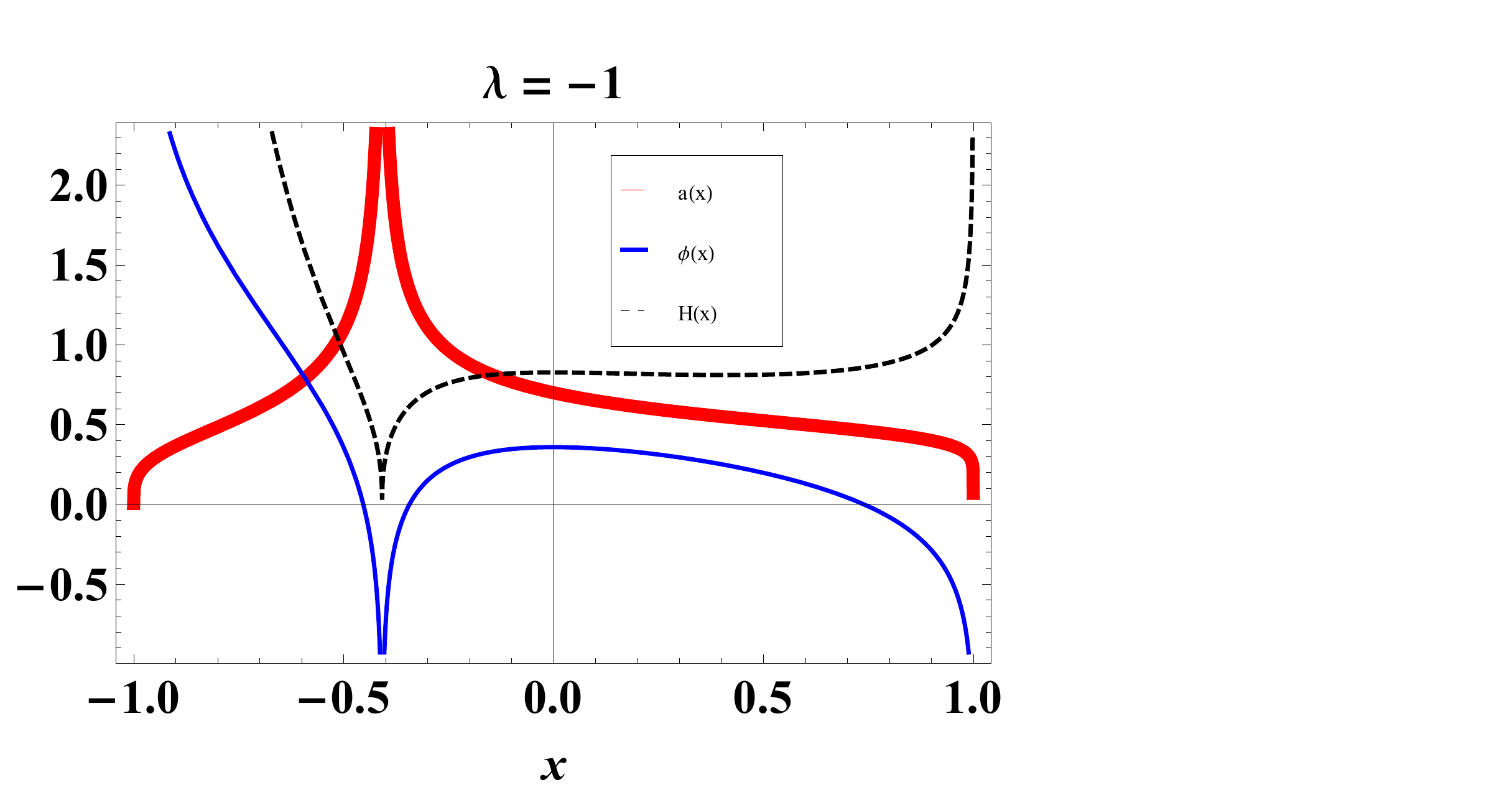} }
\subfigure[\label{T12}]{\includegraphics[scale=0.4]{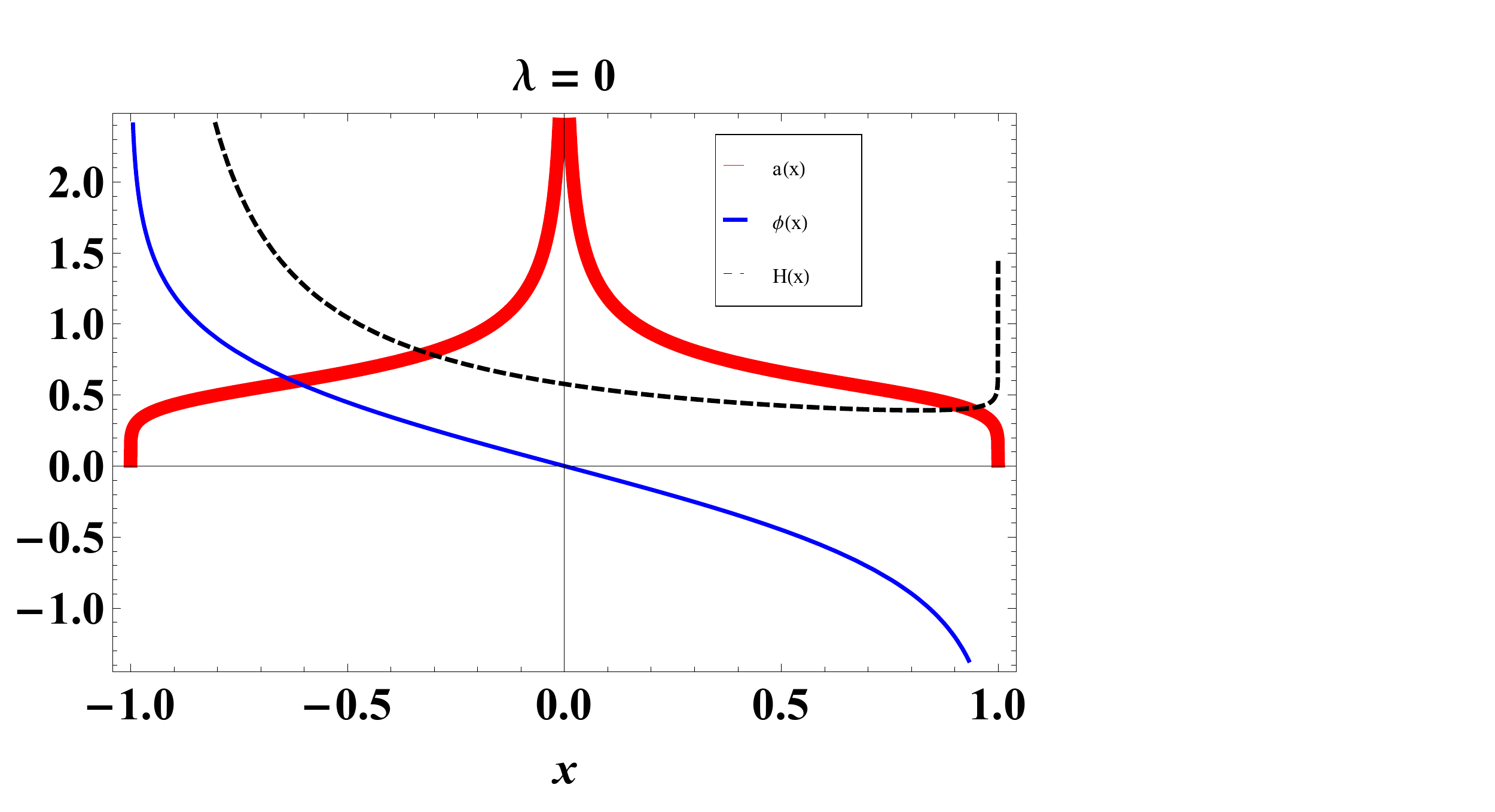}}
\subfigure[\label{T13}]{\includegraphics[scale=0.4]{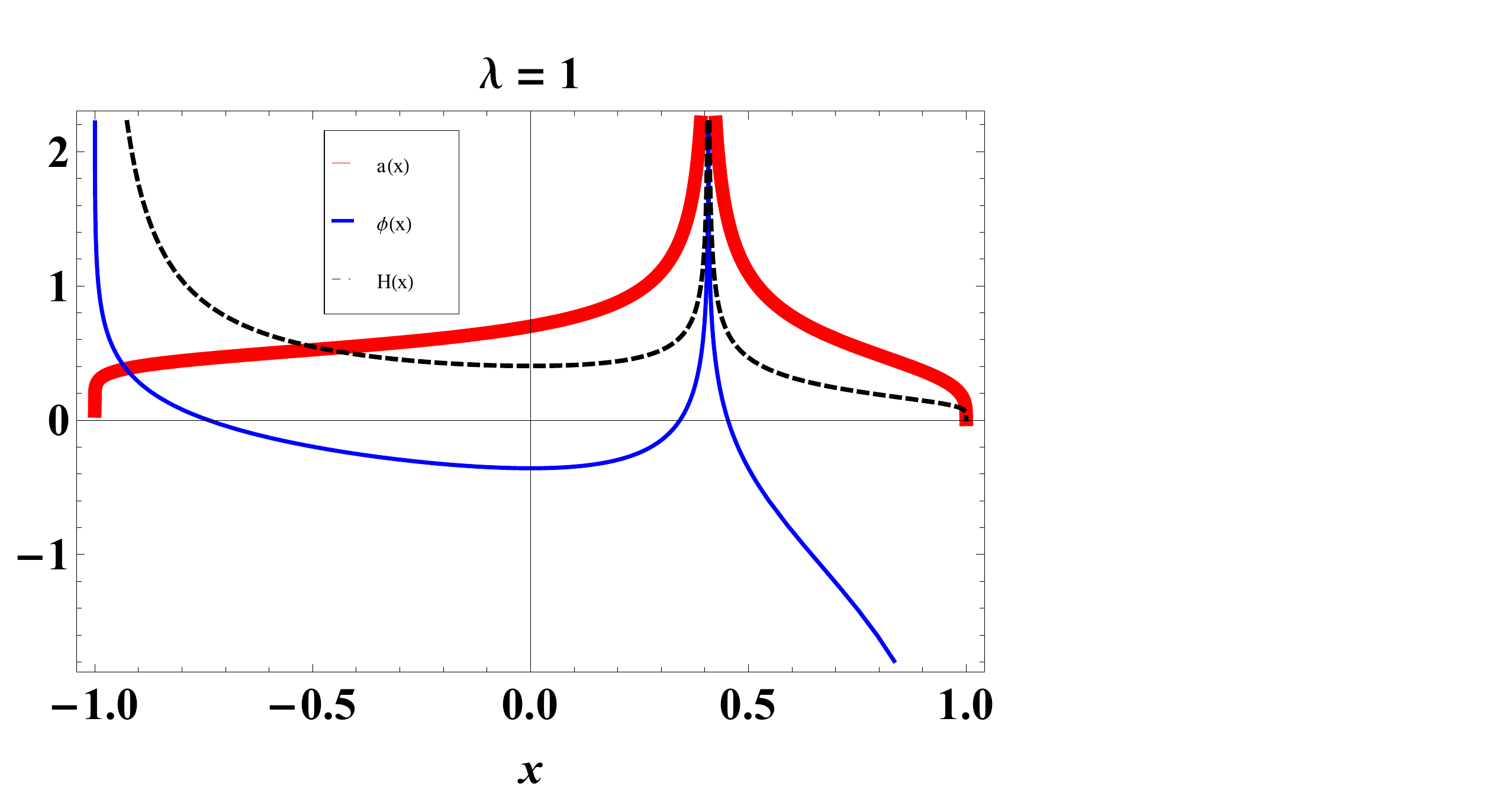} }
\subfigure[\label{T14}]{\includegraphics[scale=0.4]{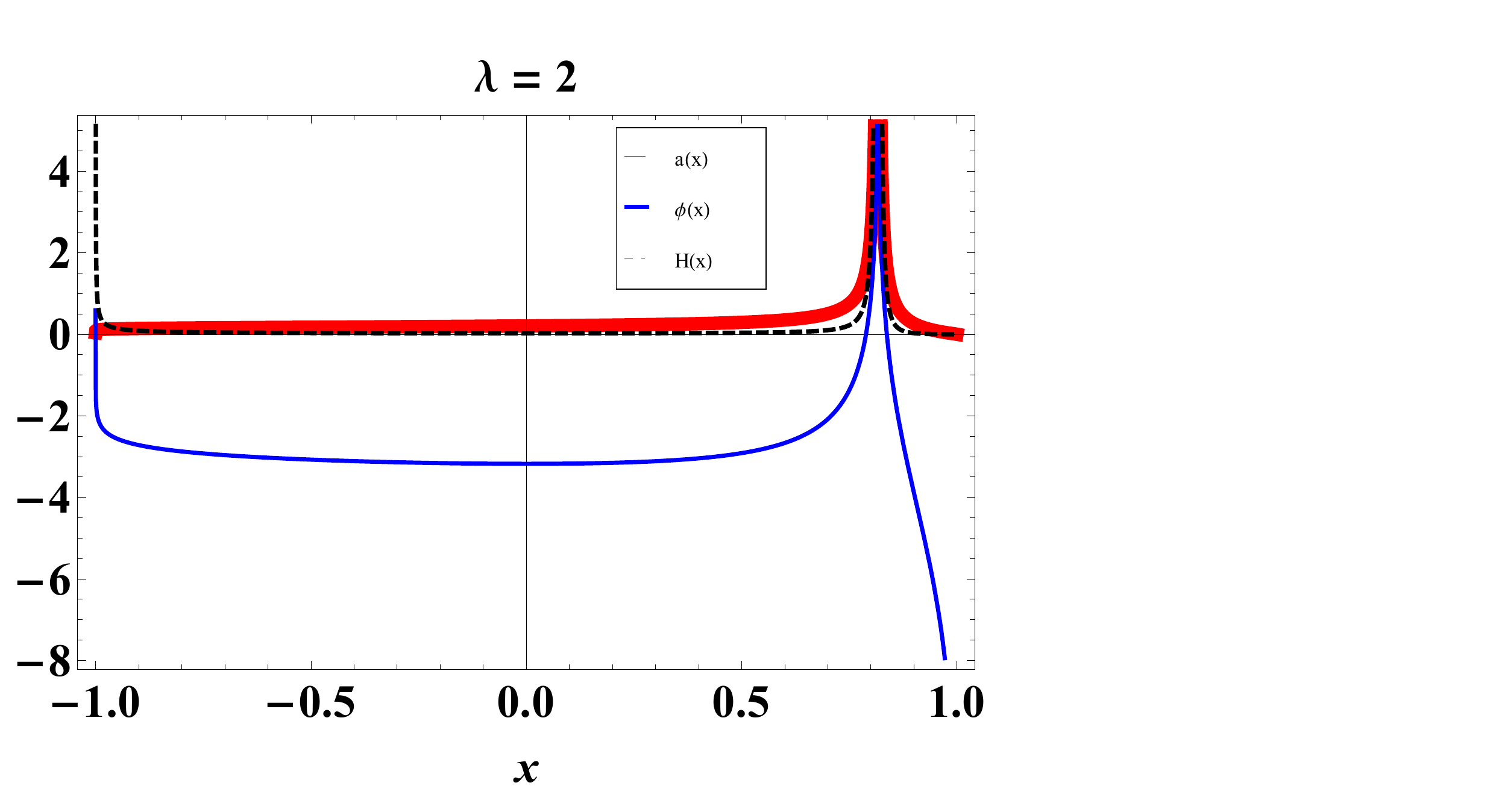}}
\end{center}\caption{\label{T11-14} $a(x)$ given by \eqref{cuadratura} with $c_1=0$, $\phi(x)$ given by \eqref{quadratura2} with  $c_2=0$ y $H(x)$ given by \eqref{quadratura3} with $c_2=0, V_0=1$, for several choices of $\lambda$.}
\end{figure}
\end{landscape}
\begin{equation}
\label{quadratura4}
   t= 2 \sqrt{3} e^{\frac{c_2 \lambda }{2}} \int \frac{(1-x)^{\frac{1}{\lambda -\sqrt{6}}} (x+1)^{\frac{1}{\lambda +\sqrt{6}}} \left| 6 x-\sqrt{6} \lambda\right| ^{-\frac{2 \lambda }{\lambda
   ^2-6}}}{\sqrt{V_0(1-x^2)} \left(6 x-\sqrt{6} \lambda \right)} \, dx.
\end{equation}
Figure \ref{T11-14} show $a(x)$, given by \eqref{cuadratura} for $c_1=0$, represented by the solid thick red line;  $\phi(x)$ given by \eqref{quadratura2} for $c_2=0$, represented by solid thinner blue line; and $H(x)$, given by \eqref{quadratura3} for $c_2=0, V_0=1$, represented by the dashed black line for several choices of $\lambda$. The cosmological time $t$ is related with   $x$ by the quadrature \eqref{quadratura4}. 
\\
Figure \ref{T11} shows: 
\begin{align}
    & \lim_{x\rightarrow -1}\left(a(x), \phi(x), H(x)\right)=(0, +\infty, +\infty),\\
    & \lim_{x\rightarrow 1}\left(a(x), \phi(x), H(x)\right)=(0,-\infty, +\infty), \\
		& \lim_{x\rightarrow \frac{\sqrt{6} \lambda}{6}}\left(a(x), \phi(x), H(x)\right)=(+\infty, -\infty, 0).
\end{align}
Figure \ref{T12} shows:
\begin{align}
    & \lim_{x\rightarrow -1}\left(a(x), \phi(x), H(x)\right)=(0,+\infty, +\infty),\\
    & \lim_{x\rightarrow 1}\left(a(x), \phi(x), H(x)\right)=(0,-\infty, +\infty), \\
		& \lim_{x\rightarrow 0}\left(a(x), \phi(x), H(x)\right)=\left(\infty, 0, \sqrt{\frac{V_0}{3}}\right).
\end{align}
Figure \ref{T13} shows:
\begin{align}
    & \lim_{x\rightarrow -1}\left(a(x), \phi(x), H(x)\right)=(0, +\infty, +\infty),\\
    & \lim_{x\rightarrow 1} \left(a(x), \phi(x), H(x)\right)=(0, -\infty, 0), \\
		& \lim_{x\rightarrow \frac{\sqrt{6} \lambda}{6}}\left(a(x), \phi(x), H(x)\right)=(+\infty, +\infty, +\infty).
		\end{align}
Figure   \ref{T14} shows:
\begin{align}
    & \lim_{x\rightarrow -1}\left(a(x), \phi(x), H(x)\right)=(0, 0, +\infty),\\
    & \lim_{x\rightarrow 1}\left(a(x), \phi(x), H(x)\right)=(0,-\infty, 0), \\
		& \lim_{x\rightarrow  \frac{\sqrt{6} \lambda}{6}}\left(a(x), \phi(x), H(x)\right)=\left(+\infty, +\infty, +\infty\right).
\end{align}
The above limits show that at the equilibrium points the scalar field eventually diverges and the Hubble parameter can be zero, infinity or it tends to a finite non-zero number.  But, when the scalar field is divergent or $H$ goes to zero, the Hubble--normalization fails. Therefore, although this procedure is well-suited to investigate local stability features of the equilibrium points, it does not provide a global description of the phase space when generically $\phi$ diverges or when $H\rightarrow 0$. In such case the method fails. For this reason, a new global systems analysis motivated by Alho \& Uggla's approach \cite{Alho:2014fha} is presented.

Using 
\begin{equation}
    T=\frac{m}{m+H}, \quad \theta=  \tan ^{-1} \left(\frac{\dot \phi}{\sqrt{2 V(\phi)}}\right),
\end{equation}
the following unconstrained dynamical system is deduced:
\begin{align}\label{syst3A}
&\frac{d T}{d{\tau}}=3 (1-T) T \sin ^2(\theta ), \quad 
\frac{d \theta}{d{\tau}}= \frac{1}{2} \cos (\theta ) \left(\sqrt{6} \lambda -6 \sin (\theta )\right),
\end{align}
defined in the finite cylinder $\mathbf{S}$ with boundaries $T=0$ and $T=1$.

\begin{table}[t]
\begin{center}
\begin{tabular}{|c|c|c|c|}
\hline
Label & $\left(T, \theta \right)$ & Existence &  Stability\\ \hline \hline
$P_1$ &  $\left(0, -\frac{\pi }{2} + 2 c_1 \pi \right)$ & $\forall \lambda$ & Saddle for $\lambda < -\sqrt{6}$. \\ 
&&& Non--hyperbolic for $\lambda = -\sqrt{6}$. \\ 
&&& Source for $\lambda > -\sqrt{6}$. \\ \hline
$P_2$ & $\left(0, \frac{\pi }{2} + 2 c_1 \pi \right)$ & $\forall \lambda$  & Saddle for $\lambda >\sqrt{6}$. \\
&&& Non--hyperbolic  for $\lambda =\sqrt{6}$.\\ 
&&& Source for $\lambda <\sqrt{6}$. \\ \hline 
$P_3$ &   $\left(0, \arcsin\left(\frac{\lambda }{\sqrt{6}}\right)\right)$ & $-\sqrt{6} \leq \lambda \leq \sqrt{6}$  & Non--hyperbolic  for  $\lambda \in \{-\sqrt{6},0, \sqrt{6}\}$. \\ 
&&& Saddle for  $-\sqrt{6}< \lambda <0$ or \\
&&& $0<\lambda <\sqrt{6}$. \\ \hline 
$P_4$ &  $\left(1,  -\frac{\pi }{2} + 2 c_1 \pi \right)$  & $\forall \lambda$ & Sink for $\lambda < -\sqrt{6}$. \\ 
&&& Non--hyperbolic  for $\lambda = -\sqrt{6}$. \\ 
&&& Saddle for $\lambda > -\sqrt{6}$. \\ \hline
$P_5$ &  $\left(1, \frac{\pi }{2}+ 2 c_1 \pi \right)$ & $\forall \lambda$  & Sink for $\lambda >\sqrt{6}$. \\
&&& Non--hyperbolic  for $\lambda =\sqrt{6}$.\\ 
&&& Saddle for $\lambda <\sqrt{6}$. \\ \hline 
$P_6$ &  $ \left( 1, \arcsin\left(\frac{\lambda }{\sqrt{6}}\right) \right)$ & $-\sqrt{6} \leq \lambda \leq \sqrt{6}$ & Non--hyperbolic  for $\lambda \in \{-\sqrt{6},0, \sqrt{6}\}$. \\ 
&&& Sink for $-\sqrt{6}< \lambda <0$ or \\
&&& $0<\lambda <\sqrt{6}$. \\ \hline 
\end{tabular}
\end{center}\caption{\label{critsyst3A} Existence conditions and stability conditions of the equilibrium points of equations  \eqref{syst3A}, $c_1\in \mathbb{Z}$.}
\end{table}

The variable $T$ is suitable for global analysis \cite{Alho:2014fha}, due to 
\begin{equation}
\label{3monotony}
\frac{d T}{d \tau}\Big|_{\sin \theta=0}=0, \quad \frac{d^2 T}{d {\tau}^2}\Big|_{\sin \theta=0}=0, \quad \quad \frac{d^3 T}{d {\tau}^3}\Big|_{\sin \theta=0}=9 \lambda ^2 (1-T) T.
\end{equation}
From the first equation of \eqref{syst3A} and equation \eqref{3monotony}, $T$ is a monotonically increasing function on $\mathbf{S}$. As a consequence, all orbits originate from the invariant subset $T=0$ (which contains the $\alpha$-limit), which is classically related to the initial singularity with $H \rightarrow \infty$, and ends on the invariant boundary subset $T=1$, which corresponds asymptotically to $H=0$. 

The equilibrium points of equations   \eqref{syst3A} are the following: 
\begin{enumerate}
\item  $P_1: (T,\theta)=\left(0, -\frac{\pi }{2} + 2 c_1 \pi \right),  c_1\in \mathbb{Z}$,  with eigenvalues $\left\{3,\sqrt{\frac{3}{2}} \lambda +3\right\}$. It represents a kinetic dominated solution with $H\rightarrow \infty$. The stability conditions of $P_1$ are the following:
        \begin{enumerate}
        \item Saddle for $\lambda < -\sqrt{6}$.
        \item Non--hyperbolic for $\lambda = -\sqrt{6}$.
        \item Source for $\lambda > -\sqrt{6}$.
        \end{enumerate} 
\begin{landscape}
 \begin{figure}[h]
\begin{center}
\includegraphics[width=8in]{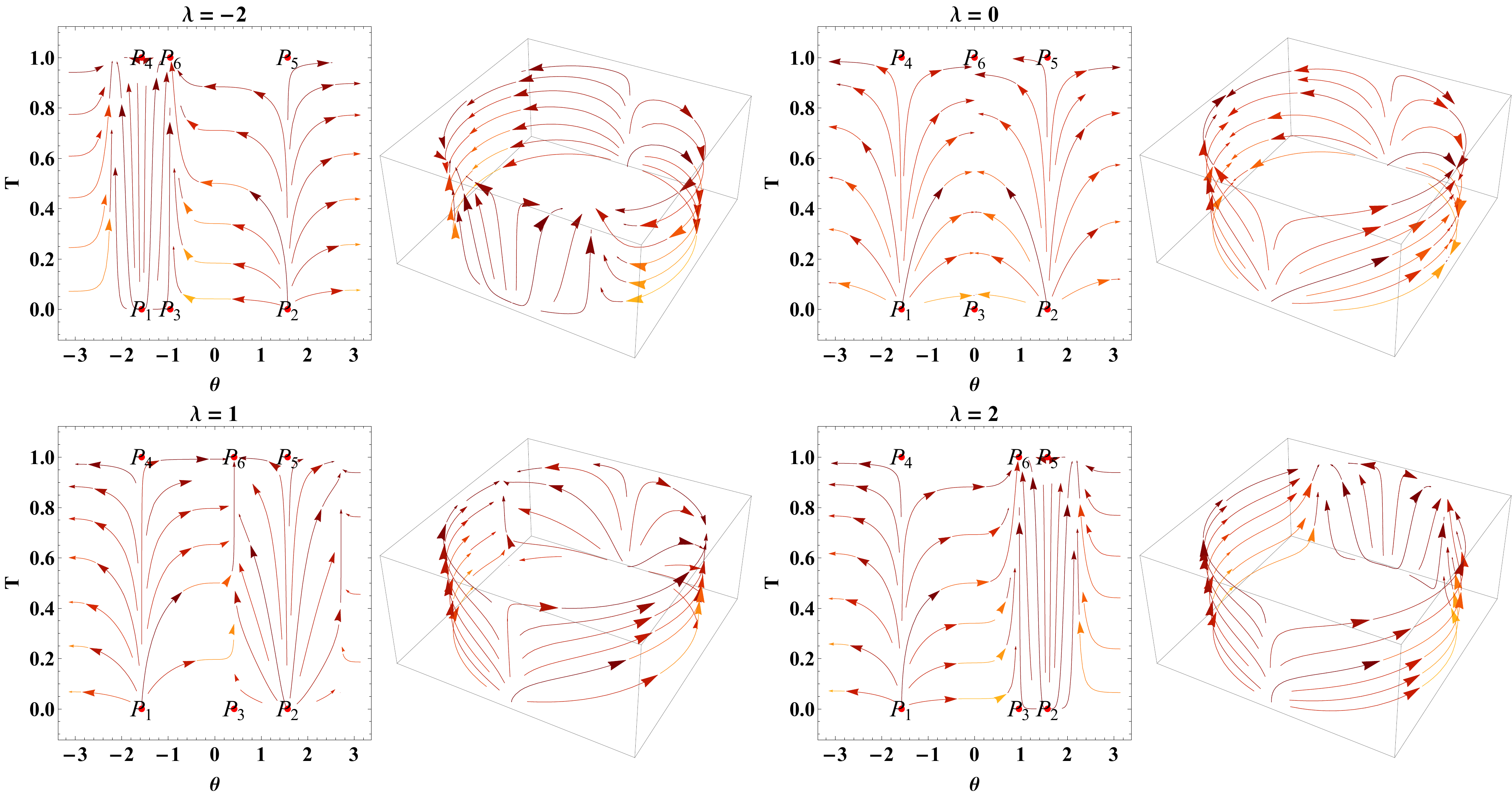}
\end{center}\caption{\label{Fig2.7} Unwrapped solution space (left panel) - Projection over the cylinder $\mathbf{S}$ (right panel) solution space of system   \eqref{syst3A}  for different values of $\lambda$.}
\end{figure}
\end{landscape}    
\item  $P_2: (T,\theta)= \left(0, \frac{\pi }{2}  + 2 c_1 \pi \right),  c_1\in \mathbb{Z}$  with eigenvalues $\left\{3,3-\sqrt{\frac{3}{2}} \lambda \right\}$. It represents a kinetic dominated solution with $H\rightarrow \infty$. The stability conditions of $P_2$ are the following: 
        \begin{enumerate}
        \item Source for $\lambda <\sqrt{6}$.
        \item Non--hyperbolic for $\lambda =\sqrt{6}$.
        \item Saddle for $\lambda >\sqrt{6}$.
        \end{enumerate}

 \item  $P_3: (T,\theta)=\left(0, \arcsin\left(\frac{\lambda }{\sqrt{6}}\right)\right)$ with eigenvalues $\left\{\frac{\lambda ^2}{2},\frac{1}{2} \left(\lambda
   ^2-6\right)\right\}$. It represents a scalar field dominated solution with $H\rightarrow \infty$. The stability conditions of $P_3$ are the following:
        \begin{enumerate} 
        \item Non--hyperbolic  for $\lambda \in \{-\sqrt{6},0, \sqrt{6}\}$.
        \item Saddle for  $-\sqrt{6}< \lambda <0$ or  $0<\lambda <\sqrt{6}$. 
        \end{enumerate}
\item  $P_4: (T,\theta)=\left(1,  -\frac{\pi }{2}  + 2 c_1 \pi \right),  c_1\in \mathbb{Z}$  with eigenvalues 
  $\left\{-3,\sqrt{\frac{3}{2}} \lambda +3\right\}$. It represents a kinetic dominated solution with $H\rightarrow 0$. The stability conditions of $P_4$ are the following: 
       \begin{enumerate}
       \item Sink for $\lambda < -\sqrt{6}$. 
       \item Non--hyperbolic  for $\lambda = -\sqrt{6}$. 
       \item Saddle for $\lambda > -\sqrt{6}$. 
       \end{enumerate}
  
\item  $P_5: (T,\theta)= \left(1, \frac{\pi }{2}  + 2 c_1 \pi \right),  c_1\in \mathbb{Z}$  with eigenvalues $\left\{-3,3-\sqrt{\frac{3}{2}} \lambda \right\}$.  It represents a kinetic dominated solution with $H\rightarrow 0$. The stability conditions of $P_5$ are the following:          \begin{enumerate}
       \item It is a sink for $\lambda >\sqrt{6}$.
       \item Non--hyperbolic  for $\lambda =\sqrt{6}$.
       \item Saddle for $\lambda <\sqrt{6}$. 
       \end{enumerate}
 
\item  $P_6: (T,\theta)=\left( 1, \arcsin\left(\frac{\lambda }{\sqrt{6}}\right)\right)$ with eigenvalues $\left\{-\frac{\lambda ^2}{2},\frac{1}{2} \left(\lambda
   ^2-6\right)\right\}$. It represents a scalar field dominated solution with $H\rightarrow 0$. The stability conditions of $P_6$ are the following:
        \begin{enumerate}
        \item Non--hyperbolic for $\lambda \in \{-\sqrt{6},0, \sqrt{6}\}$.
        \item Sink for $-\sqrt{6}< \lambda <0$ or  $0<\lambda <\sqrt{6}$.
        \end{enumerate}
\end{enumerate}

In table \ref{critsyst3A}, the existence conditions and stability conditions of the equilibrium points of equations  \eqref{syst3A} are summarized.

In the Figure \ref{Fig2.7},  some orbits of the flow of equations   \eqref{syst3A} (left panel)  and a projection over the cylinder $\mathbf{S}$ (right panel) for different values of $\lambda$ are presented.

\section{Example: a scalar field model with $E$-potential $V(\phi)= V_0\left(1-e^{-\sqrt{\frac{2}{3 \alpha}} \phi}\right)^{2 n}$ in vacuum}
\label{Section2.3B}
Considering the $E$--model with potential
$V(\phi)=V_0\left(1-e^{-\sqrt{\frac{2}{3 \alpha}} \phi}\right)^{2 n}$. This is a nonnegative potential with a single
minimum located at $(\phi,V(\phi))=(0,0)$. Therefore, the model admits a
Minkowski solution represented by the equilibrium point $(H,\dot\phi,
\phi)=(0,0,0)$. The potential has a plateau $V=V_0$, when $\phi\rightarrow
+\infty$, while $V\sim V_0 e^{-2 n \sqrt{\frac{2}{3 \alpha}} \phi}$ as $%
\phi\rightarrow -\infty$ \cite{Alho:2017opd}. At small $\phi$ the
$E$-potential behaves as $\phi^{2 n}$. 

Defining the Hubble normalized variables:
\begin{align}
x =\frac{\dot\phi}{\sqrt{6} H}, \quad Y=\left(\frac{V(\phi)}{3 H^2}\right)^{\frac{1}{2 n}}=\tilde{T}%
\left(1-e^{-\sqrt{\frac{2}{3 \alpha}} \phi}\right), \quad \tilde{T}=\left[\frac{V_0}{3 H^2}\right]^{\frac{1}{2n}},
\end{align}
the dynamical system: 
\begin{subequations}
\begin{align}
&\frac{dx}{dN}= 6 \mu Y^{2 n-1} (Y-\tilde{T})+(q-2) x, \\
&\frac{dY}{dN}=\frac{Y (-6 \mu x+q+1)+6 \mu x\tilde{T}}{n}, \\
&\frac{d\tilde{T}}{dN}=\frac{(q+1) \tilde{T}}{n},
\end{align}
where  $\mu = \frac{n}{3 \sqrt{\alpha}}$, and $q=2-3 Y^{2 n}$,  is studied.
\end{subequations}
This system has been extensively studied in \cite{Alho:2017opd} in the
context of a canonical scalar field cosmology, and it was extended in \cite{Leon:2019mbo} to Hořava–Lifshitz cosmology. Now,  the most
relevant features of the solution space according to \cite{Leon:2019mbo} will be discussed. It can be easily proven that $\tilde{T}
$ is monotonically increasing towards the future and decreasing towards the
past. The phase space is limited to the past by the invariant subset $\tilde{T}=0$  for $Y\leq 0$, and by $\tilde{T}-Y=0$  for $Y\geq 0$. The state space
is bounded when $\tilde{T}>0, \tilde{T}-Y>0$. The two past boundaries are intersected at the two
massless scalar field points $M_\pm=(\Sigma,Y)=(\pm 1, 0)$. The subset $\tilde{T}-Y=0$, on the other hand, is divided in two disconnected regions
separated by the de Sitter equilibrium point $dS=(\tilde{T},Y)=(1,1)$.

\begin{landscape}
\begin{figure}[t!]
\centering
\includegraphics[width=8in]{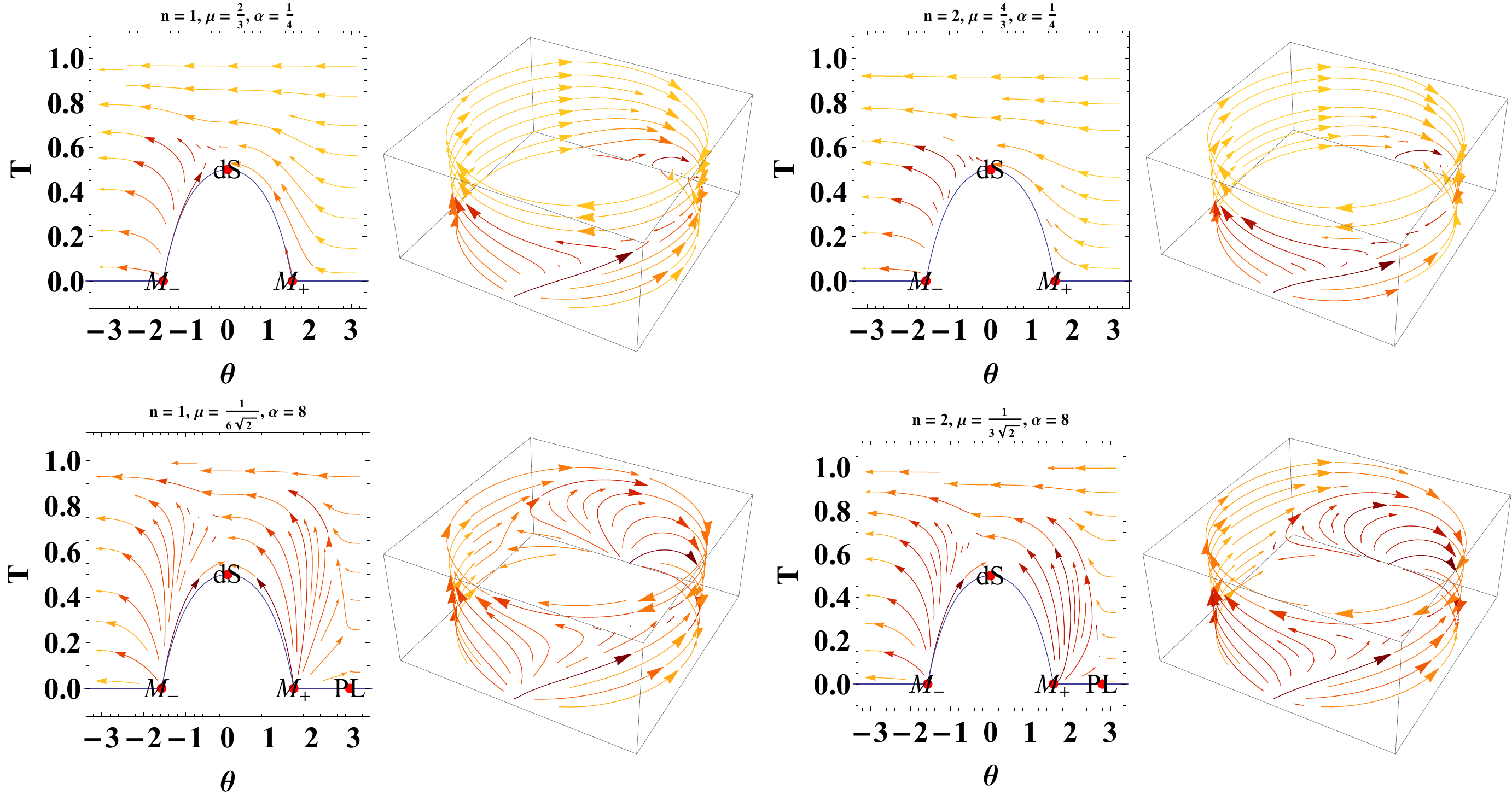} 
\caption{Unwrapped solution space (left panel).  Projection over the cylinder $\mathbf{S}$ (right panel)   of system \eqref{HLE-unwrappeda}, \eqref{HLE-unwrappedb} for some values of $n,\protect%
\mu,\protect\alpha$.}
\label{fig:Case1HLE}
\end{figure}
\end{landscape}

Introducing the complementary global transformation (which for $n=1$ is exactly the definition of $\theta$ in Section \ref{Section2.3}):
\begin{align}
&x=F(\theta ) \sin (\theta ), \quad Y=\cos (\theta ),  \quad F(\theta )=\sqrt{\frac{1-\cos ^{2 n}(\theta )}{1-\cos ^2(\theta )}}= \sqrt{{%
\sum_{k=0}^{n-1} \cos^{2k}(\theta)}},
\end{align}
the following regular unconstrained 2D dynamical system: 
\begin{subequations}
\label{HLE-unwrapped0}
\begin{align}
&\frac{d\theta}{dN}=-\frac{6 \mu F(\theta ) (\tilde{T}-\cos (\theta ))}{n}-%
\frac{3 F(\theta )^2 \sin (2\theta )}{2 n}, \label{HLE-unwrapped0a}\\
&\frac{d\tilde{T}}{dN}=\frac{3 \tilde{T} \left(1-\cos ^{2 n}(\theta )\right)%
}{n}, \label{HLE-unwrapped0b}
\end{align}
\end{subequations} is obtained, 
and the deceleration parameter becomes
\begin{equation}
q=2-3 \cos ^{2 n}(\theta ).
\end{equation}
In Figure \ref{fig:Case1HLE}, the unwrapped
solution space (left panel) and the projection over the cylinder $\mathbf{S}$ (right panel)   of the system \eqref{HLE-unwrappeda}, \eqref{HLE-unwrappedb} for some values of $n,\mu,\alpha$ are represented. This
plot clearly shows that the future boundary is $T=1$  which corresponds to $H=0$, and the final state is the Minkowski point given by a limit cycle.
Introducing the new compact variable $T=\frac{\tilde{T}}{1+\tilde{T}}$ and
the new time derivative $\frac{d{\bar{\tau}}}{d \ln a}=1+\tilde{T}=(1-T)^{-1}
$; the regular system: 
\begin{subequations}
\begin{align}
&\frac{d\theta}{d\bar{\tau}}=\frac{3 (T-1) F(\theta )^2 \sin (\theta ) \cos
(\theta )}{n}   -\frac{6 \mu F(\theta ) ((T-1) \cos (\theta )+T)}{n}, \label{HLE-unwrappeda} \\
&\frac{d T}{d\bar{\tau}}=-\frac{3 (T-1)^2 T \left(\cos ^{2 n}(\theta
)-1\right)}{n},\label{HLE-unwrappedb}
\end{align}
is obtained. 
\end{subequations}
The past boundary is attached to the phase-space, and in the new variables $(\theta, T)$  is defined by $\{T=0, \cos(\theta)\leq
0\}\cup\left\{T-(1-T)\cos\theta=0, \cos(\theta)>0\right\}$. It is also
included the future boundary $T=1$ which corresponds to $H=0$, and the final
state is the Minkowski point. 
The region  $\left\{T-(1-T)\cos\theta<0, \cos(\theta)>0\right\}$ is forbidden.
The equilibrium points of \eqref{HLE-unwrappeda}, \eqref{HLE-unwrappedb} are given by:
\begin{itemize}
\item[$M_{\pm}$:] $\tilde{T}=T=0; x=\pm 1, Y=0; \theta=\pm\frac{\pi}{2}%
+2 k \pi, k=0,1,2 \ldots$. They are massless scalar field solutions. They are saddle and source,
respectively, as it is confirmed in Figure \ref{fig:Case1HLE}.
\item[$dS$:] $\tilde{T}=1, T=\frac{1}{2}; x=0, Y=1; \theta=2 k \pi,
k=0,1,2 \ldots$. They are de Sitter solutions.
\item[$PL$:] $\tilde{T}=T=0; x=2\mu; Y=-(1-4\mu^2)^{\frac{1}{2n}};
\theta=\pm\arccos Y$. It exists for $\mu<1/2$, and corresponds to a powerlaw
selfsimilar solution for the exponential potential. 
\end{itemize}

\section{Example: a scalar-field cosmology with generalized harmonic potential $
V(\phi)= \mu^3 \left[\frac{\phi^2}{\mu} + b f \cos\left(\delta + \frac{\phi}{f}\right)\right]$, $b\neq 0$ in vacuum}
\label{Sect:2.4}

In this section, the asymptotic analysis as $\phi\rightarrow \infty$ of a scalar-field cosmology with generalized harmonic potential 
\begin{equation}
\label{harmonic1}
V(\phi)= \mu^3 \left[\frac{\phi^2}{\mu} + b f \cos\left(\delta + \frac{\phi}{f}\right)\right], \quad b\neq 0,
\end{equation} in a vacuum is presented for a flat FLRW model.  Let be $N=0, M=0, W_{\chi}\equiv 0$, $\Omega_m\equiv 0, \Omega_0\equiv 0, \rho_m\equiv 0, G_0 \equiv 0$ and $\chi\equiv 1$. \\
Furthermore, 
 \begin{equation}
\label{36-syst}
{W}_V(\phi)=
\frac{2 \phi -b \mu  \sin \left(\delta +\frac{\phi }{f}\right)}{b f \mu 
   \cos \left(\delta +\frac{\phi }{f}\right)+\phi ^2}.
\end{equation}

\subsection{Analysis as $\phi \rightarrow \infty$}
\label{Sect:2.4.1}

In this section  the dynamics as $\phi\rightarrow \infty$ of a scalar-field cosmology with generalized harmonic potential \eqref{harmonic1} in a vacuum is analyzed. 
\\
Defining the transformation:
\begin{equation}
\label{transform34}
\varphi=h(\phi)=\varphi=\left(\delta +\frac{\phi}{f}\right)^{-\frac{1}{4}},
\end{equation}
it follows that  $V(\phi)$ is 2 WBI with exponential order $N=0$.  
 \begin{equation}
\bar{W}_V(\varphi)=\left\{\begin{array}{cc}
\frac{-b \mu  \varphi^8 \sin \left(\frac{1}{\varphi^4}\right)-2 \delta  f \varphi^8+2 f \varphi^4}{f \left(b \mu  \varphi^8 \cos
   \left(\frac{1}{\varphi^4}\right)+f \left(\delta  \varphi^4-1\right)^2\right)}, & \varphi>0\\
0, & \varphi=0 \end{array}\right.,
\end{equation}
\begin{equation}
\bar{h'}(\varphi)=\left\{\begin{array}{cc}
-\frac{\varphi^5}{4 f}, & \varphi>0,\\
0,  & \varphi=0 \end{array}\right.,
\end{equation}
that satisfy the conditions (ii) and (iii) of Definition \ref{kWBI2}.
Hence,  the dynamical system: 
\begin{subequations}
\label{model1Alternative1-5syst3}
\begin{align}
& \frac{d T}{d {\tau}}=3 (1-T) T \sin ^2(\theta), \label{4syst3}\\
& \frac{d \theta}{d{\tau}}=-\frac{1}{2}  \cos (\theta ) \left(6 \sin (\theta )+\sqrt{6} \bar{W}_V(\varphi)\right),\\
& \frac{d \varphi}{d {\tau}}=\sqrt{6} \bar{h'}(\varphi) \sin (\theta ),
\end{align}
\end{subequations}
is obtained, where we have used the new time variable $\tau= \ln a$,  
defined on a phase space which consists of the vector product $\mathbf{S}\times J$ of the finite cylinder $\mathbf{S}$ with boundaries $T=0$ and $T=1$ with the interval 
$J=\left[0,  \left(\delta +\frac{\phi_0}{f}\right)^{-\frac{1}{4}}\right]$. 
The variable $T$ is suitable for global analysis \cite{Alho:2014fha}, due to:
\begin{align}
\label{4monotony}
& \frac{d T}{d \tau}\Big|_{\sin \theta=0}=0, \quad \frac{d^2 T}{d \tau^2}\Big|_{\sin \theta=0}=0, \nonumber \\
& \frac{d^3 T}{d \tau^3}\Big|_{\sin \theta=0}=\frac{9 (1-T) T v^8 \left(b \mu  v^4 \sin \left(\frac{1}{v^4}\right)+2 f \left(\delta  v^4-1\right)\right)^2}{f^2 \left(b \mu
    v^8 \cos \left(\frac{1}{v^4}\right)+f \left(\delta  v^4-1\right)^2\right)^2}. 
\end{align}
From equation \eqref{4syst3} and equation \eqref{4monotony}, $T$ is a monotonically increasing function on $\mathbf{S}\times J$. As a consequence, all orbits are originated from the invariant subset $T=0$ (which contains the $\alpha$-limit), which is classically related to the initial singularity with $H \rightarrow \infty$ and ends at the invariant boundary subset $T=1$, which corresponds to the asymptotically $H=0$. 
\\
\begin{figure}[h]
    \centering
    \includegraphics[scale=0.75]{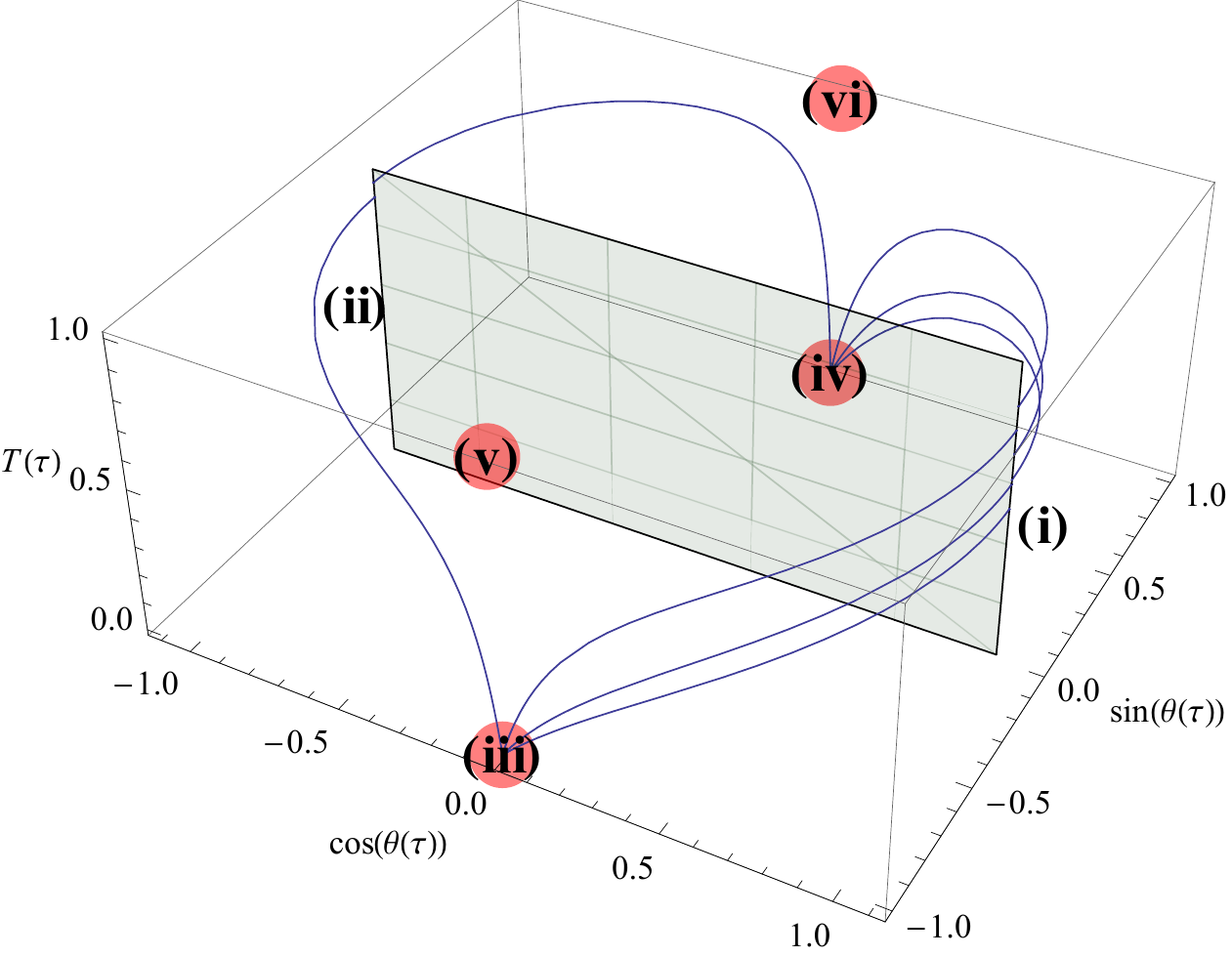}
    \caption{Evaluation of $(\cos \theta, \sin\theta, T)$ at some orbits of the system \eqref{model1Alternative1-5syst3} for $(b, f, \delta, \mu)= (0.1, 0.33, 0, 1)$. In the plot, the points (iii) and (iv) are sources; the points (v) and (vi) are saddles. The vertical plane represents the invariant set $\{(T,\theta, \varphi):\varphi=0, \sin(\theta)=0\}$. The vertical sides (i) and (ii) of this rectangle are the local sinks. }
    \label{fig:36}
\end{figure}
The curves of the equilibrium points of \eqref{model1Alternative1-5syst3} are the following:
\begin{enumerate} 
\item $({T}, \theta, \varphi)= \left(T_c, 2 n \pi, 0\right)$ with eigenvalues $\{-3,0,0\}$. It is non--hyperbolic.
\item $({T}, \theta, \varphi)= \left(T_c, \pi+ 2 n \pi, 0\right)$ with eigenvalues $\{-3,0,0\}$. It is non--hyperbolic.
\item $({T}, \theta, \varphi)= \left(0, -\frac{\pi}{2}+2 n \pi, 0\right)$ with eigenvalues $\{3,3,0\}$. It is non--hyperbolic.
\item $({T}, \theta, \varphi)= \left(0, \frac{\pi}{2}+2 n \pi, 0\right)$ with eigenvalues $\{3,3,0\}$. It is non--hyperbolic.
\item $({T}, \theta, \varphi)= \left(1, -\frac{\pi}{2}+2 n \pi, 0\right)$ with eigenvalues $\{-3,3,0\}$. It behaves as saddle. 
\item $({T}, \theta, \varphi)= \left(1, \frac{\pi}{2}+2 n \pi, 0\right)$ with eigenvalues $\{-3,3,0\}$. It behaves as saddle. 
\end{enumerate}
In Figure \ref{fig:36}, the functions $(\cos \theta, \sin\theta, T)$ are evaluated at some orbits of the system \eqref{model1Alternative1-5syst3} for $(b, f, \delta, \mu)= (0.1, 0.33, 0, 1)$. In the plot the points (iii) and (iv), which are sources, and the points (v) and (vi), which are saddles, are represented. The vertical plane represents the invariant set $\{(T,\theta, \varphi):\varphi=0, \sin(\theta)=0\}$. The vertical sides (i) and (ii) of this rectangle are the local sinks.

\subsection{Oscillating regime.}
\label{Sect:2.4.2}
In the reference \cite{Rendall:2006cq}, oscillating scalar field models with potential $\frac{1}{2}\phi^2$ and potentials $\frac{1}{2}\phi^2+W(\phi)$ with $W$ smooth and $W(\phi)=o(\phi^3)$ were studied.
Improved asymptotic expansions for the solution in homogeneous and isotropic spaces  were derived. Various generalizations  for non-linear massive scalar fields, $k$--essence models and $f(R)$-gravity were obtained. 
In this section the potential $
V(\phi)= \mu^3 \left[\frac{\phi^2}{\mu} + b f \cos\left(\delta + \frac{\phi}{f}\right)\right]$, $b\neq 0$ is investigated looking for oscillatory behavior, as expected in numerical investigations. Asymptotic expansions are derived as well.  Noticing that the cosine corrections are $O(b f \mu^3)$; therefore, they do not fall in the potential class studied by  \cite{Rendall:2006cq}. 

The pair
\begin{equation}
\left(\frac{\sqrt{2 \mu}\phi}{\sqrt{{\dot \phi}^2+2 \mu \phi^2}}, \quad \frac{\dot{\phi}}{\sqrt{{\dot{\phi}}^2+2 \mu \phi^2}}\right)
\end{equation}
defines a function that depends on the time variable $t$, with values in the unit circle. Hence, an angular function 
$\vartheta (t)$, that is unique under identification module $2\pi$, can be defined by 
\begin{equation}
\vartheta=  \tan ^{-1} \left(\frac{\dot \phi}{\sqrt{2 }\mu \phi}\right),
\end{equation}
together with the radial variable
\begin{equation}
r=\sqrt{{\dot \phi}^2+2 \mu^2 \phi^2},
\end{equation} 
which satisfy
\begin{equation}
-b f \mu ^3 \cos \left(\delta +\frac{r \cos (\vartheta )}{\sqrt{2} f \mu }\right)+3 H^2-\frac{r^2}{2}=0.
\end{equation}
The  inverse transformation $(r, \vartheta) \mapsto (\phi, \dot{\phi})$ is 
\begin{equation}
\phi = \frac{r \cos (\vartheta )}{\sqrt{2} \mu },\quad \dot\phi= r \sin (\vartheta ). 
\end{equation}

For expanding universes ($H>0$) the following holds: 
\begin{subequations}
\begin{align}
& \dot r=-\sqrt{\frac{3}{2}} r \sin ^2(\vartheta) \sqrt{2 b f \mu ^3 \cos \left(\delta +\frac{r \cos (\vartheta)}{\sqrt{2} f \mu
   }\right)+r^2} +  b \mu ^3 \sin (\vartheta) \sin \left(\delta +\frac{r \cos (\vartheta)}{\sqrt{2} f \mu }\right),\\
& \dot \vartheta=-\sqrt{2} \mu -\sqrt{\frac{3}{2}} \sin (\vartheta) \cos (\vartheta) \sqrt{2 b f \mu ^3 \cos \left(\delta +\frac{r \cos (\vartheta)}{\sqrt{2} f \mu }\right)+r^2} \nonumber \\
& +\frac{b \mu ^3 \cos (\vartheta ) \sin \left(\delta +\frac{r \cos (\vartheta
   )}{\sqrt{2} f \mu }\right)}{r}.
\end{align}
\end{subequations}
As $b\rightarrow 0$ the last equations reduce to: 
\begin{align}
&\dot r=-\sqrt{\frac{3}{2}} r^2 \sin ^2(\vartheta ), \quad
 \dot \vartheta=-\sqrt{2} \mu -\sqrt{\frac{3}{2}} r \sin (\vartheta ) \cos (\vartheta ).
\end{align}
The solutions of the limiting equations admit the asymptotic expansions \cite{Rendall:2006cq}
\begin{equation}
r(t)= \frac{4}{\sqrt{6} t}+ O(t^{-2}\ln t), \quad \vartheta(t)= -\sqrt{2} \mu  t + O(\ln t). \end{equation}
Hence, 
\begin{equation}
\phi(t)=\frac{4 \cos t}{\sqrt{6} t}+ O(t^{-2}\ln t), \quad \dot{\phi}(t)=\frac{4 \sin t}{\sqrt{6} t}+ O(t^{-2}\ln t).
\end{equation}
These expansions can be improved to the order $O(t^{-3}\ln t)$ as in \cite{Rendall:2006cq}. 
Here, asymptotic expansions of the full problem ($b\neq 0$) are derived using averaging techniques.
\\
Note that: 
\begin{subequations}
\label{syst71}
\begin{align}
& \dot r= b \mu ^3 \sin (\delta ) \sin (\vartheta )+\left(\frac{b \mu ^2 \cos (\delta ) \cos (\vartheta ) \sin (\vartheta )}{\sqrt{2} f}-\sqrt{3} \sqrt{b f \mu ^3 \cos (\delta )} \sin ^2(\vartheta )\right) r+O\left(r^2\right),\\
& \dot \vartheta +\sqrt{2} \mu = \frac{b \mu ^3 \sin (\delta ) \cos (\vartheta )}{r}+\left(\frac{b \mu ^2 \cos (\delta ) \cos ^2(\vartheta )}{\sqrt{2} f}-\sqrt{3} \sin (\vartheta ) \cos (\vartheta ) \sqrt{b f \mu ^3 \cos (\delta )}\right)\nonumber \\
& \frac{r \left(\sqrt{6} f \tan (\delta ) \sin (\vartheta ) \cos ^2(\vartheta ) \sqrt{b f \mu ^3 \cos (\delta )}-b \mu ^2 \sin (\delta ) \cos ^3(\vartheta )\right)}{4 f^2 \mu } +O\left(r^2\right).
\end{align}
\end{subequations}

\begin{figure}
\begin{center}
\subfigure[]{\includegraphics[scale=0.5]{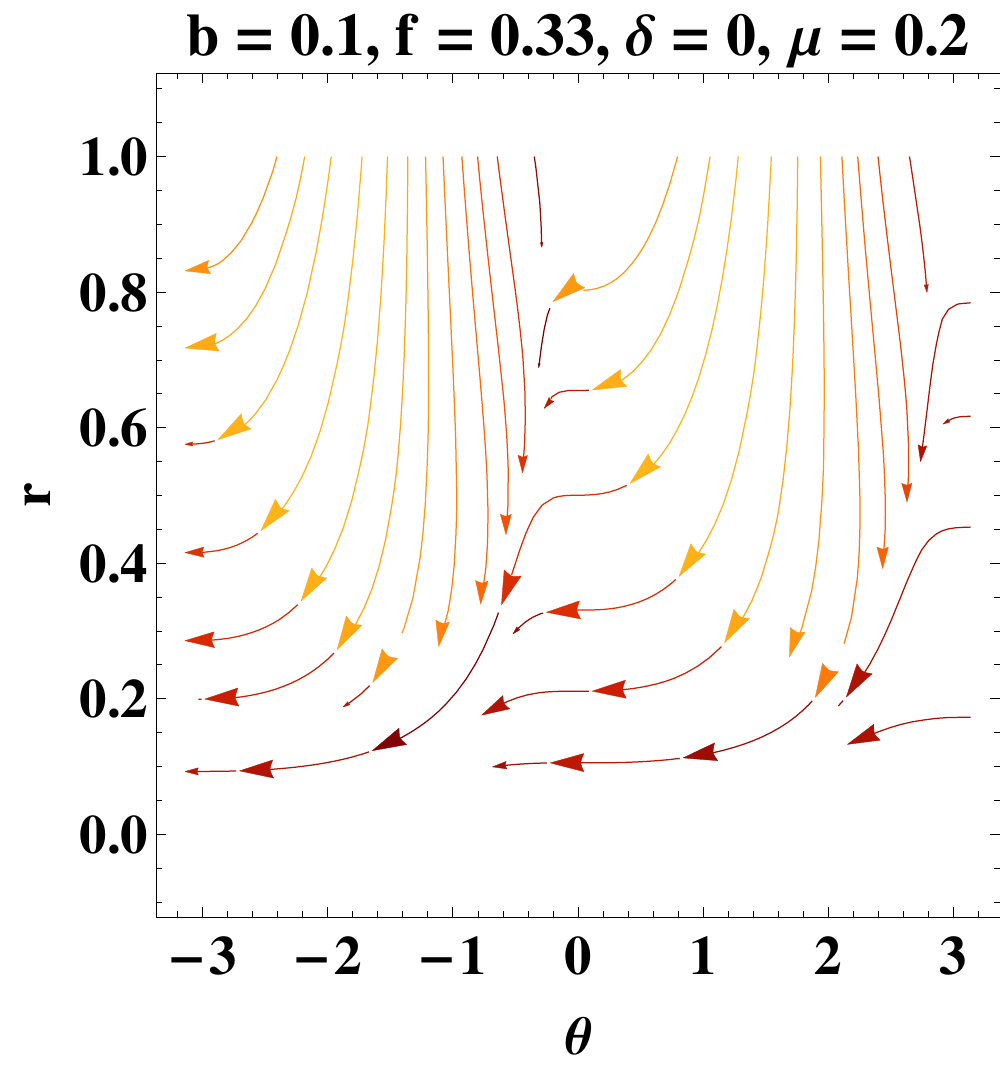} }\hspace{2cm}
\subfigure[]{\includegraphics[scale=0.55]{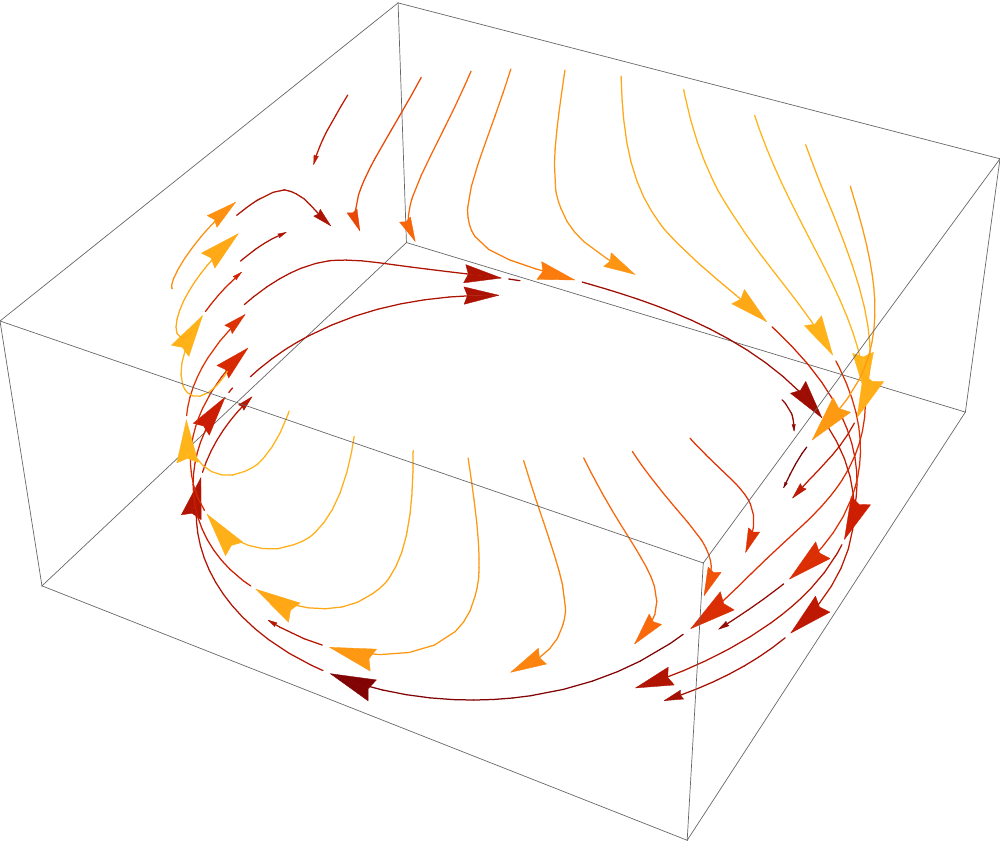}}
\subfigure[]{\includegraphics[scale=0.5]{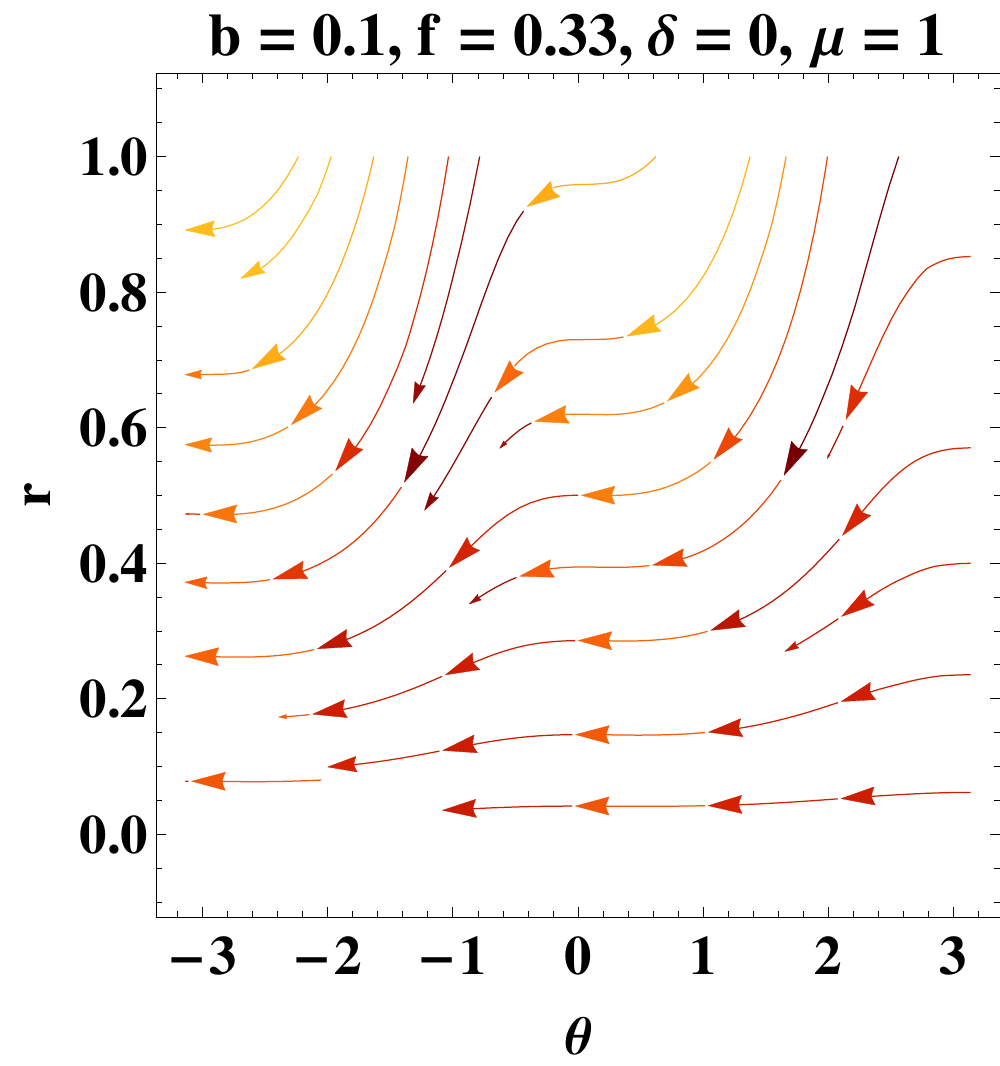} }\hspace{2cm}
\subfigure[]{\includegraphics[scale=0.55]{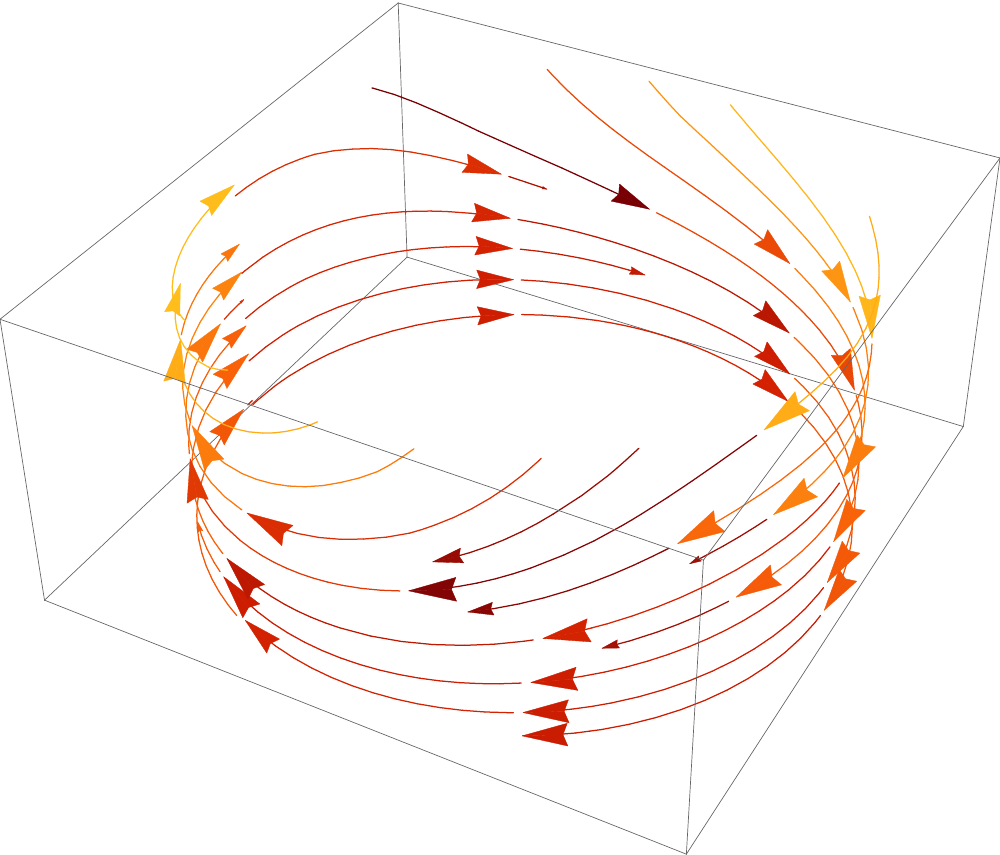}}
\end{center}\caption{\label{Fig6} Phase portrait of equations   \eqref{syst71} (left panel). Projection over the cylinder $\mathbf{S}$ (right panel) for $(b, f, \delta)= (0.1, 0.33, 0)$ and different values of $\mu$.}
\end{figure}

\begin{figure}
\begin{center}
\subfigure[]{\includegraphics[scale=0.5]{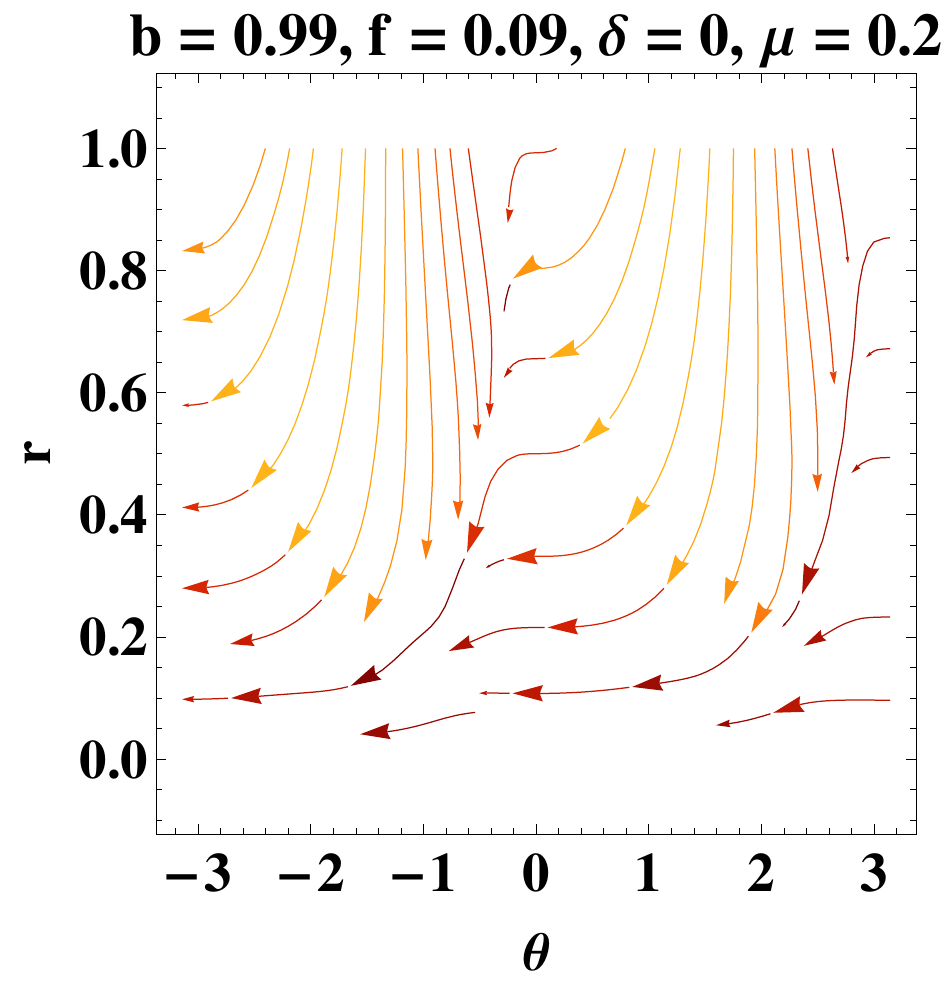} }\hspace{2cm}
\subfigure[]{\includegraphics[scale=0.55]{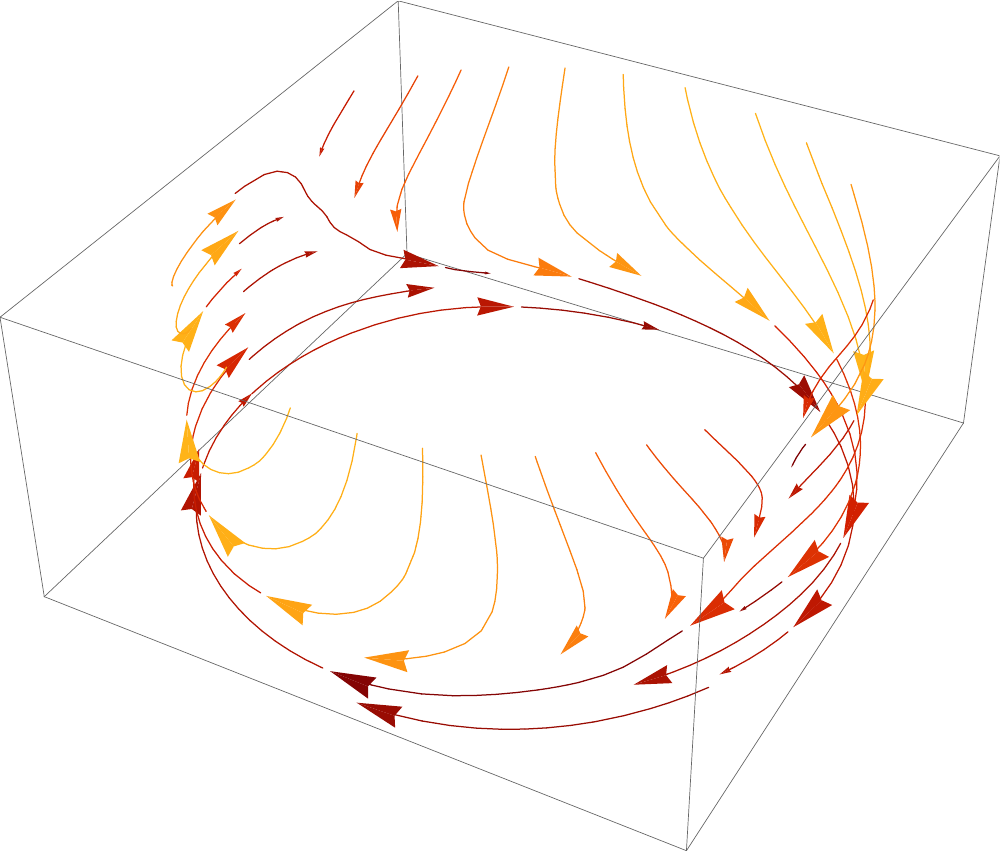}}
\subfigure[]{\includegraphics[scale=0.5]{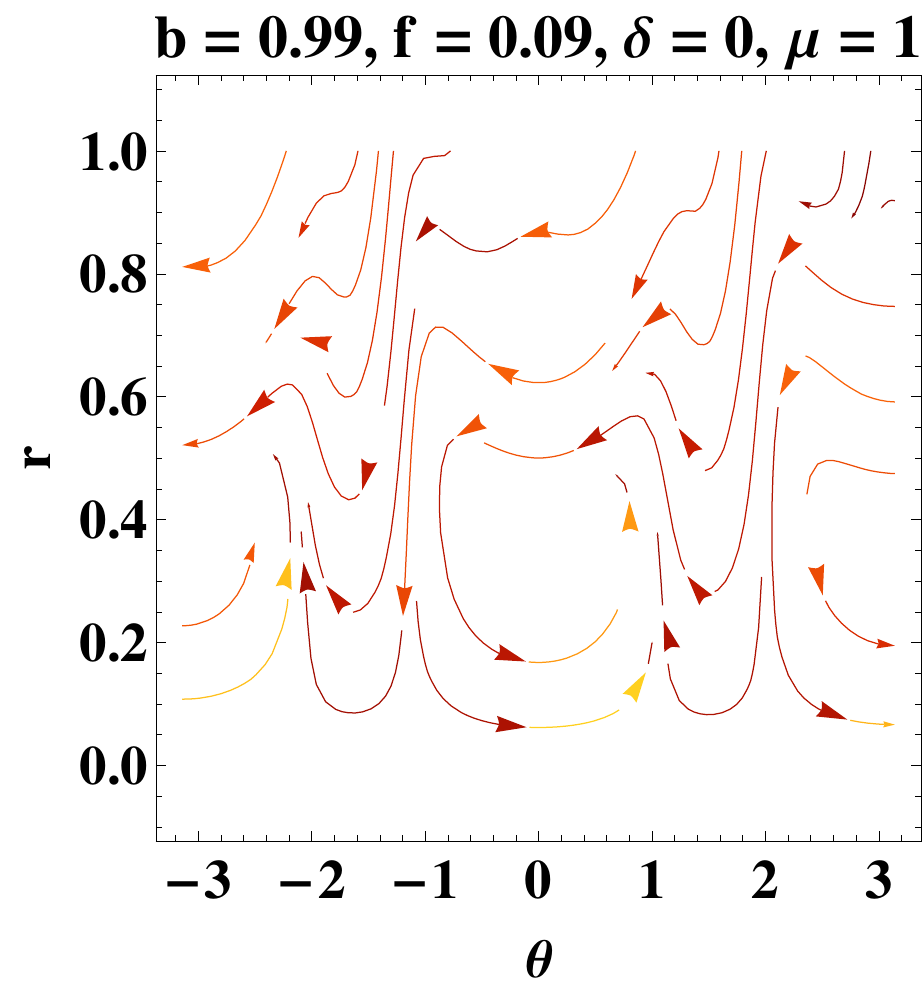} }\hspace{2cm}
\subfigure[]{\includegraphics[scale=0.55]{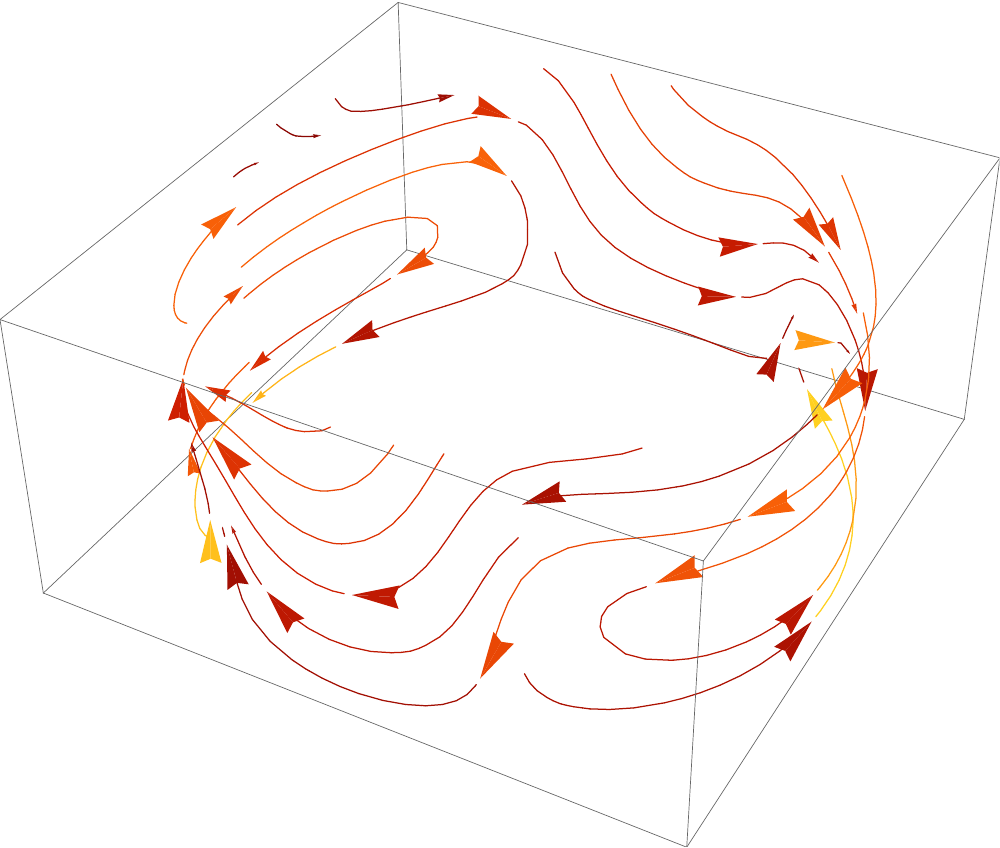}}
\end{center}\caption{\label{Fig7} Phase portrait of equations   \eqref{syst71} (left panel). Projection over the cylinder $\mathbf{S}$ (right panel) for $(b, f, \delta)= (0.99, 0.09, 0)$ and different values of $\mu$.}
\end{figure}
In order to obtain an approximated solution near the oscillatory regime, the average with respect to $\vartheta$ over any orbit of period $2 \pi$ given by 
\begin{equation}
\langle f \rangle = \frac{1}{2\pi}\int_{c}^{c +2 \pi} f(\vartheta) d \vartheta, \quad \cos(c)\geq 0,
\end{equation}
is applied to $f=(\dot r, \dot \vartheta)$, leading to 
\begin{equation}
\dot r= -\frac{1}{2} \sqrt{3} r \sqrt{b f \mu ^3 \cos (\delta )}, \quad \dot \vartheta=\frac{b \mu ^2 \cos (\delta )}{2 \sqrt{2} f}-\sqrt{2} \mu. 
\end{equation}
The averaged equations have solutions 
\begin{equation}
r(t)= r_0 e^{-\frac{1}{2} \sqrt{3} t \sqrt{b f \mu ^3 \cos (\delta )}}, \quad \vartheta (t)= \left(\frac{b \mu ^2 \cos (\delta )}{2 \sqrt{2} f}-\sqrt{2} \mu\right)t+\vartheta_0. 
\end{equation}
\newline 
Introducing along with $\vartheta$ and $r$, the new variable 
\begin{equation}
\varepsilon=\frac{H}{\mu+H}, 
\end{equation} 
with inverse
\begin{equation}
 H= \frac{\mu  \varepsilon}{1-\varepsilon},
\end{equation}
satisfying:
\begin{equation}
-b f \mu ^3 (\varepsilon -1)^2 \cos \left(\delta +\frac{r \cos (\vartheta )}{\sqrt{2} f \mu }\right)-\frac{1}{2} r^2 (\varepsilon -1)^2+3 \mu ^2 \varepsilon ^2=0,
\end{equation}
along with the time derivative  $\hat{\tau}$ given by 
\begin{equation}
\frac{d \hat{\tau}}{d t}:= \mu+ H,
\end{equation}
the following equations hold: 
\begin{subequations}
\begin{align}
& \varepsilon'=-\frac{r^2 (1-\varepsilon)^3 \sin ^2(\vartheta )}{2 \mu ^2},\\
& r'=-3 r \varepsilon  \sin ^2(\vartheta ) +b \mu ^2 (1-\varepsilon) \sin (\vartheta ) \sin \left(\delta +\frac{r \cos (\vartheta )}{\sqrt{2} f \mu }\right)\\
&\vartheta'=-\sqrt{2} (1-\varepsilon)-3 \varepsilon  \sin (\vartheta ) \cos (\vartheta )+\frac{b \mu ^2 (1-\varepsilon )
   \cos (\vartheta ) \sin \left(\delta +\frac{r \cos (\vartheta )}{\sqrt{2} f \mu }\right)}{r}.
\end{align}
\end{subequations}
To obtain an approximated solution near the oscillatory regime the average with respect to $\vartheta$ over any orbit of period $2 \pi$, 
$\langle (\varepsilon', r', \vartheta') \rangle$ is taken, leading to:
\begin{subequations}
\begin{align}
& \varepsilon'=\frac{r^2 (\varepsilon -1)^3}{4 \mu ^2},\\
& r'=-\frac{3 r \epsilon }{2},\\
& \vartheta'= \sqrt{2} (\varepsilon -1)-\Bigg\langle \frac{b \mu ^2 (\varepsilon -1) \cos (\vartheta ) \sin \left(\delta +\frac{r \cos (\vartheta )}{\sqrt{2} f \mu
   }\right)}{r}\Bigg\rangle. 
\end{align}
\end{subequations}
But, as $r\rightarrow 0$,
\begin{align}
& \frac{1}{2 \pi }\int_c^{c+2 \pi } \frac{b \mu ^2 (\varepsilon -1) \cos (\vartheta ) \sin \left(\delta +\frac{r \cos (\vartheta )}{\sqrt{2} f \mu
   }\right)}{r} \, d\vartheta\nonumber \\
	&  \sim \frac{b \mu  (\varepsilon -1) \cos (\delta )}{2 \sqrt{2} f}-\frac{r^2 (b (\varepsilon -1) \cos (\delta ))}{32 \left(\sqrt{2} f^3 \mu \right)}+O\left(r^3\right).
\end{align}
Finally, it follows that:
\begin{subequations}
\begin{align}
& \varepsilon'=-\frac{r^2 (1-\varepsilon)^3}{4 \mu ^2},\\
& r'=-\frac{3 r \varepsilon }{2},\\
& \vartheta'= \left[-\sqrt{2} + \frac{b \mu \cos (\delta )}{2 \sqrt{2} f}-\frac{r^2 (b \cos (\delta ))}{32 \left(\sqrt{2} f^3 \mu \right)}\right](1-\varepsilon), 
\end{align}
\end{subequations}
with the averaged constraint:
\begin{equation}
\mu ^2 \left(3 \varepsilon ^2-b f \mu  (1-\varepsilon)^2 \cos (\delta )\right)+\frac{r^2 (1-\varepsilon)^2 (b \mu  \cos (\delta )-4 f)}{8 f}=0,
\end{equation}
as $r\rightarrow 0$. 
\\
The above system is integrable yielding:
\begin{subequations}
\begin{align}
& r(\varepsilon )=\frac{\sqrt{2} \sqrt{c_1 (\varepsilon -1)^2+\mu ^2 (6 \varepsilon -3)}}{1-\varepsilon },\\
& \vartheta (\varepsilon )=\frac{\mu  \tanh ^{-1}\left(\frac{c_1 (\varepsilon -1)+3 \mu ^2}{\mu  \sqrt{9 \mu ^2-3 c_1}}\right) (b \mu 
   \cos (\delta )-4 f)}{f \sqrt{18 \mu ^2-6 c_1}}+\frac{b \mu  \cos (\delta )}{8 \sqrt{2} f^3 (1-\varepsilon )}+c_2, 
\end{align}
and 
\begin{align}
& 3(t-t_0)= \ln \left(\frac{(1-\varepsilon )^2 \left(\frac{3 \mu  \sqrt{3 \mu ^2-c_1}+\sqrt{3} c_1 \varepsilon -\sqrt{3} c_1+3 \sqrt{3} \mu ^2}{3 \mu  \sqrt{3 \mu ^2-c_1}-\sqrt{3} c_1 \varepsilon +\sqrt{3} c_1-3 \sqrt{3} \mu ^2}\right){}^{\frac{\mu
   }{\sqrt{\mu ^2-\frac{c_1}{3}}}}}{c_1 (\varepsilon -1)^2+\mu ^2 (6 \varepsilon -3)}\right) \nonumber \\
	& \sim \ln \left(\frac{\left(\frac{2 \mu  \left(\sqrt{9 \mu ^2-3 c_1}+3 \mu \right)-c_1}{c_1}\right){}^{\frac{\mu }{\sqrt{\mu ^2-\frac{c_1}{3}}}}}{c_1-3 \mu ^2}\right)-\frac{6 \mu ^2 \varepsilon }{c_1-3 \mu ^2}+O\left(\varepsilon ^2\right),
\end{align}
\end{subequations}
as $\varepsilon\rightarrow 0$. 

 The phase portrait of equations   \eqref{syst71} (left panel) and the projection over the cylinder $\mathbf{S}$ (right panel) for $(b, f, \delta)= (0.1, 0.33, 0)$ for different values of $\mu$ are presented in figure \ref{Fig6}. Figure \ref{Fig7} shows the phase portrait of equations  \eqref{syst71} (left panel) and the projection over the cylinder $\mathbf{S}$ (right panel) for $(b, f, \delta)= (0.99, 0.09, 0)$ and different values of $\mu$. The plots show the oscillatory behavior of the solutions.

\section{Example: a scalar-field cosmology with generalized harmonic potential $
V(\phi)= \mu ^3 \left[b f \left(\cos (\delta )-\cos \left(\delta +\frac{\phi }{f}\right)\right)+\frac{\phi ^2}{\mu}\right]
$, $b\neq 0$, in vacuum}
\label{Sect:2.5}

In this section  the asymptotic analysis as $\phi\rightarrow \infty$ of a scalar-field cosmology with generalized harmonic potential: 
\begin{equation}
\label{harmonic2}
V(\phi)= \mu ^3 \left[b f \left(\cos (\delta )-\cos \left(\delta +\frac{\phi }{f}\right)\right)+\frac{\phi ^2}{\mu}\right], \quad b\neq 0,
\end{equation} 
in a vacuum for a flat FLRW model is performed.  Let be $N=0, M=0, W_{\chi}\equiv 0$, $\Omega_m\equiv 0, \Omega_0\equiv 0, \rho_m\equiv 0, G_0 \equiv 0$ and $\chi\equiv 1$. \\
Furthermore,
 \begin{equation}
 \label{43-syst}
{W}_V(\phi)=\frac{b \mu  \sin \left(\delta +\frac{\phi }{f}\right)+2 \phi }{b f \mu \left(\cos (\delta )-\cos \left(\delta +\frac{\phi
   }{f}\right)\right)+\phi ^2}.
\end{equation}

\subsection{Analysis as $\phi \rightarrow \infty$}
\label{Sect:2.5.1}
In this section, the dynamics as $\phi\rightarrow \infty$ of a scalar-field cosmology with generalized harmonic potential \eqref{harmonic2} in a vacuum is analyzed. 
\\
Using again the transformation \eqref{transform34} and the new time variable $\tau= \ln a$, a system of the form \eqref{model1Alternative1-5syst3} holds with the definitions:
 \begin{equation}
\bar{W}_V(\varphi)=\left\{\begin{array}{cc}
\frac{b \mu  \varphi^8 \sin \left(\frac{1}{\varphi^4}\right)-2 \delta  f \varphi^8+2 f \varphi^4}{f \left(b \mu  \varphi^8 \left(\cos (\delta )-\cos
   \left(\frac{1}{\varphi^4}\right)\right)+f \left(\delta  \varphi^4-1\right)^2\right)}, & \varphi>0,\\
0, & \varphi=0 \end{array}\right.,
\end{equation}
\begin{equation}
\bar{h'}(\varphi)=\left\{\begin{array}{cc}
-\frac{\varphi^5}{4 f}, & \varphi>0,\\
0,  & \varphi=0 \end{array}\right.,
\end{equation}
that satisfy the conditions (ii) and (iii) of Definition \ref{kWBI2}.
The  phase space is the vector product $\mathbf{S}\times J$ of the finite cylinder $\mathbf{S}$ with boundaries $T=0$ and $T=1$ with the interval 
$J=\left[0,  \left(\delta +\frac{\phi_0}{f}\right)^{-\frac{1}{4}}\right]$. The variable $T$ is suitable for global analysis \cite{Alho:2014fha}, due to 
\begin{align}
\label{5monotony}
&\frac{d T}{d \tau}\Big|_{\sin \theta=0}=0, \quad \frac{d^2 T}{d \tau^2}\Big|_{\sin \theta=0}=0, \nonumber\\
& \frac{d^3 T}{d \tau^3}\Big|_{\sin \theta=0}= \frac{9 (1-T) T v^8 \left(b \mu  v^4 \sin \left(\frac{1}{v^4}\right)+2 f \left(\delta  v^4-1\right)\right)^2}{f^2 \left(b \mu
    v^8 \cos \left(\frac{1}{v^4}\right)+f \left(\delta  v^4-1\right)^2\right)^2}. 
\end{align}
As in section \ref{Sect:2.4.1},  $T$ is a monotonically increasing function on $\mathbf{S}\times J$. As a consequence, all orbits are originated from the invariant subset $T=0$ (which contains the $\alpha$-limit), which is classically related to the initial singularity with $H \rightarrow \infty$, and ends on the invariant boundary subset $T=1$, which corresponds to asymptotically $H=0$. 
\\
The curves of the equilibrium points are the same as in \eqref{model1Alternative1-5syst3} with essentially the same dynamics as in Figure \ref{fig:36}. That is, for $(b, f, \delta, \mu)= (0.1, 0.33, 0, 1)$  the points (iii) and (iv) are sources, and the points (v) and (vi)  are saddles. The vertical plane represents the invariant set $\{(T,\theta, \varphi):\varphi=0, \sin(\theta)=0\}$. The vertical sides (i) and (ii) of this rectangle are the local sinks.

\subsection{Oscillating regime.}
\label{Sect:2.5.2}

In this section, the potential 
$V(\phi)= \mu ^3 \left[b f \left(\cos (\delta )-\cos \left(\delta +\frac{\phi }{f}\right)\right)+\frac{\phi ^2}{\mu}\right]
$, $b\neq 0$ is analyzed; looking for oscillatory behavior as expected from the numerical investigations. Asymptotic expansions are derived as well. As before, the pair
\begin{equation}
\left(\frac{\sqrt{2 \mu}\phi}{\sqrt{{\dot \phi}^2+2 \mu \phi^2}}, \quad \frac{\dot{\phi}}{\sqrt{{\dot{\phi}}^2+2 \mu \phi^2}}\right),
\end{equation} defines a function of $t$ with values in the unit circle. Therefore, the angular function 
$\vartheta (t)$, which is unique under identification module $2\pi$, can be defined  by 
\begin{equation}
\vartheta=  \tan ^{-1} \left(\frac{\dot \phi}{\sqrt{2 }\mu \phi}\right),
\end{equation}
together with the radial function
\begin{equation}
r=\sqrt{{\dot \phi}^2+2 \mu^2 \phi^2},
\end{equation} 
which satisfy:
\begin{equation}
b f \mu ^3 \left(\cos \left(\delta +\frac{r \cos (\vartheta )}{\sqrt{2} f \mu }\right)-\cos (\delta )\right)+3 H^2-\frac{r^2}{2}=0.
\end{equation}
The  inverse transformation $(r, \vartheta) \mapsto (\phi, \dot{\phi})$ is 
\begin{equation}
\phi = \frac{r \cos (\vartheta )}{\sqrt{2} \mu },\quad \dot\phi= r \sin (\vartheta ). 
\end{equation}
For expanding universes ($H>0$) the equations hold:
\begin{subequations}
\label{syst104}
\begin{align}
& \dot r=-\sqrt{\frac{3}{2}} r \sin ^2(\vartheta ) \sqrt{2 b f \mu ^3 \left(\cos (\delta )-\cos \left(\delta +\frac{r \cos (\vartheta )}{\sqrt{2} f \mu }\right)\right)+r^2}\nonumber \\
& -b \mu ^3 \sin (\vartheta ) \sin \left(\delta +\frac{r \cos (\vartheta
   )}{\sqrt{2} f \mu }\right),\\
& \dot \vartheta=-\sqrt{2} \mu -\sqrt{\frac{3}{2}} \sin (\vartheta ) \cos (\vartheta ) \sqrt{2 b f \mu ^3 \left(\cos (\delta )-\cos \left(\delta +\frac{r \cos (\vartheta )}{\sqrt{2} f \mu }\right)\right)+r^2}\nonumber \\
& -\frac{b \mu ^3 \cos (\vartheta )  \sin \left(\delta +\frac{r \cos (\vartheta )}{\sqrt{2} f \mu }\right)}{r}.
\end{align}
\end{subequations}
In the limit $b\rightarrow 0$  the solutions of the limiting equation admit the asymptotic expansions \cite{Rendall:2006cq}:
\begin{equation}
\vartheta(t)= -\sqrt{2} \mu  t + O(\ln t), \quad r(t)= \frac{4}{\sqrt{6} t}+ O(t^{-2}\ln t). \end{equation}
Hence, when $b=0$: 
\begin{equation}
\phi(t)=\frac{4 \cos t}{\sqrt{6} t}+ O(t^{-2}\ln t), \quad \dot{\phi}(t)=\frac{4 \sin t}{\sqrt{6} t}+ O(t^{-2}\ln t).
\end{equation}
\begin{figure}
\begin{center}
\subfigure[]{\includegraphics[scale=0.5]{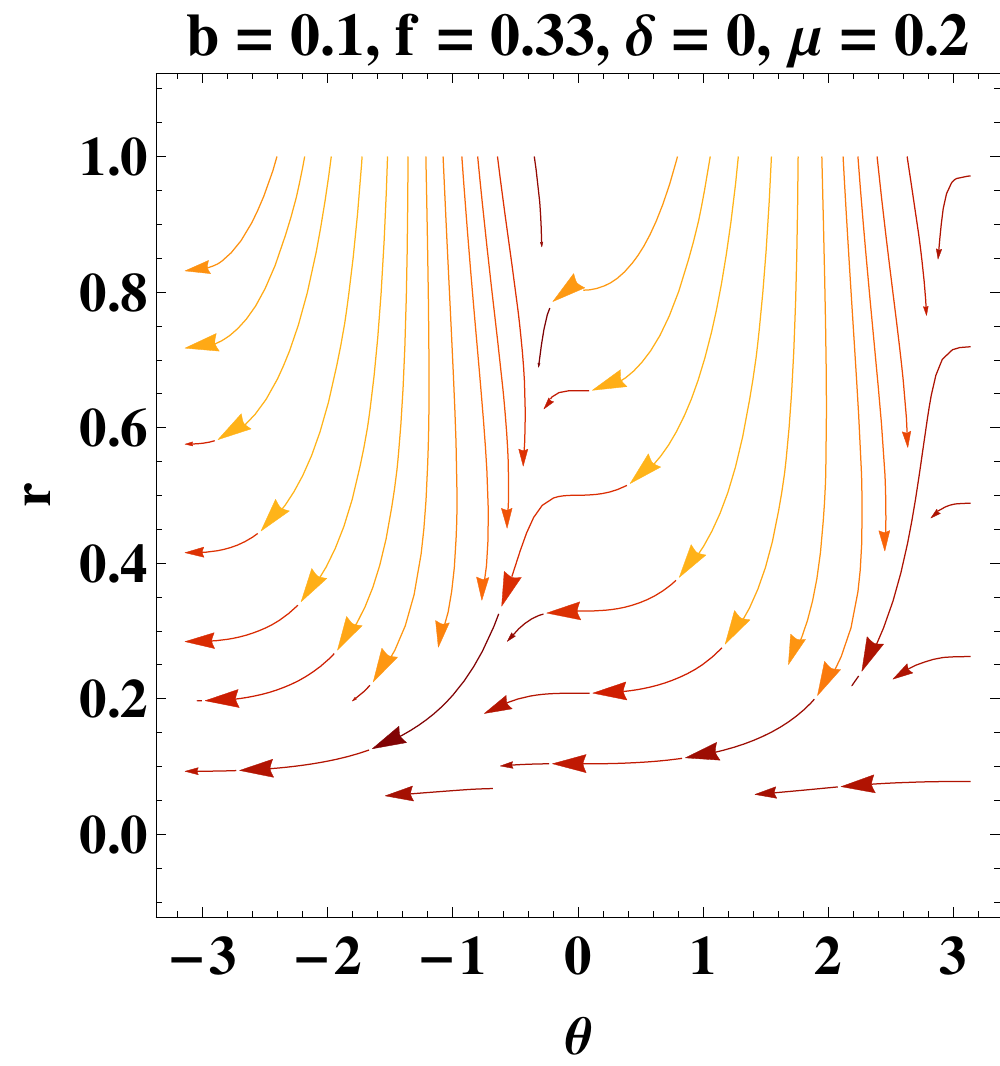} }\hspace{2cm}
\subfigure[]{\includegraphics[scale=0.55]{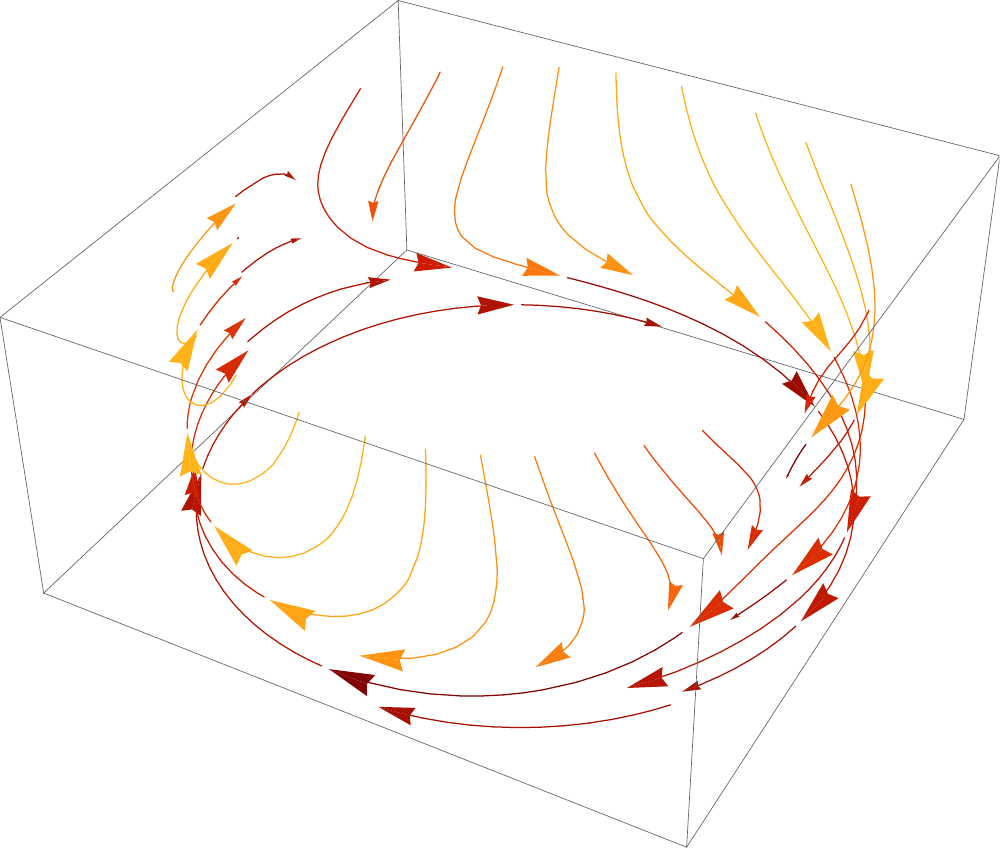}}
\subfigure[]{\includegraphics[scale=0.5]{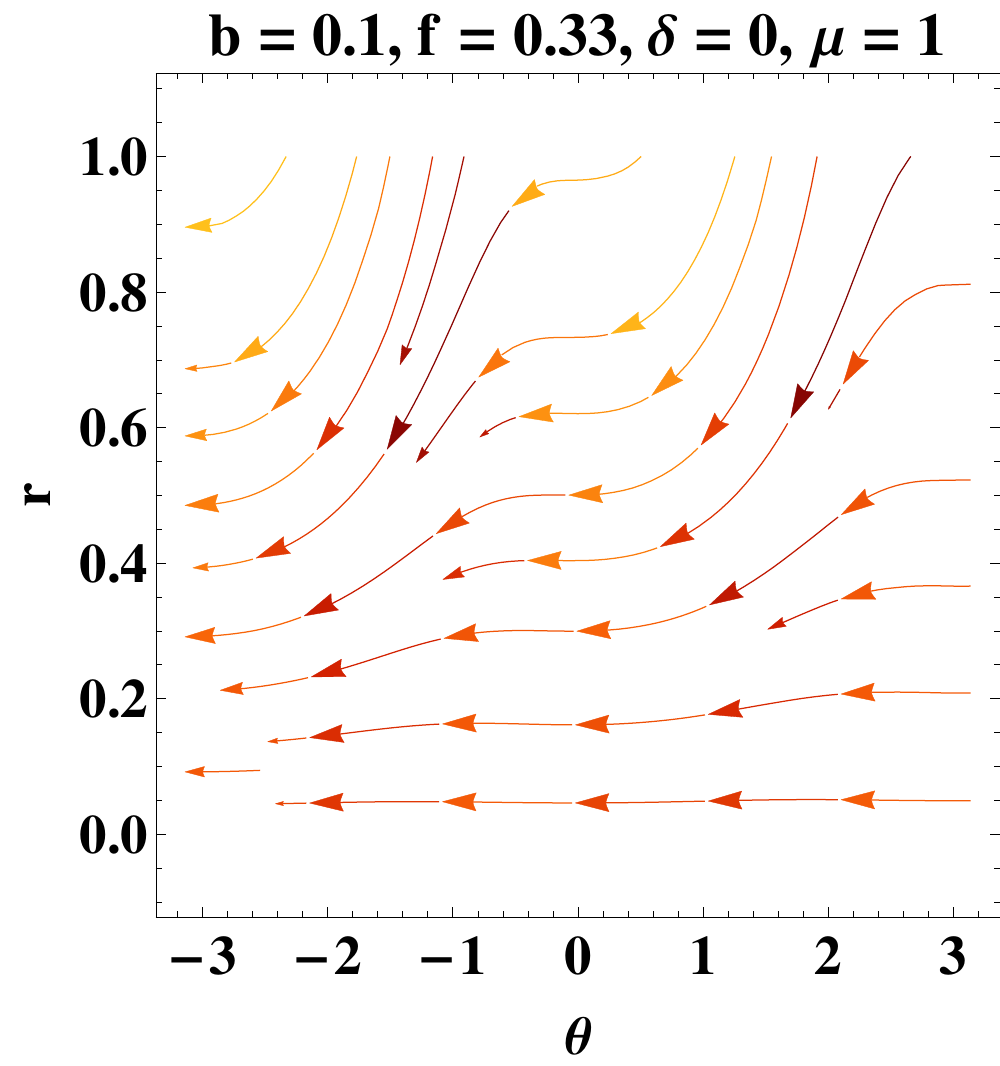} }\hspace{2cm}
\subfigure[]{\includegraphics[scale=0.55]{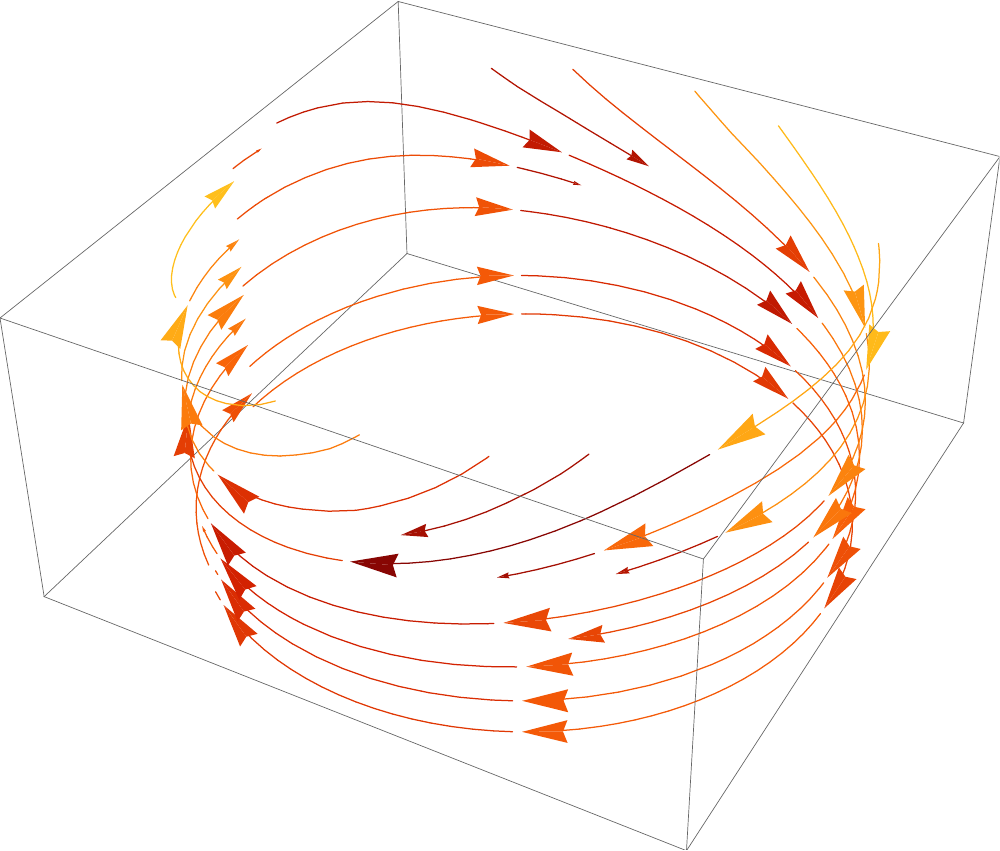}}
\end{center}\caption{\label{Fig10} Phase portrait of equations   \eqref{syst104} (left panel). Projection over the cylinder $\mathbf{S}$ (right panel) for $(b, f, \delta)= (0.1, 0.33, 0)$ and different values of $\mu$.}
\end{figure}

\begin{figure}
\begin{center}
\subfigure[]{\includegraphics[scale=0.5]{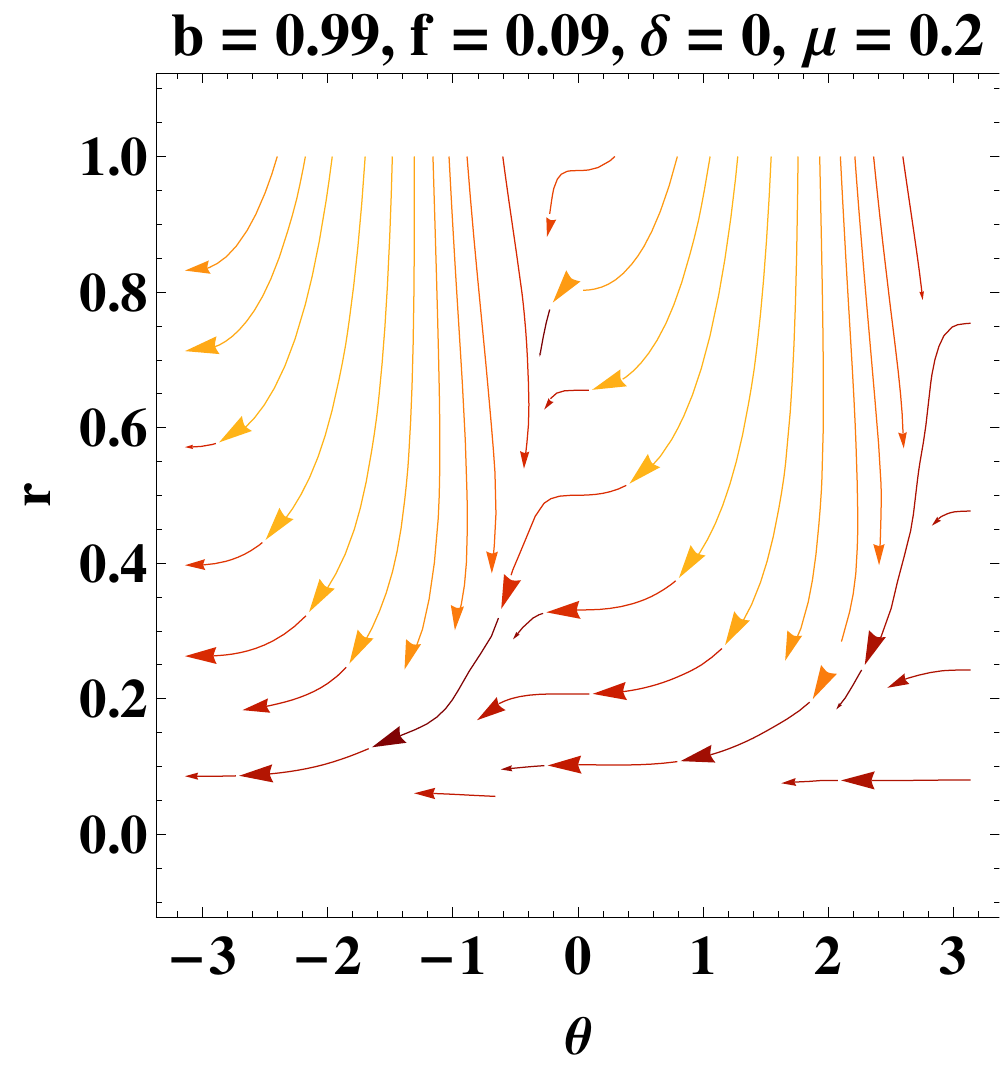} }\hspace{2cm}
\subfigure[]{\includegraphics[scale=0.55]{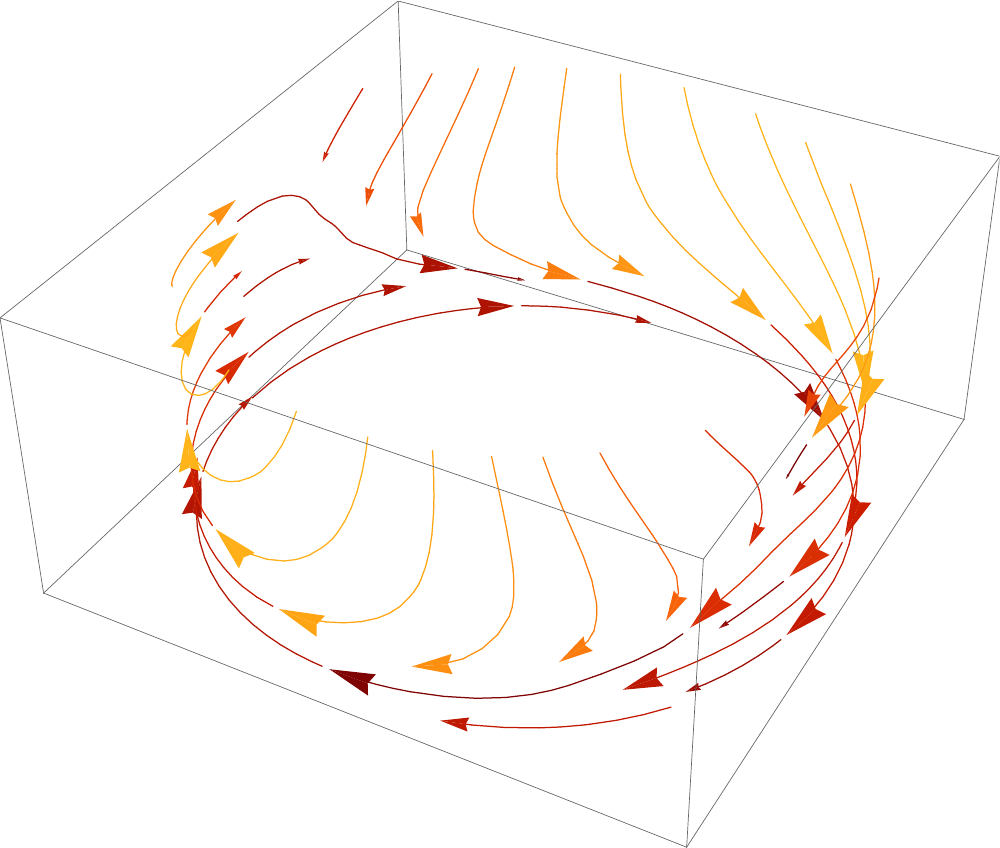}}
\subfigure[]{\includegraphics[scale=0.5]{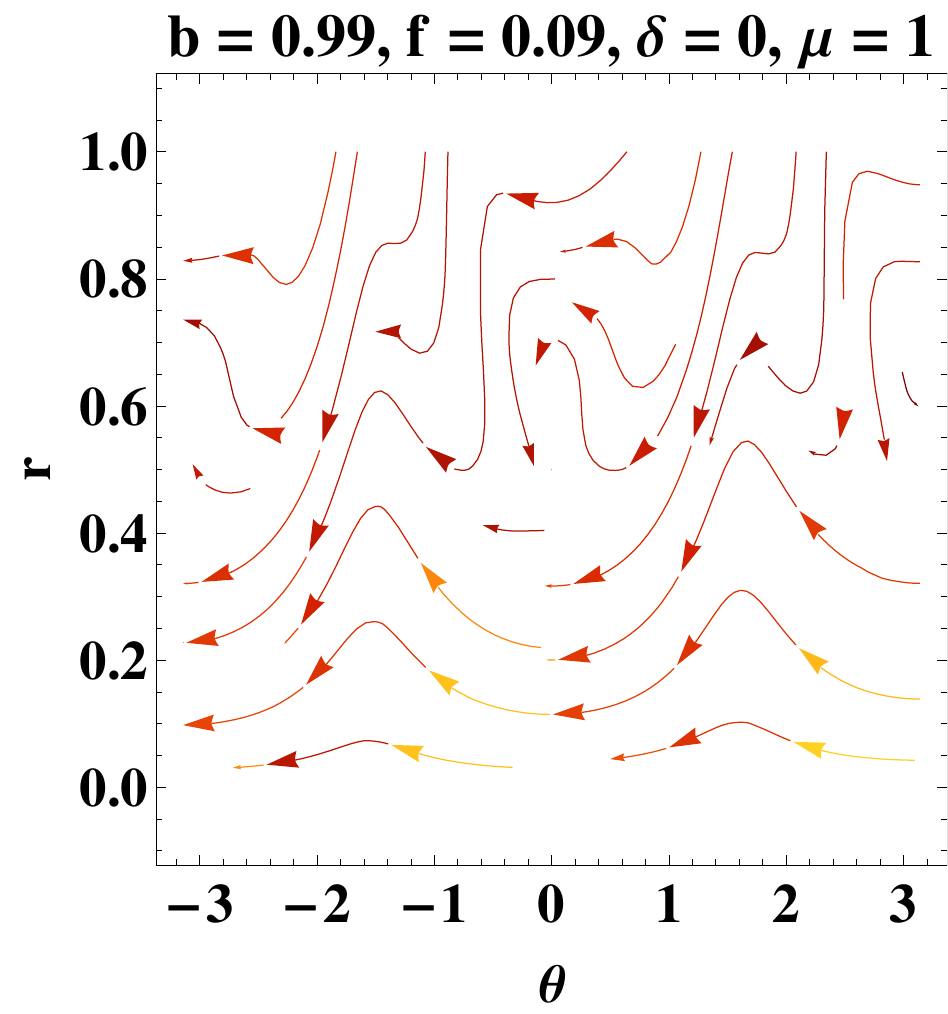} }\hspace{2cm}
\subfigure[]{\includegraphics[scale=0.55]{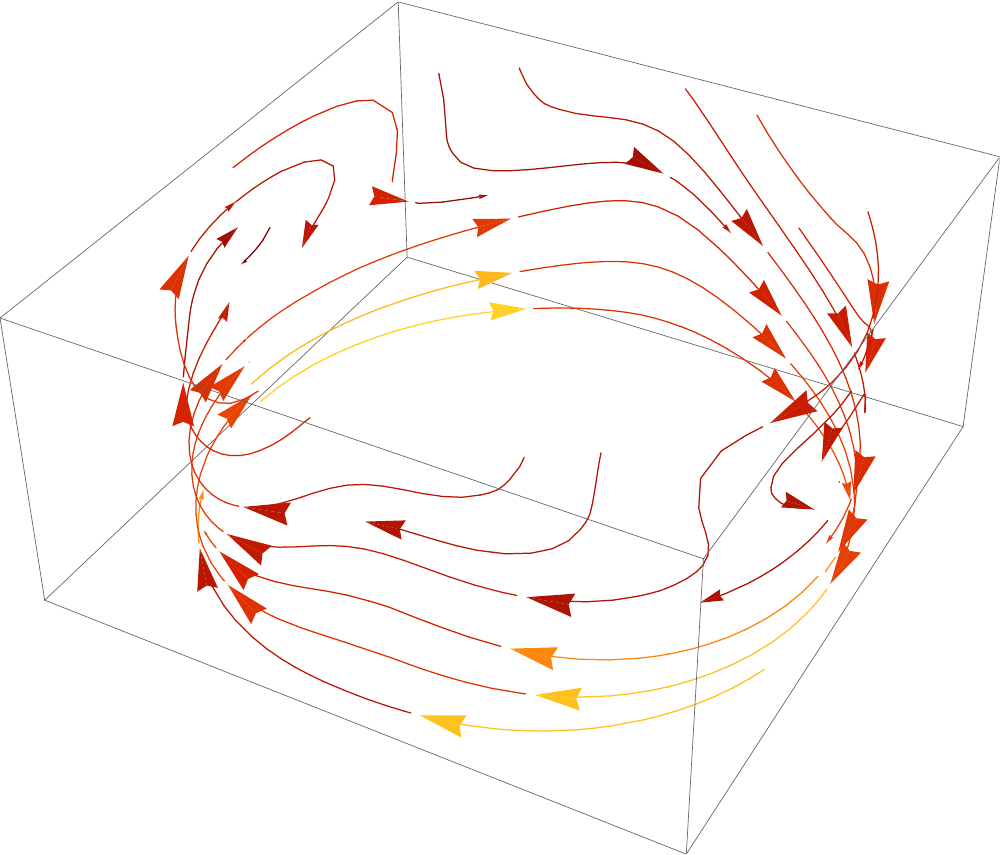}}
\end{center}\caption{\label{Fig11} Phase portrait of equations   \eqref{syst104} (left panel). Projection over the cylinder $\mathbf{S}$ (right panel) for $(b, f, \delta)= (0.99, 0.09, 0)$ and different values of $\mu$.}
\end{figure}
The asymptotic expansions of the full problem ($b\neq 0$) are derived as follows.
\\
Note that: 
\begin{subequations}
\begin{align}
& \dot r=  -b \mu ^3 \sin (\delta ) \sin (\vartheta )-\frac{b \mu ^2 \cos (\delta ) \cos (\vartheta ) \sin (\vartheta ) r}{\sqrt{2} f} \nonumber\\
& -\frac{\sqrt{3} \sqrt{b \mu ^2 \cos (\vartheta ) \sin (\delta )} \sin ^2(\vartheta )
   r^{3/2}}{\sqrt[4]{2}}+\frac{b \mu  \cos ^2(\vartheta ) \sin (\delta ) \sin (\vartheta ) r^2}{4 f^2}+O\left(r^{5/2}\right),\end{align}
	\begin{align}
& \dot \vartheta +\sqrt{2} \mu =  -\frac{\sqrt{3} \cos (\vartheta ) \sqrt{b \mu ^2 \cos (\vartheta ) \sin (\delta )} \sin (\vartheta ) \sqrt{r}}{\sqrt[4]{2}}+\frac{b \mu  \cos ^3(\vartheta ) \sin (\delta ) r}{4 f^2}
\nonumber \\
& -\frac{\left(\sqrt{3} \cos (\vartheta ) \left(b \mu  \cos
   (\delta ) \cos ^2(\vartheta )+2 f\right) \sin (\vartheta )\right) r^{3/2}}{4 \left(2^{3/4} f \sqrt{b \mu ^2 \cos (\vartheta ) \sin (\delta )}\right)}+\frac{b \cos (\delta ) \cos ^4(\vartheta ) r^2}{12 \sqrt{2}
   f^3}+O\left(r^{5/2}\right).
\end{align}
\end{subequations}
To obtain an approximated solution near the oscillatory regime the average with respect to $\vartheta$ over any orbit of period $2 \pi$, 
$\langle (\dot r, \dot \vartheta) \rangle$ is taken, leading to 
\begin{subequations}
\begin{align}
& \dot r=k r^{3/2} , \quad k=\frac{2^{3/4} \sqrt{3} \sqrt{b} \mu  \left(E\left(\left.\frac{c}{2}\right|2\right)-E\left(\left.\frac{c}{2}+\pi \right|2\right)\right) \sqrt{\sin (\delta )}}{5 \pi },\\
& \dot \vartheta +\sqrt{2} \mu=\frac{b \cos (\delta ) (r-4 f \mu ) (r+4 f \mu)}{32 \sqrt{2} f^3}. 
\end{align}
\end{subequations}

where $E\left(\left.\phi\right|m\right)$ gives the elliptic integral of the second kind:
\begin{equation}
E\left(\left.\phi\right|m\right)= \int_{0}^{\phi} \left(1- m \sin^2 (\theta)\right)^{\frac{1}{2}}d\theta, \quad -\frac{\pi}{2}<\phi<\frac{\pi}{2}.
\end{equation}
The averaged equations have solutions 
\begin{equation}
r(t)=  \frac{4}{\left(c_1+k t\right){}^2}, \vartheta (t)= -\frac{b \cos (\delta ) \left(3 f^2 \mu ^2 \left(c_1+k t\right)+\frac{1}{\left(c_1+k t\right){}^3}\right)+12 f^3 \mu  \left(c_1+k t\right)}{6 \sqrt{2} f^3
   k}+\vartheta_0. 
\end{equation}
\\
Introducing along with $\vartheta$ and $r$, the new variable 
\begin{equation}
\varepsilon=\frac{H}{\mu+H}, 
\end{equation} 
with inverse
\begin{equation}
 H= \frac{\mu  \varepsilon}{1-\varepsilon},
\end{equation}
satisfying:
\begin{equation}
b f \mu ^3 (\varepsilon -1)^2 \left(\cos \left(\delta +\frac{r \cos (\vartheta )}{\sqrt{2} f \mu }\right)-\cos (\delta )\right)-\frac{1}{2} r^2 (\varepsilon -1)^2+3 \mu ^2 \varepsilon ^2=0,
\end{equation}
along with the time derivative  $\hat{\tau}$ given by 
\begin{equation}
\frac{d \hat{\tau}}{d t} = \mu+ H,
\end{equation}
the following equations are obtained: 
\begin{subequations}
\begin{align}
& \varepsilon'=\frac{r^2 (\varepsilon -1)^3 \sin ^2(\vartheta )}{2 \mu ^2},\\
& r'=-3 r \varepsilon  \sin ^2(\vartheta )+b \mu ^2 (\varepsilon -1) \sin (\vartheta ) \sin \left(\delta +\frac{r \cos (\vartheta )}{\sqrt{2} f \mu }\right),\\
&\vartheta'=-\sqrt{2} (1-\varepsilon)-3 \varepsilon  \sin (\vartheta ) \cos (\vartheta )+\frac{b \mu ^2 (\varepsilon -1)
   \cos (\vartheta ) \sin \left(\delta +\frac{r \cos (\vartheta )}{\sqrt{2} f \mu }\right)}{r}.
\end{align}
\end{subequations}
In order to obtain an approximated solution near the oscillatory regime the average with respect to $\vartheta$ over any orbit of period $2 \pi$, 
$\langle (\varepsilon', r', \vartheta') \rangle$, is taken, leading to:
\begin{subequations}
\begin{align}
& \varepsilon'=\frac{(\varepsilon -1)^3 r^2}{4 \mu ^2},\\
& r'=-\frac{3 \varepsilon  r}{2},\\
& \vartheta'=\frac{(\varepsilon -1) (4 f+b \mu  \cos (\delta ))}{2 \sqrt{2} f}-\frac{(b (\varepsilon -1) \cos (\delta )) r^2}{32
   \left(\sqrt{2} f^3 \mu \right)}, 
\end{align}
\end{subequations}
with the averaged constraint:
\begin{equation}
3 \mu ^2 \epsilon ^2-\frac{r^2 \left((\epsilon -1)^2 (b \mu  \cos (\delta )+4 f)\right)}{8 f}=0,
\end{equation}
as $r\rightarrow 0$. 
\\
The above system is integrable yielding:
\begin{subequations}
\begin{align}
& r(\varepsilon )=\frac{\sqrt{2} \sqrt{c_1 (\varepsilon -1)^2+\mu ^2 (6 \varepsilon -3)}}{1-\varepsilon },\\
& \vartheta (\varepsilon )=\frac{\mu  \left(\frac{b \cos (\delta )}{\varepsilon -1}-\frac{8 f^2 \tanh ^{-1}\left(\frac{c_1 (\varepsilon -1)+3
   \mu ^2}{\mu  \sqrt{9 \mu ^2-3 c_1}}\right) (b \mu  \cos (\delta )+4 f)}{\sqrt{9 \mu ^2-3 c_1}}\right)}{8 \sqrt{2} f^3}+c_2, 
\end{align}
and 
\begin{align}
& 3(t-t_0)= \ln \left(\frac{(1-\varepsilon )^2 \left(\frac{3 \mu  \sqrt{3 \mu ^2-c_1}+\sqrt{3} c_1 \varepsilon -\sqrt{3} c_1+3 \sqrt{3} \mu ^2}{3 \mu  \sqrt{3 \mu ^2-c_1}-\sqrt{3} c_1 \varepsilon +\sqrt{3} c_1-3 \sqrt{3} \mu ^2}\right){}^{\frac{\mu
   }{\sqrt{\mu ^2-\frac{c_1}{3}}}}}{c_1 (\varepsilon -1)^2+\mu ^2 (6 \varepsilon -3)}\right) \nonumber \\
	& \sim \ln \left(\frac{\left(\frac{2 \mu  \left(\sqrt{9 \mu ^2-3 c_1}+3 \mu \right)-c_1}{c_1}\right){}^{\frac{\mu }{\sqrt{\mu ^2-\frac{c_1}{3}}}}}{c_1-3 \mu ^2}\right)-\frac{6 \mu ^2 \varepsilon }{c_1-3 \mu ^2}+O\left(\varepsilon ^2\right),
\end{align}
\end{subequations}
as $\varepsilon\rightarrow 0$. 

In Figure \ref{Fig10}, the phase portrait of equations   \eqref{syst104} (left panel) and the projection over the cylinder $\mathbf{S}$ (right panel) for $(b, f, \delta)= (0.1, 0.33, 0)$ and different values of $\mu$ are presented. In Figure \ref{Fig11}, the phase portrait of equations   \eqref{syst104} (left panel) and the projection over the cylinder $\mathbf{S}$ (right panel) for $(b, f, \delta)= (0.99, 0.09, 0)$ and different values of $\mu$ are shown. The plots show the periodic nature of the solutions.

\section{Discussion}
\label{discussion}
In this research, a local dynamical systems analysis for arbitrary $V(\phi)$ and $\chi(\phi)$ using Hubble normalized equations was provided. The analysis relies on two arbitrary functions $f(\lambda)$ and $g(\lambda)$ which encode the potential and the coupling function through the quadrature
\begin{equation*}
 \phi (\lambda )=\phi (1)-\int_1^{\lambda } \frac{1}{f(s)} \, ds, \quad
 V (\lambda )+\Lambda=W(1) e^{\int_1^{\lambda } \frac{s}{f(s)} \, ds}, \quad
 \chi (\lambda )=\chi(1) e^{\int_1^{\lambda } \frac{g(s)}{f(s)} \, ds}.
\end{equation*}
After that, a global dynamical systems formulation using the Alho \& Uggla's approach \cite{Alho:2014fha}  was implemented in Section \ref{SECT:3.3}. The equilibrium points that represent some solutions of cosmological interest were obtained: a matter--kinetic scaling solution, a matter--scalar field scaling solution, a kinetic dominated solution representing a stiff fluid, a solution dominated by the effective energy density of the geometric term $G_0(a)$, a scaling solution where the kinetic term and the effective energy density from $G_0(a)$ scales with the same order of magnitude, a quintessence scalar field dominated solution, the vacuum de Sitter solution associated to the minimum of the potential and a non-interacting matter dominated solution. All of which reveal a very rich cosmological behavior.

For the exponential potential $V(\phi)= V_0 e^{-\lambda \phi}$ in a vacuum analyzed in Section \ref{Section2.3}, the asymptotic states using either the Hubble--normalized equations or the  Alho \& Uggla's approach \cite{Alho:2014fha} are the same, but in some examples the asymptotic behavior can be better explained using Alho \& Uggla's approach \cite{Alho:2014fha}. E.g., for the $E$--model: $%
V(\phi )=V_{0}\left( 1-e^{-\sqrt{\frac{2}{3\alpha }}\phi }\right) ^{2n}$
 \cite{Alho:2017opd,Leon:2019mbo} or the (generalized) harmonic potential. 
Recalling in Hubble--normalized equations the evolution equation for $H$, which is given by the Raychaudhuri equation, decouples. In particular,  due it has the symmetry such that its derivative is also an exponential function, the Raychaudhuri equation always decouples for a scalar field with exponential potential.   The asymptotics of the remaining reduced system is then typically given by the equilibrium points and often can be determined by a dynamical system analysis \cite{Coley:1999uh,Coley:2003mj,wainwrightellis1997}.  For other potentials that do not satisfy the above symmetry, like  the harmonic potential $V(\phi)= \mu^2 \phi^2$, the Raychaudhuri equation fails to decouple \cite{Alho:2014fha}. The resulting Hubble normalized equations prove  to be difficult to analyze in the usual dynamical systems approach due to the oscillations entering the system via the Klein--Gordon equation \cite{Fajman:2020yjb}. 
In reference \cite{Fajman:2020yjb}, the oscillations and future asymptotics of locally rotationally symetric Bianchi type III cosmologies with a massive scalar field with potential $V(\phi)=\frac{1}{2} \phi^2$ were studied. 

According to the previous discussions the advantage of the Alho \& Uggla's approach \cite{Alho:2014fha} is taken to study scalar-field cosmologies with generalized harmonic potentials of the type $V(\phi)= \mu^2 \phi^2 + \text{cosine corrections}$.

In Section \ref{Section2.3B}, the potential of the so-called $E$--model: $V(\phi )=V_{0}\left( 1-e^{-\sqrt{\frac{2}{3\alpha }}\phi }\right) ^{2n}$
with $f(s)=-\frac{s\left( s-\sqrt{6}\mu \right) }{2n}$, $\mu =\frac{2n}{3%
\sqrt{\alpha }}$, discussed in \cite{Alho:2017opd} for a conventional scalar
field cosmology and in Hořava–Lifshitz cosmology  in \cite{Leon:2019mbo}, was studied.  Observing that the dynamics of the latter potential is
equivalent to that of the exponential potential plus a cosmological
constant, $V=V_{0}e^{-\sqrt{6}\mu \phi }+\Lambda $ having the $f$-deviser: $f(s)=-s\left( s-%
\sqrt{6}\mu \right)$, up to a rescaling in the independent variable.

In Sections \ref{Sect:2.4} and \ref{Sect:2.5}, scalar field cosmologies under the potentials $V_1(\phi)= \mu^3 \left[\frac{\phi^2}{\mu} + b f \cos\left(\delta + \frac{\phi}{f}\right)\right]$, $b\neq 0$ and 
$V_2(\phi)= \mu ^3 \left[b f \left(\cos (\delta )-\cos \left(\delta +\frac{\phi }{f}\right)\right)+\frac{\phi ^2}{\mu}\right]
$, $b\neq 0$ were respectively studied. The Alho \& Uggla's approach \cite{Alho:2014fha}
was used later in Sections \ref{Sect:2.4.1} and \ref{Sect:2.5.1} to find qualitative features and also to present the asymptotic analysis as $\phi\rightarrow \infty$ for the harmonic potentials $V_1(\phi)$ and 
$V_2(\phi)$ in a vacuum, respectively.  In Section \ref{Sect:2.4.2},  the oscillatory regime for the scalar field under the potential $V_1(\phi)$  was investigated.  Meanwhile, in Section \ref{Sect:2.5.2}, the oscillations of the scalar field under the potential $V_2(\phi)$ were studied.

\section{Conclusions}
\label{Sect:7}

In this paper, both local and global phase-space descriptions and averaging methods were used to find qualitative features of solutions for the FLRW and the Bianchi I metrics in the context of scalar field cosmologies with arbitrary potentials and arbitrary couplings to matter. The stability of the equilibrium points in a phase-space, as well as the dynamics in the regime where the scalar field diverges was studied. Equilibrium points that represent some solutions of cosmological interest such as: several types of scaling solutions, a kinetic dominated solution representing a stiff fluid, a solution dominated by an effective energy density of geometric origin, a quintessence scalar field dominated solution, the vacuum de Sitter solution associated to the minimum of the potential and a non-interacting matter dominated solution were found. All these reveal a very rich cosmological phenomenology. 

The preliminary analysis of oscillations in scalar-field cosmologies with generalized harmonic potentials of type $V(\phi)= \mu^2 \phi^2 + \text{cosine corrections}$ (implemented in Sections \ref{Sect:2.4.2} and \ref{Sect:2.5.2}) will be improved in a forthcoming paper using averaging techniques similar to those used  in \cite{Fajman:2020yjb} for a family of generalized  harmonic  potentials when $H$ monotonically tends to zero. In this approach, the Hubble scalar  plays the role of a time dependent perturbation parameter which controls the magnitude of error between full and the time-averaged solutions. The oscillations can be viewed as perturbations that can be smoothed out; and the Raychaudhuri equation of the averaged equations decouples. At the end, the analysis of the system  is  reduced to study the corresponding averaged equations. 

\ack

This research was funded by  Agencia Nacional de Investigaci\'on y Desarrollo- ANID  through the program FONDECYT Iniciaci\'on grant no.
11180126 and by Vicerrector\'{\i}a de Investigaci\'on y Desarrollo Tecnológico at
Universidad Cat\'olica del Norte. Ellen de los M. Fern\'andez Flores, Alfredo Millano, and Joey latta are acknowledged for proofreading this manuscript and improving the English. Thanks to Alan Coley for his encouraging suggestions.

\appendix

\section{Existence and stability conditions of the equilibrium points of the system \eqref{systH1} as $\phi \rightarrow \infty$}
\label{AppA}
 The equilibrium points of the system \eqref{systH1} are the following: 
\begin{enumerate}
    \item[$A_1(\hat{\lambda})$:] $(x, \Omega_m, \Omega_0, \lambda)=\left(\frac{(4-3 \gamma ) g(\hat{\lambda})}{\sqrt{6} (\gamma -2)},
   1-\frac{(4-3 \gamma )^2 g(\hat{\lambda})^2}{6 (\gamma
   -2)^2}, 0, \hat{\lambda}\right)$, where   $\hat{\lambda}$ denotes the values of $\lambda$ for which  $f(\lambda)=0$. It exists for $1-\frac{(4-3 \gamma )^2 g(\hat{\lambda })^2}{6 (\gamma -2)^2}\geq 0$. 
   The eigenvalues are $\scriptstyle\Bigg\{\frac{3 (\gamma -2)}{2}-\frac{(4-3 \gamma )^2 g(\hat{\lambda
   })^2}{4 (\gamma -2)}, 3 \gamma -\frac{(4-3 \gamma )^2
   g(\hat{\lambda})^2}{2 (\gamma -2)}-p, \frac{6 (\gamma -2) \gamma +(3 \gamma -4) g(\hat{\lambda}) \left((4-3 \gamma )
   g(\hat{\lambda})+2 \hat{\lambda }\right)}{2 (\gamma
   -2)},\frac{(3 \gamma -4) g(\hat{\lambda})
   f'(\hat{\lambda})}{\gamma -2}\Bigg\}$.  \\ For $1\leq \gamma< 2$, it follows that  $A_1(\hat{\lambda})$ is a sink for:
   \begin{enumerate}
    \item $3<p\leq 4, \;  1\leq \gamma <\frac{p}{3}, \;  -\sqrt{2} \sqrt{\frac{(\gamma -2) (3 \gamma -p)}{(3 \gamma -4)^2}}<g(\hat{\lambda })<0, \;  f'(\hat{\lambda })>0, \;  \hat{\lambda }>\left(\frac{3 \gamma
   }{2}-2\right) g(\hat{\lambda })-\frac{3 (\gamma -2) \gamma }{(3 \gamma -4) g(\hat{\lambda })}$, or 
   \item $3<p\leq 4, \;  1\leq \gamma <\frac{p}{3}, \;  0<g(\hat{\lambda })<\sqrt{2}
   \sqrt{\frac{(\gamma -2) (3 \gamma -p)}{(3 \gamma -4)^2}}, \;  f'(\hat{\lambda })<0, \;  \hat{\lambda }<\left(\frac{3 \gamma }{2}-2\right) g(\hat{\lambda })-\frac{3 (\gamma -2) \gamma }{(3 \gamma -4)
   g(\hat{\lambda })}$, or 
   \item $4<p\leq 6, \;  1\leq \gamma <\frac{4}{3}, \;  -\sqrt{2} \sqrt{\frac{(\gamma -2) (3 \gamma -p)}{(3 \gamma -4)^2}}<g(\hat{\lambda })<0, \;  f'\left(\hat{\lambda
   }\right)>0, \;  \hat{\lambda }>\left(\frac{3 \gamma }{2}-2\right) g(\hat{\lambda })-\frac{3 (\gamma -2) \gamma }{(3 \gamma -4) g(\hat{\lambda })}$, or 
   \item $4<p\leq 6, \;  1\leq \gamma
   <\frac{4}{3}, \;  0<g(\hat{\lambda })<\sqrt{2} \sqrt{\frac{(\gamma -2) (3 \gamma -p)}{(3 \gamma -4)^2}}, \;  f'(\hat{\lambda })<0, \;  \hat{\lambda }<\left(\frac{3 \gamma }{2}-2\right) g\left(\hat{\lambda
   }\right)-\frac{3 (\gamma -2) \gamma }{(3 \gamma -4) g(\hat{\lambda })}$, or 
   \item $4<p\leq 6, \;  \frac{4}{3}<\gamma <\frac{p}{3}, \;  -\sqrt{2} \sqrt{\frac{(\gamma -2) (3 \gamma -p)}{(3 \gamma
   -4)^2}}<g(\hat{\lambda })<0, \;  f'(\hat{\lambda })<0, \;  \hat{\lambda }<\left(\frac{3 \gamma }{2}-2\right) g(\hat{\lambda })-\frac{3 (\gamma -2) \gamma }{(3 \gamma -4) g\left(\hat{\lambda
   }\right)}$, or 
   \item $4<p\leq 6, \;  \frac{4}{3}<\gamma <\frac{p}{3}, \;  0<g(\hat{\lambda })<\sqrt{2} \sqrt{\frac{(\gamma -2) (3 \gamma -p)}{(3 \gamma -4)^2}}, \;  f'(\hat{\lambda })>0, \; 
   \hat{\lambda }>\left(\frac{3 \gamma }{2}-2\right) g(\hat{\lambda })-\frac{3 (\gamma -2) \gamma }{(3 \gamma -4) g(\hat{\lambda })}$, or 
   \item $p>6, \;  1\leq \gamma <\frac{4}{3}, \;  -\frac{\sqrt{6}
   (\gamma -2)}{3 \gamma -4}<g(\hat{\lambda })<0, \;  f'(\hat{\lambda })>0, \;  \hat{\lambda }>\left(\frac{3 \gamma }{2}-2\right) g(\hat{\lambda })-\frac{3 (\gamma -2) \gamma }{(3 \gamma -4)
   g(\hat{\lambda })}$, or 
   \item $p>6, \;  1\leq \gamma <\frac{4}{3}, \;  0<g(\hat{\lambda })<\frac{\sqrt{6} (\gamma -2)}{3 \gamma -4}, \;  f'(\hat{\lambda })<0, \;  \hat{\lambda
   }<\left(\frac{3 \gamma }{2}-2\right) g(\hat{\lambda })-\frac{3 (\gamma -2) \gamma }{(3 \gamma -4) g(\hat{\lambda })}$, or 
   \item $p>6, \;  \frac{4}{3}<\gamma <2, \;  \frac{\sqrt{6} (\gamma -2)}{3
   \gamma -4}<g(\hat{\lambda })<0, \;  f'(\hat{\lambda })<0, \;  \hat{\lambda }<\left(\frac{3 \gamma }{2}-2\right) g(\hat{\lambda })-\frac{3 (\gamma -2) \gamma }{(3 \gamma -4) g\left(\hat{\lambda
   }\right)}$, or 
   \item $p>6, \;  \frac{4}{3}<\gamma <2, \;  0<g(\hat{\lambda })<-\frac{\sqrt{6} (\gamma -2)}{3 \gamma -4}, \;  f'(\hat{\lambda })>0, \;  \hat{\lambda }>\left(\frac{3 \gamma
   }{2}-2\right) g(\hat{\lambda })-\frac{3 (\gamma -2) \gamma }{(3 \gamma -4) g(\hat{\lambda })}$.
   \end{enumerate}
If it exists, it will never be a source. 
   
   \item[$A_2(\hat{\lambda})$:] $(x, \Omega_m, \Omega_0, \lambda)=\left(\frac{\sqrt{\frac{2}{3}} (p-3 \gamma )}{(3 \gamma -4) g\left(\hat{\lambda
   }\right)}, \frac{2 (6-p) (p-3 \gamma )}{3 (4-3 \gamma )^2
   g(\hat{\lambda})^2}, \frac{2 (p-3 \gamma ) (\gamma
   -2)}{(4-3 \gamma )^2 g(\hat{\lambda})^2}+1, \hat{\lambda}\right)$. It exists for $\frac{2 (6-p) (p-3 \gamma )}{3 (4-3 \gamma )^2
   g(\hat{\lambda})^2}\geq 0$. The eigenvalues are \\
   $\scriptstyle \Bigg\{-\frac{1}{4} \left(6-p-\frac{\sqrt{p-6} \sqrt{3 (4-3 \gamma )^2
   g(\hat{\lambda})^2 (-8 \gamma +3 p-2)+16 (\gamma -2) (p-3
   \gamma )^2}}{(3 \gamma -4) g\left(\hat{\lambda
   }\right)}\right)$,\\
   $\scriptstyle -\frac{1}{4} \left(6-p+\frac{\sqrt{p-6} \sqrt{3 (4-3 \gamma )^2
   g(\hat{\lambda})^2 (-8 \gamma +3 p-2)+16 (\gamma -2) (p-3
   \gamma )^2}}{(3 \gamma -4) g\left(\hat{\lambda
   }\right)}\right), p-\frac{2 \hat{\lambda } (p-3 \gamma )}{(3 \gamma
   -4) g(\hat{\lambda})},-\frac{2 (p-3 \gamma )
   f'(\hat{\lambda})}{(3 \gamma -4) g\left(\hat{\lambda
   }\right)}\Bigg\}$. 
\\ For $1\leq \gamma< 2$, it follows that 
$A_2(\hat{\lambda})$ is a sink for:
\begin{enumerate}
    \item $1<\gamma <\frac{4}{3}, \;  3 \gamma <p\leq \frac{2}{3} (4 \gamma +1), \;  g(\hat{\lambda })<-\sqrt{2} \sqrt{\frac{(\gamma -2) (3 \gamma -p)}{(3 \gamma -4)^2}}, \;  f'(\hat{\lambda })>0, \; 
   \hat{\lambda }>\frac{(3 \gamma -4) p g(\hat{\lambda })}{2 (p-3 \gamma )}$, or 
   \item $1<\gamma <\frac{4}{3}, \;  3 \gamma <p\leq \frac{2}{3} (4 \gamma +1), \;  g(\hat{\lambda })>\sqrt{2}
   \sqrt{\frac{(\gamma -2) (3 \gamma -p)}{(3 \gamma -4)^2}}, \;  f'(\hat{\lambda })<0, \;  \hat{\lambda }<\frac{(3 \gamma -4) p g(\hat{\lambda })}{2 (p-3 \gamma )}$, or 
   \item $1<\gamma <\frac{4}{3}, \; 
   \frac{2}{3} (4 \gamma +1)<p<6, \;  -\frac{4 \sqrt{-\frac{(\gamma -2) (p-3 \gamma )^2}{(3 \gamma -4)^2 (-8 \gamma +3 p-2)}}}{\sqrt{3}}\leq g(\hat{\lambda })<-\sqrt{2} \sqrt{\frac{(\gamma -2) (3 \gamma -p)}{(3 \gamma
   -4)^2}}, \;  f'(\hat{\lambda })>0, \;  \hat{\lambda }>\frac{(3 \gamma -4) p g(\hat{\lambda })}{2 (p-3 \gamma )}$, or 
   \item $1<\gamma <\frac{4}{3}, \;  \frac{2}{3} (4 \gamma +1)<p<6, \;  \sqrt{2}
   \sqrt{\frac{(\gamma -2) (3 \gamma -p)}{(3 \gamma -4)^2}}<g(\hat{\lambda })\leq \frac{4 \sqrt{-\frac{(\gamma -2) (p-3 \gamma )^2}{(3 \gamma -4)^2 (-8 \gamma +3 p-2)}}}{\sqrt{3}}, \;  f'\left(\hat{\lambda
   }\right)<0, \;  \hat{\lambda }<\frac{(3 \gamma -4) p g(\hat{\lambda })}{2 (p-3 \gamma )}$, or 
   \item $\frac{4}{3}<\gamma <2, \;  3 \gamma <p\leq \frac{2}{3} (4 \gamma +1), \;  g\left(\hat{\lambda
   }\right)<-\sqrt{2} \sqrt{\frac{(\gamma -2) (3 \gamma -p)}{(3 \gamma -4)^2}}, \;  f'(\hat{\lambda })<0, \;  \hat{\lambda }<\frac{(3 \gamma -4) p g(\hat{\lambda })}{2 (p-3 \gamma )}$, or
   \item $\frac{4}{3}<\gamma <2, \;  3 \gamma <p\leq \frac{2}{3} (4 \gamma +1), \;  g(\hat{\lambda })>\sqrt{2} \sqrt{\frac{(\gamma -2) (3 \gamma -p)}{(3 \gamma -4)^2}}, \;  f'(\hat{\lambda })>0, \; 
   \hat{\lambda }>\frac{(3 \gamma -4) p g(\hat{\lambda })}{2 (p-3 \gamma )}$, or 
   \item $\frac{4}{3}<\gamma <2, \;  \frac{2}{3} (4 \gamma +1)<p<6, \;  -\frac{4 \sqrt{-\frac{(\gamma -2) (p-3 \gamma )^2}{(3 \gamma
   -4)^2 (-8 \gamma +3 p-2)}}}{\sqrt{3}}\leq g(\hat{\lambda })<-\sqrt{2} \sqrt{\frac{(\gamma -2) (3 \gamma -p)}{(3 \gamma -4)^2}}, \;  f'(\hat{\lambda })<0, \;  \hat{\lambda }<\frac{(3 \gamma -4) p
   g(\hat{\lambda })}{2 (p-3 \gamma )}$, or 
   \item $\frac{4}{3}<\gamma <2, \;  \frac{2}{3} (4 \gamma +1)<p<6, \;  \sqrt{2} \sqrt{\frac{(\gamma -2) (3 \gamma -p)}{(3 \gamma -4)^2}}<g(\hat{\lambda })\leq
   \frac{4 \sqrt{-\frac{(\gamma -2) (p-3 \gamma )^2}{(3 \gamma -4)^2 (-8 \gamma +3 p-2)}}}{\sqrt{3}}, \;  f'(\hat{\lambda })>0, \;  \hat{\lambda }>\frac{(3 \gamma -4) p g(\hat{\lambda })}{2 (p-3 \gamma)}$.
  
\end{enumerate}
It is non--hyperbolic for $p=6$, saddle for $p=2$. If it exists, it will never be a source. 
   
   \item[$A_3(\hat{\lambda})$:] $(x, \Omega_m, \Omega_0, \lambda)=\left(\frac{\sqrt{6} \gamma }{(4-3 \gamma ) g(\hat{\lambda})+2
   \hat{\lambda }}, \frac{2 (4-3 \gamma ) g(\hat{\lambda})
   \hat{\lambda }+4 \left(\hat{\lambda }^2-3 \gamma \right)}{\left((3
   \gamma -4) g(\hat{\lambda})-2 \hat{\lambda }\right)^2}, 0
   , \hat{\lambda}\right)$. Assuming $1\leq \gamma \leq 2$, it exists for:
   \begin{enumerate}
       \item $\scriptstyle 1\leq \gamma \leq 2, \; p>0, \;  \hat{\lambda }\leq \frac{1}{4} \left(-4 g(\hat{\lambda })-\sqrt{48 \gamma +9 \gamma ^2 g(\hat{\lambda })^2-24 \gamma  g(\hat{\lambda })^2+16 g(\hat{\lambda })^2}+3 \gamma 
   g(\hat{\lambda })\right)$, or 
    \item $\scriptstyle 1\leq \gamma \leq 2, \; p>0, \; \hat{\lambda }\geq \frac{1}{4} \left(-4 g(\hat{\lambda })+\sqrt{48 \gamma +9 \gamma ^2 g(\hat{\lambda })^2-24 \gamma 
   g(\hat{\lambda })^2+16 g(\hat{\lambda })^2}+3 \gamma  g(\hat{\lambda })\right)$.
   \end{enumerate}
  The eigenvalues are
   ${\left\{\frac{p (4-3 \gamma ) g(\hat{\lambda})+2 (p-3 \gamma ) \hat{\lambda }}{(-4+3 \gamma ) g(\hat{\lambda})-2
\hat{\lambda }},\frac{1}{(-8+6 \gamma ) g(\hat{\lambda})-4 \hat{\lambda }}\left((12-9 \gamma ) g(\hat{\lambda})-\right.\right.}\\
{3 (-2+\gamma ) \hat{\lambda }-\sqrt{3} \surd \left(2 (-4+3 \gamma )^3 g(\hat{\lambda})^3 \hat{\lambda }+(4-3 \gamma )^2 g(\hat{\lambda})^2 \left(3+12 \gamma -8 \hat{\lambda }^2\right)-\right.}\\
{\left.\left.2 (-4+3 \gamma ) g(\hat{\lambda}) \hat{\lambda } \left(6-3 \gamma +6 \gamma ^2-4 \hat{\lambda }^2\right)-3 (-2+\gamma
) \left(24 \gamma ^2+(2-9 \gamma ) \hat{\lambda }^2\right)\right)\right),}\\
{\frac{1}{(-8+6 \gamma ) g(\hat{\lambda})-4 \hat{\lambda }}\left((12-9 \gamma ) g(\hat{\lambda})-3 (-2+\gamma ) \hat{\lambda
}+\right.}\\
{\sqrt{3} \surd \left(2 (-4+3 \gamma )^3 g(\hat{\lambda})^3 \hat{\lambda }+(4-3 \gamma )^2 g(\hat{\lambda})^2 \left(3+12
\gamma -8 \hat{\lambda }^2\right)-2 (-4+3 \gamma ) g(\hat{\lambda}) \hat{\lambda } \right.}\\
{\left.\left.\left.\left(6-3 \gamma +6 \gamma ^2-4 \hat{\lambda }^2\right)-3 (-2+\gamma ) \left(24 \gamma ^2+(2-9 \gamma ) \hat{\lambda }^2\right)\right)\right),\frac{6
\gamma  f'(\hat{\lambda})}{(-4+3 \gamma ) g(\hat{\lambda})-2 \hat{\lambda }}\right\}}$. \\
In this case using a semi-analytically procedure for non-minimal coupling $g\equiv 0$  the eigenvalues are reduced to \\
$\Bigg\{-\frac{-3 \sqrt{(2-\gamma ) \left(24 \gamma ^2+(2-9 \gamma ) \hat{\lambda }^2\right)}-3 (\gamma -2) \hat{\lambda }}{4 \hat{\lambda }},-\frac{3 \sqrt{(2-\gamma ) \left(24 \gamma ^2+(2-9 \gamma ) \hat{\lambda
   }^2\right)}-3 (\gamma -2) \hat{\lambda }}{4 \hat{\lambda }}$,\\
   $3 \gamma -p, -\frac{3 \gamma  f'(\hat{\lambda })}{\hat{\lambda }}\Bigg\}$. 
Hence, assuming that $1\leq \gamma<2$, $A_3(\hat{\lambda})$ is a sink for:
 \begin{enumerate}
     \item $1\leq \gamma <2, p>3 \gamma, -\frac{2 \sqrt{6} \gamma }{\sqrt{9 \gamma -2}}\leq \hat{\lambda }<-\sqrt{3} \sqrt{\gamma }, f'(\hat{\lambda })<0$, or 
     \item $1\leq \gamma <2, p>3 \gamma,
   \sqrt{3} \sqrt{\gamma }<\hat{\lambda }\leq \frac{2 \sqrt{6} \gamma }{\sqrt{9 \gamma -2}}, f'(\hat{\lambda })>0$. 
 \end{enumerate}
  Otherwise, it is a saddle. 
 For non-minimal coupling the analysis is more complicated, so a numerical study is more reliable. 
  
   \item[$A_4(\hat{\lambda})$:] $(x, \Omega_m, \Omega_0, \lambda)=\left(-1, 0, 0,\hat{\lambda}\right)$. The eigenvalues are\\
   $\left\{6-p,-3 \gamma +\sqrt{\frac{3}{2}} (3 \gamma -4)
   g(\hat{\lambda})+6,\sqrt{6} \hat{\lambda }+6,\sqrt{6}
   f'(\hat{\lambda})\right\}$. 
   
   $A_4(\hat{\lambda})$ is a source for:
   \begin{enumerate}
       \item $1\leq \gamma <\frac{4}{3},  0\leq p<6,  g(\hat{\lambda })<\frac{\sqrt{6} (\gamma -2)}{3 \gamma -4},  f'(\hat{\lambda })>0,  \hat{\lambda }>-\sqrt{6}$, or 
       \item $\gamma =\frac{4}{3},   0\leq p<6,  f'(\hat{\lambda })>0,  \hat{\lambda }>-\sqrt{6}$, or 
   \item $\frac{4}{3}<\gamma <2,  0\leq p<6,  g(\hat{\lambda })>\frac{\sqrt{6} (\gamma -2)}{3 \gamma -4}, 
   f'(\hat{\lambda })>0,  \hat{\lambda }>-\sqrt{6}$.
   \end{enumerate}

   $A_4(\hat{\lambda})$ is a sink for
   \begin{enumerate}
       \item $1\leq \gamma <\frac{4}{3}, p>6, g(\hat{\lambda })>\frac{\sqrt{6} (\gamma -2)}{3 \gamma -4}, f'(\hat{\lambda })<0, \hat{\lambda }<-\sqrt{6}$, or 
       \item $\frac{4}{3}<\gamma <2, p>6, g(\hat{\lambda })<\frac{\sqrt{6} (\gamma -2)}{3 \gamma -4}, f'(\hat{\lambda })<0, \hat{\lambda }<-\sqrt{6}$.
   \end{enumerate}
   
   \item[$A_5(\hat{\lambda})$:] $(x, \Omega_m, \Omega_0, \lambda)=\left(0, 0, 1, \hat{\lambda}\right)$. The eigenvalues are 
   $\left\{0,\frac{p-6}{2},p,p-3 \gamma \right\}$.  It has a non--hyperbolic 3D unstable manifold for $p>6$.
   
   \item[$A_6(\hat{\lambda})$:] $(x, \Omega_m, \Omega_0, \lambda)=\left(1, 0, 0, \hat{\lambda}\right)$. The eigenvalues are \\
   $\left\{6-p,-3 \gamma +\sqrt{\frac{3}{2}} (4-3 \gamma )
   g(\hat{\lambda})+6,6-\sqrt{6} \hat{\lambda },-\sqrt{6}
   f'(\hat{\lambda})\right\}$.
   
   $A_6(\hat{\lambda})$ is a source for:
   \begin{enumerate}
       \item $1\leq \gamma <\frac{4}{3},  p<6,  g(\hat{\lambda })>-\frac{\sqrt{6} (\gamma -2)}{3 \gamma -4},  f'(\hat{\lambda })<0,  \hat{\lambda }<\sqrt{6}$, or 
       \item $\gamma =\frac{4}{3},  p<6,    f'(\hat{\lambda })<0,  \hat{\lambda }<\sqrt{6}$, or 
   \item $\frac{4}{3}<\gamma<2,  p<6,  g(\hat{\lambda })<-\frac{\sqrt{6} (\gamma -2)}{3 \gamma -4},  f'(\hat{\lambda })<0, 
   \hat{\lambda }<\sqrt{6}$.
   \end{enumerate}
   
   $A_6(\hat{\lambda})$ is a  sink for:
    \begin{enumerate}
        \item $1\leq \gamma <\frac{4}{3},  p>6,  g(\hat{\lambda })<-\frac{\sqrt{6} (\gamma -2)}{3 \gamma -4},  f'(\hat{\lambda })>0,  \hat{\lambda }>\sqrt{6}$, or 
        \item $1\leq \gamma >\frac{4}{3},  p>6,   g(\hat{\lambda })>-\frac{\sqrt{6} (\gamma -2)}{3 \gamma -4},  f'(\hat{\lambda })>0,  \hat{\lambda }>\sqrt{6}$.
    \end{enumerate}
   
   \item[$A_7(\hat{\lambda})$:] $(x, \Omega_m, \Omega_0, \lambda)=\left(\frac{p}{\sqrt{6} \hat{\lambda }}, 0, 1-\frac{p}{\hat{\lambda }^2} , \hat{\lambda}\right)$. It exists for $\hat{\lambda }<0, 0\leq p\leq 6$, or $\hat{\lambda }>0, 0\leq p\leq 6$.  The eigenvalues are $\Bigg\{-\frac{1}{4} \left(6-p-\frac{\sqrt{\left(p-6\right)\left(\hat{\lambda }^2 (9 p-6)-8 p^2\right)}}{\hat{\lambda
   }}\right)$,\\
   $-\frac{1}{4} \left(6-p+\frac{\sqrt{\left(p-6\right)\left(\hat{\lambda }^2 (9 p-6)-8 p^2\right)}}{\hat{\lambda
   }}\right), -3 \gamma +\frac{(4-3 \gamma ) p g(\hat{\lambda})}{2
   \hat{\lambda }}+p,-\frac{p
   f'(\hat{\lambda})}{\hat{\lambda }}\Bigg\}$.

For $1\leq \gamma\leq 2$, $A_7(\hat{\lambda})$, it is a sink for:
   \begin{enumerate}
   \item $1\leq \gamma <\frac{4}{3}, \;  \hat{\lambda }\leq -\sqrt{6}, \;  0<p\leq \frac{1}{16} \left(9 \hat{\lambda }^2-\sqrt{3} \sqrt{\hat{\lambda }^2 \left(27 \hat{\lambda }^2-64\right)}\right), \;  g(\hat{\lambda})>\frac{2 \hat{\lambda } (p-3 \gamma )}{(3 \gamma -4) p}, \;  f'(\hat{\lambda })<0$, or 
   \item $1\leq \gamma <\frac{4}{3}, \;  \hat{\lambda }\geq \sqrt{6}, \;  0<p\leq \frac{1}{16} \left(9 \hat{\lambda
   }^2-\sqrt{3} \sqrt{\hat{\lambda }^2 \left(27 \hat{\lambda }^2-64\right)}\right), \;  g(\hat{\lambda })<\frac{2 \hat{\lambda } (p-3 \gamma )}{(3 \gamma -4) p}, \;  f'(\hat{\lambda })>0$, or
   \item $1\leq \gamma <\frac{4}{3}, \;  \hat{\lambda }>\frac{8}{3 \sqrt{3}}, \;  0<p\leq \frac{1}{16} \left(9 \hat{\lambda }^2-\sqrt{3} \sqrt{\hat{\lambda }^2 \left(27 \hat{\lambda }^2-64\right)}\right), \;  g(\hat{\lambda})<\frac{2 \hat{\lambda } (p-3 \gamma )}{(3 \gamma -4) p}, \;  f'(\hat{\lambda })>0$, or 
   \item $1\leq \gamma <\frac{4}{3}, \;  -\sqrt{6}<\hat{\lambda }<-\frac{8}{3 \sqrt{3}}, \;  0<p\leq \frac{1}{16}
   \left(9 \hat{\lambda }^2-\sqrt{3} \sqrt{\hat{\lambda }^2 \left(27 \hat{\lambda }^2-64\right)}\right), \;  g(\hat{\lambda })>\frac{2 \hat{\lambda } (p-3 \gamma )}{(3 \gamma -4) p}, \;  f'\left(\hat{\lambda
   }\right)<0$, or 
   \item $1\leq \gamma <\frac{4}{3}, \;  -\frac{8}{3 \sqrt{3}}\leq \hat{\lambda }<0, \;  0<p<\hat{\lambda }^2, \;  g(\hat{\lambda })>\frac{2 \hat{\lambda } (p-3 \gamma )}{(3 \gamma -4) p}, \; 
   f'(\hat{\lambda })<0$, or 
   \item $1\leq \gamma <\frac{4}{3}, \;  -\sqrt{6}<\hat{\lambda }<-\frac{8}{3 \sqrt{3}}, \;  \frac{1}{16} \left(9 \hat{\lambda }^2+\sqrt{3} \sqrt{\hat{\lambda }^2 \left(27 \hat{\lambda
   }^2-64\right)}\right)\leq p<\hat{\lambda }^2, \;  g(\hat{\lambda })>\frac{2 \hat{\lambda } (p-3 \gamma )}{(3 \gamma -4) p}, \;  f'(\hat{\lambda })<0$, or 
   \item $1\leq \gamma <\frac{4}{3}, \; 
   0<\hat{\lambda }\leq \frac{8}{3 \sqrt{3}}, \;  0<p<\hat{\lambda }^2, \;  g(\hat{\lambda })<\frac{2 \hat{\lambda } (p-3 \gamma )}{(3 \gamma -4) p}, \;  f'(\hat{\lambda })>0$, or 
   \item $1\leq \gamma
   <\frac{4}{3}, \;  \frac{8}{3 \sqrt{3}}<\hat{\lambda }<\sqrt{6}, \;  \frac{1}{16} \left(9 \hat{\lambda }^2+\sqrt{3} \sqrt{\hat{\lambda }^2 \left(27 \hat{\lambda }^2-64\right)}\right)\leq p<\hat{\lambda }^2, \; 
   g(\hat{\lambda })<\frac{2 \hat{\lambda } (p-3 \gamma )}{(3 \gamma -4) p}, \;  f'(\hat{\lambda })>0$, or 
   \item $\gamma =\frac{4}{3}, \;  \hat{\lambda }\geq \frac{8}{\sqrt{15}}, \;  0<p\leq
   \frac{1}{16} \left(9 \hat{\lambda }^2-\sqrt{3} \sqrt{\hat{\lambda }^2 \left(27 \hat{\lambda }^2-64\right)}\right), \;  f'(\hat{\lambda })>0$, or 
   \item $\gamma =\frac{4}{3}, \;  \hat{\lambda }>2, \;  0<p\leq
   \frac{1}{16} \left(9 \hat{\lambda }^2-\sqrt{3} \sqrt{\hat{\lambda }^2 \left(27 \hat{\lambda }^2-64\right)}\right), \;  f'(\hat{\lambda })>0$, or 
   \item $\gamma =\frac{4}{3}, \;  \hat{\lambda }\leq
   -\frac{8}{\sqrt{15}}, \;  0<p\leq \frac{1}{16} \left(9 \hat{\lambda }^2-\sqrt{3} \sqrt{\hat{\lambda }^2 \left(27 \hat{\lambda }^2-64\right)}\right), \;  f'(\hat{\lambda })<0$, or 
   \item $\gamma =\frac{4}{3}, \; 
   \frac{8}{3 \sqrt{3}}<\hat{\lambda }\leq 2, \;  0<p\leq \frac{1}{16} \left(9 \hat{\lambda }^2-\sqrt{3} \sqrt{\hat{\lambda }^2 \left(27 \hat{\lambda }^2-64\right)}\right), \;  f'(\hat{\lambda })>0$, or \item $\gamma =\frac{4}{3}, \;  0<\hat{\lambda }\leq \frac{8}{3 \sqrt{3}}, \;  0<p<\hat{\lambda }^2, \;  f'(\hat{\lambda })>0$, or 
   \item $\gamma =\frac{4}{3}, \;  2<\hat{\lambda }<\frac{8}{\sqrt{15}}, \; 
   \frac{1}{16} \left(9 \hat{\lambda }^2+\sqrt{3} \sqrt{\hat{\lambda }^2 \left(27 \hat{\lambda }^2-64\right)}\right)\leq p<4, \;  f'(\hat{\lambda })>0$, or 
   \item $\gamma =\frac{4}{3}, \;  \frac{8}{3
   \sqrt{3}}<\hat{\lambda }\leq 2, \;  \frac{1}{16} \left(9 \hat{\lambda }^2+\sqrt{3} \sqrt{\hat{\lambda }^2 \left(27 \hat{\lambda }^2-64\right)}\right)\leq p<\hat{\lambda }^2, \;  f'(\hat{\lambda })>0$, or
   \item $\gamma =\frac{4}{3}, \;  -\frac{8}{\sqrt{15}}<\hat{\lambda }<-2, \;  0<p\leq \frac{1}{16} \left(9 \hat{\lambda }^2-\sqrt{3} \sqrt{\hat{\lambda }^2 \left(27 \hat{\lambda }^2-64\right)}\right), \;  f'(\hat{\lambda})<0$, or 
   \item $\gamma =\frac{4}{3}, \;  -2\leq \hat{\lambda }<-\frac{8}{3 \sqrt{3}}, \;  0<p\leq \frac{1}{16} \left(9 \hat{\lambda }^2-\sqrt{3} \sqrt{\hat{\lambda }^2 \left(27 \hat{\lambda
   }^2-64\right)}\right), \;  f'(\hat{\lambda })<0$, or 
   \item $\gamma =\frac{4}{3}, \;  -\frac{8}{3 \sqrt{3}}\leq \hat{\lambda }<0, \;  0<p<\hat{\lambda }^2, \;  f'(\hat{\lambda })<0$, or
   \item $\gamma =\frac{4}{3}, \;  -\frac{8}{\sqrt{15}}<\hat{\lambda }<-2, \;  \frac{1}{16} \left(9 \hat{\lambda }^2+\sqrt{3} \sqrt{\hat{\lambda }^2 \left(27 \hat{\lambda }^2-64\right)}\right)\leq p<4, \;  f'\left(\hat{\lambda
   }\right)<0$, or 
   \item $\gamma =\frac{4}{3}, \;  -2\leq \hat{\lambda }<-\frac{8}{3 \sqrt{3}}, \;  \frac{1}{16} \left(9 \hat{\lambda }^2+\sqrt{3} \sqrt{\hat{\lambda }^2 \left(27 \hat{\lambda }^2-64\right)}\right)\leq
   p<\hat{\lambda }^2, \;  f'(\hat{\lambda })<0$, or 
   \item $\frac{4}{3}<\gamma <2, \;  \hat{\lambda }\geq \sqrt{6}, \;  0<p\leq \frac{1}{16} \left(9 \hat{\lambda }^2-\sqrt{3} \sqrt{\hat{\lambda }^2 \left(27
   \hat{\lambda }^2-64\right)}\right), \;  g(\hat{\lambda })>\frac{2 \hat{\lambda } (p-3 \gamma )}{(3 \gamma -4) p}, \;  f'(\hat{\lambda })>0$, or 
   \item $\frac{4}{3}<\gamma <2, \;  \hat{\lambda }=2
   \sqrt{6} \sqrt{\frac{\gamma ^2}{9 \gamma -2}}, \;  0<p\leq \frac{1}{16} \left(9 \hat{\lambda }^2-\sqrt{3} \sqrt{\hat{\lambda }^2 \left(27 \hat{\lambda }^2-64\right)}\right), \;  g(\hat{\lambda })>\frac{2
   \hat{\lambda } (p-3 \gamma )}{(3 \gamma -4) p}, \;  f'(\hat{\lambda })>0$, or 
   \item $\frac{4}{3}<\gamma <2, \;  \hat{\lambda }>2 \sqrt{6} \sqrt{\frac{\gamma ^2}{9 \gamma -2}}, \;  0<p\leq \frac{1}{16} \left(9
   \hat{\lambda }^2-\sqrt{3} \sqrt{\hat{\lambda }^2 \left(27 \hat{\lambda }^2-64\right)}\right), \;  g(\hat{\lambda })>\frac{2 \hat{\lambda } (p-3 \gamma )}{(3 \gamma -4) p}, \;  f'\left(\hat{\lambda
   }\right)>0$, or 
   \item $\frac{4}{3}<\gamma <2, \;  \hat{\lambda }=2 \sqrt{6} \sqrt{\frac{\gamma ^2}{9 \gamma -2}}, \;  3 \gamma \leq p<\hat{\lambda }^2, \;  g(\hat{\lambda })>\frac{2 \hat{\lambda } (p-3 \gamma
   )}{(3 \gamma -4) p}, \;  f'(\hat{\lambda })>0$, or 
   \item $\frac{4}{3}<\gamma <2, \;  \hat{\lambda }\leq -\sqrt{6}, \;  0<p\leq \frac{1}{16} \left(9 \hat{\lambda }^2-\sqrt{3} \sqrt{\hat{\lambda }^2 \left(27
   \hat{\lambda }^2-64\right)}\right), \;  g(\hat{\lambda })<\frac{2 \hat{\lambda } (p-3 \gamma )}{(3 \gamma -4) p}, \;  f'(\hat{\lambda })<0$, or 
   \item $\frac{4}{3}<\gamma <2, \;  \hat{\lambda }=-2
   \sqrt{6} \sqrt{\frac{\gamma ^2}{9 \gamma -2}}, \;  0<p\leq \frac{1}{16} \left(9 \hat{\lambda }^2-\sqrt{3} \sqrt{\hat{\lambda }^2 \left(27 \hat{\lambda }^2-64\right)}\right), \;  g(\hat{\lambda })<\frac{2
   \hat{\lambda } (p-3 \gamma )}{(3 \gamma -4) p}, \;  f'(\hat{\lambda })<0$, or 
   \item $\frac{4}{3}<\gamma <2, \;  \hat{\lambda }=-2 \sqrt{6} \sqrt{\frac{\gamma ^2}{9 \gamma -2}}, \;  3 \gamma \leq p<\hat{\lambda
   }^2, \;  g(\hat{\lambda })<\frac{2 \hat{\lambda } (p-3 \gamma )}{(3 \gamma -4) p}, \;  f'(\hat{\lambda })<0$, or 
   \item $\frac{4}{3}<\gamma <2, \;  \frac{8}{3 \sqrt{3}}<\hat{\lambda }<2 \sqrt{6}
   \sqrt{\frac{\gamma ^2}{9 \gamma -2}}, \;  0<p\leq \frac{1}{16} \left(9 \hat{\lambda }^2-\sqrt{3} \sqrt{\hat{\lambda }^2 \left(27 \hat{\lambda }^2-64\right)}\right), \;  g(\hat{\lambda })>\frac{2 \hat{\lambda } (p-3
   \gamma )}{(3 \gamma -4) p}, \;  f'(\hat{\lambda })>0$, or 
   \item $\frac{4}{3}<\gamma <2, \;  0<\hat{\lambda }\leq \frac{8}{3 \sqrt{3}}, \;  0<p<\hat{\lambda }^2, \;  g(\hat{\lambda })>\frac{2
   \hat{\lambda } (p-3 \gamma )}{(3 \gamma -4) p}, \;  f'(\hat{\lambda })>0$, or 
   \item $\frac{4}{3}<\gamma <2, \;  \frac{8}{3 \sqrt{3}}<\hat{\lambda }<2 \sqrt{6} \sqrt{\frac{\gamma ^2}{9 \gamma -2}}, \; 
   \frac{1}{16} \left(9 \hat{\lambda }^2+\sqrt{3} \sqrt{\hat{\lambda }^2 \left(27 \hat{\lambda }^2-64\right)}\right)\leq p<\hat{\lambda }^2, \;  g(\hat{\lambda })>\frac{2 \hat{\lambda } (p-3 \gamma )}{(3 \gamma -4)
   p}, \;  f'(\hat{\lambda })>0$, or 
   \item $\frac{4}{3}<\gamma <2, \;  2 \sqrt{6} \sqrt{\frac{\gamma ^2}{9 \gamma -2}}<\hat{\lambda }<\sqrt{6}, \;  \frac{1}{16} \left(9 \hat{\lambda }^2+\sqrt{3} \sqrt{\hat{\lambda
   }^2 \left(27 \hat{\lambda }^2-64\right)}\right)\leq p<\hat{\lambda }^2, \;  g(\hat{\lambda })>\frac{2 \hat{\lambda } (p-3 \gamma )}{(3 \gamma -4) p}, \;  f'(\hat{\lambda })>0$, or
   \item $\frac{4}{3}<\gamma <2, \;  -\sqrt{6}<\hat{\lambda }<-2 \sqrt{6} \sqrt{\frac{\gamma ^2}{9 \gamma -2}}, \;  0<p\leq \frac{1}{16} \left(9 \hat{\lambda }^2-\sqrt{3} \sqrt{\hat{\lambda }^2 \left(27 \hat{\lambda
   }^2-64\right)}\right), \;  g(\hat{\lambda })<\frac{2 \hat{\lambda } (p-3 \gamma )}{(3 \gamma -4) p}, \;  f'(\hat{\lambda })<0$, or 
   \item $\frac{4}{3}<\gamma <2, \;  -2 \sqrt{6} \sqrt{\frac{\gamma
   ^2}{9 \gamma -2}}<\hat{\lambda }<-\frac{8}{3 \sqrt{3}}, \;  0<p\leq \frac{1}{16} \left(9 \hat{\lambda }^2-\sqrt{3} \sqrt{\hat{\lambda }^2 \left(27 \hat{\lambda }^2-64\right)}\right), \;  g(\hat{\lambda })<\frac{2
   \hat{\lambda } (p-3 \gamma )}{(3 \gamma -4) p}, \;  f'(\hat{\lambda })<0$, or 
   \item $\frac{4}{3}<\gamma <2, \;  -\frac{8}{3 \sqrt{3}}\leq \hat{\lambda }<0, \;  0<p<\hat{\lambda }^2, \;  g\left(\hat{\lambda
   }\right)<\frac{2 \hat{\lambda } (p-3 \gamma )}{(3 \gamma -4) p}, \;  f'(\hat{\lambda })<0$, or 
   \item $\frac{4}{3}<\gamma <2, \;  -\sqrt{6}<\hat{\lambda }<-2 \sqrt{6} \sqrt{\frac{\gamma ^2}{9 \gamma -2}}, \; 
   \frac{1}{16} \left(9 \hat{\lambda }^2+\sqrt{3} \sqrt{\hat{\lambda }^2 \left(27 \hat{\lambda }^2-64\right)}\right)\leq p<\hat{\lambda }^2, \;  g(\hat{\lambda })<\frac{2 \hat{\lambda } (p-3 \gamma )}{(3 \gamma -4)
   p}, \;  f'(\hat{\lambda })<0$, or 
   \item $\frac{4}{3}<\gamma <2, \;  -2 \sqrt{6} \sqrt{\frac{\gamma ^2}{9 \gamma -2}}<\hat{\lambda }<-\frac{8}{3 \sqrt{3}}, \;  \frac{1}{16} \left(9 \hat{\lambda }^2+\sqrt{3}
   \sqrt{\hat{\lambda }^2 \left(27 \hat{\lambda }^2-64\right)}\right)\leq p<\hat{\lambda }^2, \;  g(\hat{\lambda })<\frac{2 \hat{\lambda } (p-3 \gamma )}{(3 \gamma -4) p}, \;  f'(\hat{\lambda })<0$, or
   \item $\gamma =2, \;  \hat{\lambda }\geq \sqrt{6}, \;  0<p\leq \frac{1}{16} \left(9 \hat{\lambda }^2-\sqrt{3} \sqrt{\hat{\lambda }^2 \left(27 \hat{\lambda }^2-64\right)}\right), \;  g(\hat{\lambda})>\frac{\hat{\lambda } (p-6)}{p}, \;  f'(\hat{\lambda })>0$, or 
   \item $\gamma =2, \;  \hat{\lambda }>\frac{8}{3 \sqrt{3}}, \;  0<p\leq \frac{1}{16} \left(9 \hat{\lambda }^2-\sqrt{3} \sqrt{\hat{\lambda
   }^2 \left(27 \hat{\lambda }^2-64\right)}\right), \;  g(\hat{\lambda })>\frac{\hat{\lambda } (p-6)}{p}, \;  f'(\hat{\lambda })>0$, or 
   \item $\gamma =2, \;  \hat{\lambda }\leq -\sqrt{6}, \;  0<p\leq
   \frac{1}{16} \left(9 \hat{\lambda }^2-\sqrt{3} \sqrt{\hat{\lambda }^2 \left(27 \hat{\lambda }^2-64\right)}\right), \;  g(\hat{\lambda })<\frac{\hat{\lambda } (p-6)}{p}, \;  f'(\hat{\lambda })<0$, or
   \item $\gamma =2, \;  0<\hat{\lambda }\leq \frac{8}{3 \sqrt{3}}, \;  0<p<\hat{\lambda }^2, \;  g(\hat{\lambda })>\frac{\hat{\lambda } (p-6)}{p}, \;  f'(\hat{\lambda })>0$, or 
   \item $\gamma =2, \; 
   \frac{8}{3 \sqrt{3}}<\hat{\lambda }<\sqrt{6}, \;  \frac{1}{16} \left(9 \hat{\lambda }^2+\sqrt{3} \sqrt{\hat{\lambda }^2 \left(27 \hat{\lambda }^2-64\right)}\right)\leq p<\hat{\lambda }^2, \;  g\left(\hat{\lambda
   }\right)>\frac{\hat{\lambda } (p-6)}{p}, \;  f'(\hat{\lambda })>0$, or 
   \item $\gamma =2, \;  -\sqrt{6}<\hat{\lambda }<-\frac{8}{3 \sqrt{3}}, \;  0<p\leq \frac{1}{16} \left(9 \hat{\lambda }^2-\sqrt{3}
   \sqrt{\hat{\lambda }^2 \left(27 \hat{\lambda }^2-64\right)}\right), \;  g(\hat{\lambda })<\frac{\hat{\lambda } (p-6)}{p}, \;  f'(\hat{\lambda })<0$, or 
   \item $\gamma =2, \;  -\frac{8}{3 \sqrt{3}}\leq
   \hat{\lambda }<0, \;  0<p<\hat{\lambda }^2, \;  g(\hat{\lambda })<\frac{\hat{\lambda } (p-6)}{p}, \;  f'(\hat{\lambda })<0$, or 
   \item $\gamma =2, \;  -\sqrt{6}<\hat{\lambda }<-\frac{8}{3
   \sqrt{3}}, \;  \frac{1}{16} \left(9 \hat{\lambda }^2+\sqrt{3} \sqrt{\hat{\lambda }^2 \left(27 \hat{\lambda }^2-64\right)}\right)\leq p<\hat{\lambda }^2, \;  g(\hat{\lambda })<\frac{\hat{\lambda } (p-6)}{p}, \;    f'(\hat{\lambda })<0$.
   \end{enumerate}
   It will never be a source.
   
   \item[$A_8(\hat{\lambda})$:] $(x, \Omega_m, \Omega_0, \lambda)=\left(\frac{\hat{\lambda }}{\sqrt{6}}, 0, 0, \hat{\lambda}\right)$. It exists for $\hat{\lambda }^2\leq 6$. The eigenvalues are \\
   $\left\{\frac{1}{2} \left(\hat{\lambda }^2-6\right),\hat{\lambda }^2-p,-3
   \gamma +\frac{1}{2} (4-3 \gamma ) \hat{\lambda } g(\hat{\lambda
   })+\hat{\lambda }^2,-\hat{\lambda } f'(\hat{\lambda
   })\right\}$.
   
  $A_8(\hat{\lambda})$ is a sink for:
   \begin{enumerate}
       \item $1\leq \gamma <\frac{4}{3},  p>\hat{\lambda }^2,  f'(\hat{\lambda })>0,  0<\hat{\lambda }<\sqrt{6},  g(\hat{\lambda })<\frac{6 \gamma -2 \hat{\lambda }^2}{4 \hat{\lambda }-3 \gamma  \hat{\lambda}}$, or 
   \item $1\leq \gamma <\frac{4}{3},  p>\hat{\lambda }^2,  g(\hat{\lambda })>\frac{6 \gamma -2 \hat{\lambda }^2}{4 \hat{\lambda }-3 \gamma  \hat{\lambda }},  -\sqrt{6}<\hat{\lambda }<0, 
   f'(\hat{\lambda })<0$, or 
   \item $\gamma=\frac{4}{3},  p>\hat{\lambda }^2,  -2<\hat{\lambda }<0,  f'(\hat{\lambda })<0$, or 
   \item $\gamma=\frac{4}{3},  p>\hat{\lambda }^2,  f'\hat{\lambda})>0,  0<\hat{\lambda }<2$, or 
   \item $\frac{4}{3}<\gamma <2,  p>\hat{\lambda }^2,  g(\hat{\lambda })>\frac{6 \gamma -2 \hat{\lambda }^2}{4 \hat{\lambda }-3 \gamma  \hat{\lambda }}, 
   f'(\hat{\lambda })>0,  0<\hat{\lambda }<\sqrt{6}$, or 
   \item $\frac{4}{3}<\gamma <2,  p>\hat{\lambda }^2,  -\sqrt{6}<\hat{\lambda }<0,  g(\hat{\lambda })<\frac{6 \gamma -2 \hat{\lambda
   }^2}{4 \hat{\lambda }-3 \gamma  \hat{\lambda }},  f'(\hat{\lambda })<0$.
   \end{enumerate}
 It will never be a source.
   
   \item[$A_9$:] $(x, \Omega_m, \Omega_0, \lambda)=\left(0, 0, 0, 0\right)$. The eigenvalues are\\
   $\left\{-p,-3 \gamma ,-\frac{1}{2} \left(3+\sqrt{9-12 f(0)}\right),-\frac{1}{2} \left(3-\sqrt{9-12 f(0)}\right)\right\}$. Sink for $p>0, \gamma>0, f(0)>0$. Otherwise, it is a saddle. 
   
   \item[$A_{10}(\tilde{\lambda})$:] $(x, \Omega_m, \Omega_0, \lambda)=\left(0, 1, 0, \tilde{\lambda}\right)$, where we denote by  $\tilde{\lambda}$, the values of $\lambda$ for which  $g(\lambda)=0$. The eigenvalues are
   $\Bigg\{\frac{1}{4} \left(3 \gamma -6-\sqrt{24 (4-3 \gamma )
   f(\tilde{\lambda }) g'(\tilde{\lambda })+9
   (\gamma -2)^2}\right)$,\\
   $\frac{1}{4} \left(3 \gamma -6+\sqrt{24 (4-3
   \gamma ) f(\tilde{\lambda }) g'(\tilde{\lambda
   })+9 (\gamma -2)^2}\right), 3 \gamma ,3 \gamma -p\Bigg\}$. It is a saddle. 
\end{enumerate}

\section{Existence and stability conditions of the equilibrium points of system \eqref{31-syst} as  $\phi\rightarrow \infty$}
\label{AppB}
The equilibrium points of system \eqref{31-syst} with $\varphi=0$  (i.e., corresponding to $\phi\rightarrow \infty$) are the following:
\begin{enumerate}
    \item[$B_1$:] $\left(0,\tan ^{-1}\left[\sqrt{1-\frac{N^2}{6}},-\frac{N}{\sqrt{6}}\right]+2 \pi  c_1,0,0,0\right), c_1\in \mathbb{Z}$ with eigenvalues \\ $\left\{0,\frac{N^2}{2},\frac{1}{2} \left(N^2-6\right),N^2-p,N (2 M+N)-\frac{3}{2} \gamma  (M N+2)\right\}$. It exists for $N^2\leq 6$. It is always a non--hyperbolic saddle for $N^2\leq 6$.
    
    \item[$B_2$:] $\left(1,\tan ^{-1}\left[\sqrt{1-\frac{N^2}{6}},-\frac{N}{\sqrt{6}}\right]+2 \pi  c_1,0,0,0\right), c_1\in \mathbb{Z}$ with eigenvalues \\ $\left\{0,-\frac{N^2}{2},\frac{1}{2} \left(N^2-6\right),N^2-p,N (2 M+N)-\frac{3}{2} \gamma  (M N+2)\right\}$. It exists for $N^2\leq 6$. The case of physical interest is when it is non--hyperbolic with a 4D stable manifold for:
    \begin{enumerate}
        \item $M\in \mathbb{R},   -2<N<0,   p>N^2,   \gamma =\frac{4}{3}$, or 
        \item $M\in \mathbb{R},   0<N<2,   p>N^2,   \gamma =\frac{4}{3}$, or
        \item $-\sqrt{6}<N\leq -2,   p>N^2,   1\leq \gamma <\frac{4}{3},   M>\frac{2 (N^2-3\gamma)}{N(3 \gamma  -4)}$, or
        \item $-2<N<0,   p>N^2,   1\leq \gamma <\frac{4}{3},   M>\frac{2 (N^2-3\gamma)}{N(3 \gamma  -4)}$, or 
        \item $0<N<2,   p>N^2,   \frac{4}{3}<\gamma \leq 2,   M>\frac{2 (N^2-3\gamma)}{N(3 \gamma  -4)}$, or 
        \item $2\leq N<\sqrt{6},   p>N^2,   \frac{4}{3}<\gamma \leq 2,   M>\frac{2 (N^2-3\gamma)}{N(3 \gamma  -4)}$, or
        \item $-\sqrt{6}<N\leq -2,   p>N^2,   \frac{4}{3}<\gamma \leq 2,   M<\frac{2 (N^2-3\gamma)}{N(3 \gamma  -4)}$, or 
        \item $0<N<2,   p>N^2,   1\leq \gamma <\frac{4}{3},   M<\frac{2 (N^2-3\gamma)}{N(3 \gamma  -4)}$, or 
        \item $2\leq N<\sqrt{6},   p>N^2,   1\leq \gamma <\frac{4}{3},   M<\frac{2 (N^2-3\gamma)}{N(3 \gamma  -4)}$, or 
        \item $-2<N<0,   p>N^2,   \frac{4}{3}<\gamma \leq 2,   M<\frac{2 (N^2-3 \gamma)}{N (3 \gamma-4)}$.
    \end{enumerate}
    
        \item[$B_3$:] $\left(0,2 \pi  c_1,0,1,0\right), c_1\in \mathbb{Z}$ with eigenvalues  $\left\{0,\frac{p-6}{2},\frac{p}{2},p,p-3 \gamma \right\}$. The physical interesting situation is when it is non--hyperbolic with a 4D unstable manifold for $p>6, 1\leq \gamma \leq 2$. It is a non--hyperbolic saddle otherwise. 
        
        \item[$B_4$:] $\left(1,2 \pi  c_1,0,1,0\right), c_1\in \mathbb{Z}$ with eigenvalues $\left\{0,\frac{p-6}{2},-\frac{p}{2},p,p-3 \gamma \right\}$.  It is a non--hyperbolic saddle. 
    
        \item[$B_5$:] $\left(0,\tan ^{-1}\left[\sqrt{1-\frac{p^2}{6 N^2}},-\frac{p}{\sqrt{6} N}\right]+2 \pi  c_1,0,1-\frac{p}{N^2},0\right), c_1\in \mathbb{Z}$ with eigenvalues \\ $\Big\{0,\frac{p}{2},\frac{p ((4-3 \gamma ) M+2 N)}{2 N}-3 \gamma, \frac{1}{4} \left(p-6+\frac{\sqrt{\left(p-6 \right)\left(N^2 (9 p-6)-8 p^2\right)}}{N}\right)$, \\
    $\frac{1}{4} \left(p-6-\frac{\sqrt{\left(p-6 \right)\left(N^2 (9 p-6)-8 p^2\right)}}{N}\right) \Big\}$. It exists for $p>0, N^2>\frac{p^2}{6}$. Furthermore, it satisfies $\Omega_0\geq 0$  for $0<p<6, N^2\geq p$, or $p\geq 6, N^2>\frac{p^2}{6}$. It is a non--hyperbolic saddle. 
    
        \item[$B_6$:] $\left(1,\tan ^{-1}\left[\sqrt{1-\frac{p^2}{6 N^2}},-\frac{p}{\sqrt{6} N}\right]+2 \pi  c_1,0,1-\frac{p}{N^2},0\right), c_1\in \mathbb{Z}$ with eigenvalues \\
   $\Big\{0,-\frac{p}{2},\frac{p ((4-3 \gamma ) M+2 N)}{2 N}-3 \gamma,\frac{1}{4} \left(p-6+\frac{\sqrt{\left(p-6\right)\left(N^2 (9 p-6)-8 p^2\right)}}{N}\right)$,\\
   $\frac{1}{4} \left(p-6-\frac{\sqrt{\left(p-6\right)\left(N^2 (9 p-6)-8 p^2\right)}}{N}\right) \Big\}$. It exists for $p>0, N^2>\frac{p^2}{6}$. Furthermore, it satisfies $\Omega_0\geq 0$  for $0<p<6, N^2\geq p$, or $p\geq 6, N^2>\frac{p^2}{6}$. The situation of physical interest is when it is non--hyperbolic with a 4D stable manifold for:
   \begin{enumerate}
    \item $N>\frac{8}{\sqrt{15}}, \; 0<p\leq \frac{1}{16} \left(9 N^2-\sqrt{3} \sqrt{N^2 \left(27 N^2-64\right)}\right), \; \gamma =\frac{4}{3}$, or 
       
    \item $N<-\frac{8}{3 \sqrt{3}}, \; 0<p\leq \frac{1}{16} \left(9 N^2-\sqrt{3}
   \sqrt{N^2 \left(27 N^2-64\right)}\right), \; \gamma =\frac{4}{3}$, or 
   
   \item $N=\frac{8}{\sqrt{15}}, \; 0<p\leq \frac{4}{5}, \; \gamma =\frac{4}{3}$, or 
   
   \item $0<N\leq \frac{8}{3 \sqrt{3}}, \; 0<p<N^2, \; \gamma
   =\frac{4}{3}$, or 
   
   \item $-\frac{8}{3 \sqrt{3}}\leq N<0, \; 0<p<N^2, \; \gamma =\frac{4}{3}$, or 
   
   \item $\frac{8}{3 \sqrt{3}}<N<\frac{8}{\sqrt{15}}, \; 0<p\leq \frac{1}{16} \left(9 N^2-\sqrt{3} \sqrt{N^2 \left(27
   N^2-64\right)}\right), \; \gamma =\frac{4}{3}$, or 
   
   \item $\frac{8}{3 \sqrt{3}}<N\leq 2, \; \frac{1}{16} \left(9 N^2+\sqrt{3} \sqrt{N^2 \left(27 N^2-64\right)}\right)\leq p<N^2, \; \gamma =\frac{4}{3}$, or 
   
   \item $
   -2<N<-\frac{8}{3 \sqrt{3}}, \; \frac{1}{16} \left(9 N^2+\sqrt{3} \sqrt{N^2 \left(27 N^2-64\right)}\right)\leq p<N^2, \; \gamma =\frac{4}{3}$, or 
   
   \item $2<N<\frac{8}{\sqrt{15}}, \; \frac{1}{16} \left(9
   N^2+\sqrt{3} \sqrt{N^2 \left(27 N^2-64\right)}\right)\leq p<4, \; \gamma =\frac{4}{3}$, or 
   
   \item $-\frac{8}{\sqrt{15}}<N\leq -2, \; \frac{1}{16} \left(9 N^2+\sqrt{3} \sqrt{N^2 \left(27 N^2-64\right)}\right)\leq p<4, \;
   \gamma =\frac{4}{3}$, or 
   
   \item $\frac{8}{3 \sqrt{3}}<N<\frac{8}{\sqrt{15}}, \; 0<p\leq \frac{1}{16} \left(9 N^2-\sqrt{3} \sqrt{N^2 \left(27 N^2-64\right)}\right), \; 1\leq \gamma <\frac{4}{3}, \; M<\frac{2 N (p-3
   \gamma )}{(3 \gamma -4) p}$, or 
   
   \item $\frac{8}{3 \sqrt{3}}<N<\frac{8}{\sqrt{15}}, \; 0<p\leq \frac{1}{16} \left(9 N^2-\sqrt{3} \sqrt{N^2 \left(27 N^2-64\right)}\right), \; \frac{4}{3}<\gamma \leq 2, \; M>\frac{2 N
   (p-3 \gamma )}{(3 \gamma -4) p}$, or 
   
   \item $N<-\frac{8}{3 \sqrt{3}}, \; 0<p\leq \frac{1}{16} \left(9 N^2-\sqrt{3} \sqrt{N^2 \left(27 N^2-64\right)}\right), \; 1\leq \gamma <\frac{4}{3}, \; M>\frac{2 N (p-3 \gamma )}{(3
   \gamma -4) p}$, or 
   
   \item $N<-\frac{8}{3 \sqrt{3}}, \; 0<p\leq \frac{1}{16} \left(9 N^2-\sqrt{3} \sqrt{N^2 \left(27 N^2-64\right)}\right), \; \frac{4}{3}<\gamma \leq 2, \; M<\frac{2 N (p-3 \gamma )}{(3 \gamma -4)
   p}$, or 
   
   \item $N>\frac{8}{\sqrt{15}}, \; 0<p\leq \frac{1}{16} \left(9 N^2-\sqrt{3} \sqrt{N^2 \left(27 N^2-64\right)}\right), \; 1\leq \gamma <\frac{4}{3}, \; M<\frac{2 N (p-3 \gamma )}{(3 \gamma -4) p}$, or 
   
   \item $N>\frac{8}{\sqrt{15}}, \; 0<p\leq \frac{1}{16} \left(9 N^2-\sqrt{3} \sqrt{N^2 \left(27 N^2-64\right)}\right), \; \frac{4}{3}<\gamma \leq 2, \; M>\frac{2 N (p-3 \gamma )}{(3 \gamma -4) p}$, or 
   
   \item $N=\frac{8}{\sqrt{15}}, \; 0<p\leq \frac{4}{5}, \; 1\leq \gamma <\frac{4}{3}, \; M<\frac{16 (p-3 \gamma )}{\sqrt{15} (3 \gamma -4) p}$, or 
   
   \item $N=\frac{8}{\sqrt{15}}, \; 0<p\leq \frac{4}{5}, \;
   \frac{4}{3}<\gamma \leq 2, \; M>\frac{16 (p-3 \gamma )}{\sqrt{15} (3 \gamma -4) p}$, or 
   
   \item $N=\frac{8}{\sqrt{15}}, \; 4\leq p<\frac{64}{15}, \; 1\leq \gamma <\frac{4}{3}, \; M<\frac{16 (p-3 \gamma )}{\sqrt{15} (3
   \gamma -4) p}$, or 
   
   \item $N=\frac{8}{\sqrt{15}}, \; 4\leq p<\frac{64}{15}, \; \frac{4}{3}<\gamma \leq 2, \; M>\frac{16 (p-3 \gamma )}{\sqrt{15} (3 \gamma -4) p}$, or 
   
   \item $-2<N<-\frac{8}{3 \sqrt{3}}, \;
   \frac{1}{16} \left(9 N^2+\sqrt{3} \sqrt{N^2 \left(27 N^2-64\right)}\right)\leq p<N^2, \; 1\leq \gamma <\frac{4}{3}, \; M>\frac{2 N (p-3 \gamma )}{(3 \gamma -4) p}$, or 
   
   \item $-\sqrt{6}<N\leq -\frac{8}{\sqrt{15}}, \;
   \frac{1}{16} \left(9 N^2+\sqrt{3} \sqrt{N^2 \left(27 N^2-64\right)}\right)\leq p<N^2, \; 1\leq \gamma <\frac{4}{3}, \; M>\frac{2 N (p-3 \gamma )}{(3 \gamma -4) p}$, or 
   
   \item $\frac{8}{3 \sqrt{3}}<N\leq 2, \;
   \frac{1}{16} \left(9 N^2+\sqrt{3} \sqrt{N^2 \left(27 N^2-64\right)}\right)\leq p<N^2, \; \frac{4}{3}<\gamma \leq 2, \; M>\frac{2 N (p-3 \gamma )}{(3 \gamma -4) p}$, or 
   
   \item $\frac{8}{\sqrt{15}}<N<\sqrt{6}, \;
   \frac{1}{16} \left(9 N^2+\sqrt{3} \sqrt{N^2 \left(27 N^2-64\right)}\right)\leq p<N^2, \; \frac{4}{3}<\gamma \leq 2, \; M>\frac{2 N (p-3 \gamma )}{(3 \gamma -4) p}$, or 
   
   \item $\frac{8}{3 \sqrt{3}}<N\leq 2, \;
   \frac{1}{16} \left(9 N^2+\sqrt{3} \sqrt{N^2 \left(27 N^2-64\right)}\right)\leq p<N^2, \; 1\leq \gamma <\frac{4}{3}, \; M<\frac{2 N (p-3 \gamma )}{(3 \gamma -4) p}$, or 
   
   \item $\frac{8}{\sqrt{15}}<N<\sqrt{6}, \;
   \frac{1}{16} \left(9 N^2+\sqrt{3} \sqrt{N^2 \left(27 N^2-64\right)}\right)\leq p<N^2, \; 1\leq \gamma <\frac{4}{3}, \; M<\frac{2 N (p-3 \gamma )}{(3 \gamma -4) p}$, or 
   
   \item $-2<N<-\frac{8}{3 \sqrt{3}}, \; \frac{1}{16}
   \left(9 N^2+\sqrt{3} \sqrt{N^2 \left(27 N^2-64\right)}\right)\leq p<N^2, \; \frac{4}{3}<\gamma \leq 2, \; M<\frac{2 N (p-3 \gamma )}{(3 \gamma -4) p}$, or 
   
   \item $-\sqrt{6}<N\leq -\frac{8}{\sqrt{15}}, \; \frac{1}{16}
   \left(9 N^2+\sqrt{3} \sqrt{N^2 \left(27 N^2-64\right)}\right)\leq p<N^2, \; \frac{4}{3}<\gamma \leq 2, \; M<\frac{2 N (p-3 \gamma )}{(3 \gamma -4) p}$, or 
   
   \item $2<N<\frac{8}{\sqrt{15}}, \; \frac{1}{16} \left(9
   N^2+\sqrt{3} \sqrt{N^2 \left(27 N^2-64\right)}\right)\leq p<4, \; \frac{4}{3}<\gamma \leq 2, \; M>\frac{2 N (p-3 \gamma )}{(3 \gamma -4) p}$, or 
   
   \item $-\frac{8}{\sqrt{15}}<N\leq -2, \; \frac{1}{16} \left(9
   N^2+\sqrt{3} \sqrt{N^2 \left(27 N^2-64\right)}\right)\leq p<4, \; 1\leq \gamma <\frac{4}{3}, \; M>\frac{2 N (p-3 \gamma )}{(3 \gamma -4) p}$, or 
   
   \item $2<N<\frac{8}{\sqrt{15}}, \; \frac{1}{16} \left(9 N^2+\sqrt{3}
   \sqrt{N^2 \left(27 N^2-64\right)}\right)\leq p<4, \; 1\leq \gamma <\frac{4}{3}, \; M<\frac{2 N (p-3 \gamma )}{(3 \gamma -4) p}$, or 
   
   \item $-\frac{8}{\sqrt{15}}<N\leq -2, \; \frac{1}{16} \left(9 N^2+\sqrt{3} \sqrt{N^2
   \left(27 N^2-64\right)}\right)\leq p<4, \; \frac{4}{3}<\gamma \leq 2, \; M<\frac{2 N (p-3 \gamma )}{(3 \gamma -4) p}$, or 
   
   \item $-\frac{8}{3 \sqrt{3}}\leq N<0, \; 0<p<N^2, \; 1\leq \gamma <\frac{4}{3}, \; M>\frac{2 N
   (p-3 \gamma )}{(3 \gamma -4) p}$, or 
   
   \item $0<N\leq \frac{8}{3 \sqrt{3}}, \; 0<p<N^2, \; \frac{4}{3}<\gamma \leq 2, \; M>\frac{2 N (p-3 \gamma )}{(3 \gamma -4) p}$, or 
   
   \item $2<N<\frac{8}{\sqrt{15}}, \; 4\leq
   p<N^2, \; \frac{4}{3}<\gamma \leq 2, \; M>\frac{2 N (p-3 \gamma )}{(3 \gamma -4) p}$, or 
   
   \item $-\frac{8}{\sqrt{15}}<N<-2, \; 4\leq p<N^2, \; 1\leq \gamma <\frac{4}{3}, \; M>\frac{2 N (p-3 \gamma )}{(3 \gamma -4)
   p}$, or 
   
   \item $-\frac{8}{3 \sqrt{3}}\leq N<0, \; 0<p<N^2, \; \frac{4}{3}<\gamma \leq 2, \; M<\frac{2 N (p-3 \gamma )}{(3 \gamma -4) p}$, or 
   
   \item $0<N\leq \frac{8}{3 \sqrt{3}}, \; 0<p<N^2, \; 1\leq \gamma
   <\frac{4}{3}, \; M<\frac{2 N (p-3 \gamma )}{(3 \gamma -4) p}$, or \item $2<N<\frac{8}{\sqrt{15}}, \; 4\leq p<N^2, \; 1\leq \gamma <\frac{4}{3}, \; M<\frac{2 N (p-3 \gamma )}{(3 \gamma -4) p}$, or 
   
   \item $-\frac{8}{\sqrt{15}}<N<-2, \; 4\leq p<N^2, \; \frac{4}{3}<\gamma \leq 2, \; M<\frac{2 N (p-3 \gamma )}{(3 \gamma -4) p}$.
   
   \end{enumerate}
    
    \item[$B_7$:] $ \left(0,\tan ^{-1}\left[\sqrt{1-\frac{2 (p-3 \gamma )^2}{3 M^2(4-3 \gamma )^2}},\frac{2(p-3 \gamma)}{\sqrt{6}M(4 -3 \gamma)}\right]+2 \pi  c_1,\frac{2 (6-p) (p-3 \gamma )}{3 (4-3 \gamma )^2 M^2},\frac{(4-3 \gamma )^2
   M^2+2 (\gamma -2) (p-3 \gamma )}{(4-3 \gamma )^2 M^2},0\right), c_1\in \mathbb{Z}$, with eigenvalues \\ $\Big\{0, \frac{p}{2}, \frac{-3 \gamma  M p+4 M p-6 \gamma  N+2 N p}{M(4-3\gamma)},  \frac{1}{4} \left(p-6+\frac{\sqrt{\left(p-6\right) \left(3 (4-3 \gamma )^2 M^2 (-8 \gamma +3 p-2)+16 (\gamma -2) (p-3 \gamma )^2\right)}}{(3 \gamma -4) M}\right)$, \\ $\frac{1}{4} \left(p-6-\frac{\sqrt{\left(p-6\right) \left(3 (4-3 \gamma )^2 M^2 (-8 \gamma +3 p-2)+16 (\gamma -2) (p-3 \gamma )^2\right)}}{(3 \gamma -4) M}\right)\Big\}$. It exists for:
   \begin{enumerate}
    \item $1\leq \gamma <\frac{4}{3}, \; p=3 \gamma , \; M>0$, or 
    \item $\frac{4}{3}<\gamma \leq 2, \; p=3 \gamma , \; M>0$, or 
    \item $1\leq \gamma <\frac{4}{3}, \; p=3 \gamma , \; M>0$, or
    \item $\frac{4}{3}<\gamma \leq 2, \; p=3 \gamma , \; M>0$, or 
    \item $1\leq \gamma <\frac{4}{3}, \; p=3 \gamma , \; M<0$, or 
    \item $\frac{4}{3}<\gamma \leq 2, \; p=3 \gamma , \; M<0$, or 
    \item $1\leq \gamma
   <\frac{4}{3}, \; p=3 \gamma , \; M<0$, or 
   \item $\frac{4}{3}<\gamma \leq 2, \; p=3 \gamma , \; M<0$, or 
   \item $1\leq \gamma <\frac{4}{3}, \; 3 \gamma <p\leq 6, \; M\geq \frac{\sqrt{\frac{2}{3}} (3 \gamma -p)}{3
   \gamma -4}$, or 
   \item $\frac{4}{3}<\gamma <2, \; 3 \gamma <p\leq 6, \; M\geq -\frac{\sqrt{\frac{2}{3}} (3 \gamma -p)}{3 \gamma -4}$, or 
   \item $1\leq \gamma <\frac{4}{3}, \; 3 \gamma <p\leq 6, \; M\geq
   \frac{\sqrt{\frac{2}{3}} (3 \gamma -p)}{3 \gamma -4}$, or 
   \item $\frac{4}{3}<\gamma <2, \; 3 \gamma <p\leq 6, \; M\geq -\frac{\sqrt{\frac{2}{3}} (3 \gamma -p)}{3 \gamma -4}$, or 
   \item $1\leq \gamma
   <\frac{4}{3}, \; 3 \gamma <p\leq 6, \; M\leq -\frac{\sqrt{\frac{2}{3}} (3 \gamma -p)}{3 \gamma -4}$, or 
   \item $\frac{4}{3}<\gamma <2, \; 3 \gamma <p\leq 6, \; M\leq \frac{\sqrt{\frac{2}{3}} (3 \gamma -p)}{3 \gamma
   -4}$, or 
   \item $1\leq \gamma <\frac{4}{3}, \; 3 \gamma <p\leq 6, \; M\leq -\frac{\sqrt{\frac{2}{3}} (3 \gamma -p)}{3 \gamma -4}$, or 
   \item $\frac{4}{3}<\gamma <2, \; 3 \gamma <p\leq 6, \; M\leq
   \frac{\sqrt{\frac{2}{3}} (3 \gamma -p)}{3 \gamma -4}$.
   \end{enumerate}
 It is a non--hyperbolic saddle.

    \item[$B_8$:] $\left(1,\tan ^{-1}\left[\sqrt{1-\frac{2 (p-3 \gamma )^2}{3 M^2(4-3 \gamma )^2}},\frac{2(p-3 \gamma)}{\sqrt{6}M(4 -3 \gamma)}\right]+2 \pi  c_1,\frac{2 (6-p) (p-3 \gamma )}{3 (4-3 \gamma )^2 M^2},\frac{(4-3 \gamma )^2
   M^2+2 (\gamma -2) (p-3 \gamma )}{(4-3 \gamma )^2 M^2},0\right), c_1\in \mathbb{Z}$, with eigenvalues \\
   $\Big\{0,-\frac{p}{2},\frac{-3 \gamma  M p+4 M p-6 \gamma  N+2 N p}{M(4-3\gamma)},\frac{1}{4} \left(p-6+\frac{\sqrt{\left(p-6\right) \left(3 (4-3 \gamma )^2 M^2 (-8 \gamma +3 p-2)+16 (\gamma -2) (p-3 \gamma )^2\right)}}{(3 \gamma -4)
   M}\right)$,\\
   $\frac{1}{4} \left(p-6-\frac{\sqrt{\left(p-6\right) \left(3 (4-3 \gamma )^2 M^2 (-8 \gamma +3 p-2)+16 (\gamma -2) (p-3 \gamma )^2\right)}}{(3 \gamma -4) M}\right)\Big\}$.
   It exists for:
   \begin{enumerate}
    \item $1\leq \gamma <\frac{4}{3}, \; p=3 \gamma , \; M>0$, or 
    \item $\frac{4}{3}<\gamma \leq 2, \; p=3 \gamma , \; M>0$, or 
    \item $1\leq \gamma <\frac{4}{3}, \; p=3 \gamma , \; M>0$, or
    \item $\frac{4}{3}<\gamma \leq 2, \; p=3 \gamma , \; M>0$, or 
    \item $1\leq \gamma <\frac{4}{3}, \; p=3 \gamma , \; M<0$, or 
    \item $\frac{4}{3}<\gamma \leq 2, \; p=3 \gamma , \; M<0$, or 
    \item $1\leq \gamma
   <\frac{4}{3}, \; p=3 \gamma , \; M<0$, or 
   \item $\frac{4}{3}<\gamma \leq 2, \; p=3 \gamma , \; M<0$, or 
   \item $1\leq \gamma <\frac{4}{3}, \; 3 \gamma <p\leq 6, \; M\geq \frac{\sqrt{\frac{2}{3}} (3 \gamma -p)}{3
   \gamma -4}$, or 
   \item $\frac{4}{3}<\gamma <2, \; 3 \gamma <p\leq 6, \; M\geq -\frac{\sqrt{\frac{2}{3}} (3 \gamma -p)}{3 \gamma -4}$, or 
   \item $1\leq \gamma <\frac{4}{3}, \; 3 \gamma <p\leq 6, \; M\geq
   \frac{\sqrt{\frac{2}{3}} (3 \gamma -p)}{3 \gamma -4}$, or 
   \item $\frac{4}{3}<\gamma <2, \; 3 \gamma <p\leq 6, \; M\geq -\frac{\sqrt{\frac{2}{3}} (3 \gamma -p)}{3 \gamma -4}$, or 
   \item $1\leq \gamma
   <\frac{4}{3}, \; 3 \gamma <p\leq 6, \; M\leq -\frac{\sqrt{\frac{2}{3}} (3 \gamma -p)}{3 \gamma -4}$, or 
   \item $\frac{4}{3}<\gamma <2, \; 3 \gamma <p\leq 6, \; M\leq \frac{\sqrt{\frac{2}{3}} (3 \gamma -p)}{3 \gamma
   -4}$, or 
   \item $1\leq \gamma <\frac{4}{3}, \; 3 \gamma <p\leq 6, \; M\leq -\frac{\sqrt{\frac{2}{3}} (3 \gamma -p)}{3 \gamma -4}$, or 
   \item $\frac{4}{3}<\gamma <2, \; 3 \gamma <p\leq 6, \; M\leq
   \frac{\sqrt{\frac{2}{3}} (3 \gamma -p)}{3 \gamma -4}$.
   \end{enumerate}
 The situation of physical interest is when it is non--hyperbolic with a 4D stable manifold for:
   \begin{enumerate}
    \item $N<-3 \sqrt{6}, \; p=\frac{1}{16} \left(27 N^2-\sqrt{3} \sqrt{N^2 \left(243 N^2-1216\right)}\right), \; \frac{4}{3}<\gamma <\frac{2 p \left(N^2-p\right)}{6 N^2-p^2}, \; \frac{2 N (p-3 \gamma )}{(3 \gamma -4)
   p}<M<-\sqrt{2} \sqrt{\frac{(\gamma -2) (3 \gamma -p)}{(3 \gamma -4)^2}}$, or 
   \item $N<-3 \sqrt{6}, \; N^2-\sqrt{N^2 \left(N^2-6\right)}<p\leq \frac{1}{8} \left(9 N^2-\sqrt{3} \sqrt{N^2 \left(27
   N^2-160\right)}\right), \; \gamma <\frac{2 p \left(N^2-p\right)}{6 N^2-p^2}, \; -\sqrt{2} \sqrt{\frac{(\gamma -2) (3 \gamma -p)}{(3 \gamma -4)^2}}<M<\frac{2 N (p-3 \gamma )}{(3 \gamma -4) p}$, or 
   \item $N<-3
   \sqrt{6}, \; \frac{1}{2} \left(3 N^2-\sqrt{3} \sqrt{N^2 \left(3 N^2-16\right)}\right)<p<\frac{1}{16} \left(27 N^2-\sqrt{3} \sqrt{N^2 \left(243 N^2-1216\right)}\right), \; \frac{4}{3}<\gamma <\frac{2 p \left(N^2-p\right)}{6
   N^2-p^2}, \; \frac{2 N (p-3 \gamma )}{(3 \gamma -4) p}<M<-\sqrt{2} \sqrt{\frac{(\gamma -2) (3 \gamma -p)}{(3 \gamma -4)^2}}$, or 
   \item $N<-3 \sqrt{6}, \; \frac{1}{8} \left(9 N^2-\sqrt{3} \sqrt{N^2 \left(27
   N^2-160\right)}\right)<p\leq \frac{1}{2} \left(3 N^2-\sqrt{3} \sqrt{N^2 \left(3 N^2-16\right)}\right), \; \frac{9 N^2 p-6 N^2-8 p^2}{4 \left(6 N^2-p^2\right)}\leq \gamma <\frac{2 p \left(N^2-p\right)}{6 N^2-p^2}, \;
   -\sqrt{2} \sqrt{\frac{(\gamma -2) (3 \gamma -p)}{(3 \gamma -4)^2}}<M<\frac{2 N (p-3 \gamma )}{(3 \gamma -4) p}$, or 
   \item $-3 \sqrt{6}\leq N<-\sqrt{6}, \; p=\frac{1}{16} \left(27 N^2-\sqrt{3} \sqrt{N^2 \left(243
   N^2-1216\right)}\right), \; \frac{4}{3}<\gamma <\frac{2 p \left(N^2-p\right)}{6 N^2-p^2}, \; \frac{2 N (p-3 \gamma )}{(3 \gamma -4) p}<M<-\sqrt{2} \sqrt{\frac{(\gamma -2) (3 \gamma -p)}{(3 \gamma -4)^2}}$, or 
   \item $-3
   \sqrt{6}\leq N<-\sqrt{6}, \; N^2-\sqrt{N^2 \left(N^2-6\right)}<p\leq \frac{1}{8} \left(9 N^2-\sqrt{3} \sqrt{N^2 \left(27 N^2-160\right)}\right), \; \gamma <\frac{2 p \left(N^2-p\right)}{6 N^2-p^2}, \; -\sqrt{2}
   \sqrt{\frac{(\gamma -2) (3 \gamma -p)}{(3 \gamma -4)^2}}<M<\frac{2 N (p-3 \gamma )}{(3 \gamma -4) p}$, or 
   \item $-3 \sqrt{6}\leq N<-\sqrt{6}, \; \frac{1}{2} \left(3 N^2-\sqrt{3} \sqrt{N^2 \left(3
   N^2-16\right)}\right)<p<\frac{1}{16} \left(27 N^2-\sqrt{3} \sqrt{N^2 \left(243 N^2-1216\right)}\right), \; \frac{4}{3}<\gamma <\frac{2 p \left(N^2-p\right)}{6 N^2-p^2}, \; \frac{2 N (p-3 \gamma )}{(3 \gamma -4)
   p}<M<-\sqrt{2} \sqrt{\frac{(\gamma -2) (3 \gamma -p)}{(3 \gamma -4)^2}}$, or 
   \item $-3 \sqrt{6}\leq N<-\sqrt{6}, \; \frac{1}{8} \left(9 N^2-\sqrt{3} \sqrt{N^2 \left(27 N^2-160\right)}\right)<p<\frac{1}{2} \left(3
   N^2-\sqrt{3} \sqrt{N^2 \left(3 N^2-16\right)}\right), \; \frac{9 N^2 p-6 N^2-8 p^2}{4 \left(6 N^2-p^2\right)}\leq \gamma <\frac{2 p \left(N^2-p\right)}{6 N^2-p^2}, \; -\sqrt{2} \sqrt{\frac{(\gamma -2) (3 \gamma -p)}{(3
   \gamma -4)^2}}<M<\frac{2 N (p-3 \gamma )}{(3 \gamma -4) p}$, or 
   \item $N>\sqrt{6}, \; p=\frac{1}{16} \left(27 N^2-\sqrt{3} \sqrt{N^2 \left(243 N^2-1216\right)}\right), \; \frac{4}{3}<\gamma <\frac{2 p
   \left(N^2-p\right)}{6 N^2-p^2}, \; \sqrt{2} \sqrt{\frac{(\gamma -2) (3 \gamma -p)}{(3 \gamma -4)^2}}<M<\frac{2 N (p-3 \gamma )}{(3 \gamma -4) p}$, or 
   \item $N>\sqrt{6}, \; N^2-\sqrt{N^2 \left(N^2-6\right)}<p\leq
   \frac{1}{8} \left(9 N^2-\sqrt{3} \sqrt{N^2 \left(27 N^2-160\right)}\right), \; \gamma <\frac{2 p \left(N^2-p\right)}{6 N^2-p^2}, \; \frac{2 N (p-3 \gamma )}{(3 \gamma -4) p}<M<\sqrt{2} \sqrt{\frac{(\gamma -2) (3 \gamma
   -p)}{(3 \gamma -4)^2}}$, or 
   \item $N>\sqrt{6}, \; \frac{1}{2} \left(3 N^2-\sqrt{3} \sqrt{N^2 \left(3 N^2-16\right)}\right)<p<\frac{1}{16} \left(27 N^2-\sqrt{3} \sqrt{N^2 \left(243 N^2-1216\right)}\right), \;
   \frac{4}{3}<\gamma <\frac{2 p \left(N^2-p\right)}{6 N^2-p^2}, \; \sqrt{2} \sqrt{\frac{(\gamma -2) (3 \gamma -p)}{(3 \gamma -4)^2}}<M<\frac{2 N (p-3 \gamma )}{(3 \gamma -4) p}$.
   \end{enumerate}
   
      \item[$B_9$:] $\left(0,\tan ^{-1}\left[ \frac{\sqrt{6 (2-\gamma )^2-(4-3 \gamma )^2 M^2}}{\sqrt{6}(2-\gamma )},\frac{(4-3 \gamma ) M}{\sqrt{6}(2-\gamma)}\right]+2 \pi  c_1,\frac{6 (2-\gamma )^2-(4-3 \gamma )^2 M^2}{6 (2-\gamma )^2},0,0\right), c_1\in \mathbb{Z}$, with eigenvalues \\
   $\Big\{0,\frac{6 (\gamma -2)^2-(4-3 \gamma )^2 M^2}{4 (\gamma -2)},-\frac{(4-3 \gamma )^2 M^2-6 (\gamma -2) \gamma }{4 (\gamma -2)},\gamma  \left(3-\frac{9 M^2}{2}\right)+\frac{2 M (N-M)}{\gamma -2}+3 M (M+N),-\frac{(4-3 \gamma
   )^2 M^2+2 (\gamma -2) (p-3 \gamma )}{2 (\gamma -2)}\Big\}$. It exists for:
   \begin{enumerate}
       \item $\gamma =\frac{4}{3}, \;  p>0$, or 
       \item $1\leq \gamma <\frac{4}{3}, \;  \frac{2 \sqrt{6}-\sqrt{6} \gamma }{3 \gamma -4}\leq M\leq \frac{\sqrt{6} \gamma -2 \sqrt{6}}{3 \gamma -4}, \;  p>0$, or
       \item $\frac{4}{3}<\gamma <2, \;  \frac{\sqrt{6} \gamma -2 \sqrt{6}}{3 \gamma -4}\leq M\leq \frac{2 \sqrt{6}-\sqrt{6} \gamma }{3 \gamma -4}, \;  p>0$.
   \end{enumerate}
 It is a non--hyperbolic saddle. 
   
      \item[$B_{10}$:] $\left(1,\tan ^{-1}\left[ \frac{\sqrt{6 (2-\gamma )^2-(4-3 \gamma )^2 M^2}}{\sqrt{6}(2-\gamma )},\frac{(4-3 \gamma ) M}{\sqrt{6}(2-\gamma)}\right]+2 \pi  c_1,\frac{6 (2-\gamma )^2-(4-3 \gamma )^2 M^2}{6 (2-\gamma )^2},0,0\right), c_1\in \mathbb{Z}$ with eigenvalues 
     $\Big\{0,\frac{6 (\gamma -2)^2-(4-3 \gamma )^2 M^2}{4 (\gamma -2)},\frac{(4-3
   \gamma )^2 M^2-6 (\gamma -2) \gamma }{4 (\gamma -2)},\gamma  \left(3-\frac{9 M^2}{2}\right)+\frac{2 M (N-M)}{\gamma -2}+3 M (M+N), -\frac{(4-3 \gamma )^2 M^2+2 (\gamma -2) (p-3 \gamma )}{2 (\gamma -2)}\Big\}$.
    It exists for:
   \begin{enumerate}
       \item $\gamma =\frac{4}{3}, \;  p>0$, or 
       \item $1\leq \gamma <\frac{4}{3}, \;  \frac{2 \sqrt{6}-\sqrt{6} \gamma }{3 \gamma -4}\leq M\leq \frac{\sqrt{6} \gamma -2 \sqrt{6}}{3 \gamma -4}, \;  p>0$, or
       \item $\frac{4}{3}<\gamma <2, \;  \frac{\sqrt{6} \gamma -2 \sqrt{6}}{3 \gamma -4}\leq M\leq \frac{2 \sqrt{6}-\sqrt{6} \gamma }{3 \gamma -4}, \;  p>0$.
   \end{enumerate}
  The situation of physical interest is when it is non--hyperbolic with a 4D stable manifold for:
   \begin{enumerate}
    \item $1\leq \gamma <\frac{4}{3}, \;  N<-\sqrt{6}, \;  \frac{N}{3 \gamma -4}-\sqrt{\frac{6 \gamma ^2-12 \gamma +N^2}{(3 \gamma -4)^2}}<M<\frac{\sqrt{6} \gamma -2 \sqrt{6}}{3 \gamma -4}, \;  p>\frac{6 \gamma ^2-12 \gamma -9 \gamma
   ^2 M^2+24 \gamma  M^2-16 M^2}{2 \gamma -4}$, or 
   \item $\frac{4}{3}<\gamma <2, \;  N<-\sqrt{6}, \;  \frac{\sqrt{6} \gamma -2 \sqrt{6}}{3 \gamma -4}<M<\sqrt{\frac{6 \gamma ^2-12 \gamma +N^2}{(3 \gamma -4)^2}}+\frac{N}{3
   \gamma -4}, \;  p>\frac{6 \gamma ^2-12 \gamma -9 \gamma ^2 M^2+24 \gamma  M^2-16 M^2}{2 \gamma -4}$, or 
   \item $1\leq \gamma <\frac{4}{3}, \;  N>\sqrt{6}, \;  \frac{2 \sqrt{6}-\sqrt{6} \gamma }{3 \gamma -4}<M<\sqrt{\frac{6
   \gamma ^2-12 \gamma +N^2}{(3 \gamma -4)^2}}+\frac{N}{3 \gamma -4}, \;  p>\frac{6 \gamma ^2-12 \gamma -9 \gamma ^2 M^2+24 \gamma  M^2-16 M^2}{2 \gamma -4}$, or 
   \item $\frac{4}{3}<\gamma <2, \;  N>\sqrt{6}, \;  \frac{N}{3
   \gamma -4}-\sqrt{\frac{6 \gamma ^2-12 \gamma +N^2}{(3 \gamma -4)^2}}<M<\frac{2 \sqrt{6}-\sqrt{6} \gamma }{3 \gamma -4}, \;  p>\frac{6 \gamma ^2-12 \gamma -9 \gamma ^2 M^2+24 \gamma  M^2-16 M^2}{2 \gamma -4}$.
   \end{enumerate}
     
      \item[$B_{11}$:] $\left(0,\tan^{-1}\left[\sqrt{1-\frac{6 \gamma ^2}{\left(2 N+M (4-3 \gamma )\right)^2}},-\frac{\sqrt{6}
\gamma }{2 N+M (4-3 \gamma )}\right]+2 \pi  c_1, \frac{4 N (2 M+N)-6 (2+M N) \gamma }{(2 N+M (4-3 \gamma ))^2},0,0\right), c_1\in \mathbb{Z}$, with eigenvalues 
   ${\left\{0,\frac{3 N \gamma }{4 M+2 N-3 M \gamma },\frac{M p (4-3 \gamma )+2 N (p-3 \gamma )}{-2 N+M (-4+3 \gamma )},\right.}\\
{\frac{1}{2 (2 N+M (4-3 \gamma ))^2}(3 (2 N+M (4-3 \gamma )) (N (-2+\gamma )+M (-4+3 \gamma ))+}\\
{\sqrt{3} \surd \left((2 N+M (4-3 \gamma ))^2 \left(-M^2 \left(-3+8 N^2-12 \gamma \right) (4-3 \gamma )^2+2 M^3 N (-4+3 \gamma )^3+\right.\right.}\\
{\left.\left.\left.2 M N (-4+3 \gamma ) \left(-6+4 N^2+3 \gamma -6 \gamma ^2\right)-3 (-2+\gamma ) \left(N^2 (2-9 \gamma )+24 \gamma ^2\right)\right)\right)\right),}\\
{\frac{1}{2 (2 N+M (4-3 \gamma ))^2}(3 (2 N+M (4-3 \gamma )) (N (-2+\gamma )+M (-4+3 \gamma ))-}\\
{\sqrt{3} \surd \left((2 N+M (4-3 \gamma ))^2 \left(-M^2 \left(-3+8 N^2-12 \gamma \right) (4-3 \gamma )^2+2 M^3 N (-4+3 \gamma )^3+\right.\right.}\\
{\left.\left.\left.\left.2 M N (-4+3 \gamma ) \left(-6+4 N^2+3 \gamma -6 \gamma ^2\right)-3 (-2+\gamma ) \left(N^2 (2-9 \gamma )+24 \gamma ^2\right)\right)\right)\right)\right\}}$. It exists for:
\begin{enumerate}
   \item $M\in \mathbb{R}, \; \gamma =\frac{4}{3}, \; N<-2$, or 
   \item $M\in \mathbb{R}, \; \gamma =\frac{4}{3}, \; N>2$, or 
   \item $\frac{4}{3}<\gamma \leq 2, \; N=-\sqrt{6}, \; M>\frac{\sqrt{6} (\gamma -2)}{3 \gamma
   -4}$, or 
   \item $\frac{4}{3}<\gamma \leq 2, \; 0<N\leq \sqrt{6}, \; M<\frac{2 \left(N^2-3 \gamma \right)}{(3 \gamma -4) N}$, or 
   \item $\frac{4}{3}<\gamma \leq 2, \; N<-\sqrt{6}, \; M\geq \frac{\sqrt{6} \gamma +2
   N}{3 \gamma -4}$, or 
   \item $\frac{4}{3}<\gamma \leq 2, \; -\sqrt{6}<N<0, \; M>\frac{2 \left(N^2-3 \gamma \right)}{(3 \gamma -4) N}$, or 
   \item $\frac{4}{3}<\gamma \leq 2, \; N>\sqrt{6}, \; M\leq -\frac{\sqrt{6}
   \gamma -2 N}{3 \gamma -4}$, or 
   \item $1\leq \gamma <\frac{4}{3}, \; N=-\sqrt{6}, \; M<\frac{\sqrt{6} (\gamma -2)}{3 \gamma -4}$, or 
   \item $1\leq \gamma <\frac{4}{3}, \; N>\sqrt{6}, \; M\geq -\frac{\sqrt{6} \gamma
   -2 N}{3 \gamma -4}$, or 
   \item $1\leq \gamma <\frac{4}{3}, \; N<-\sqrt{6}, \; M\leq \frac{\sqrt{6} \gamma +2 N}{3 \gamma -4}$, or 
   \item $1\leq \gamma <\frac{4}{3}, \; 0<N\leq \sqrt{6}, \; M>\frac{2 \left(N^2-3
   \gamma \right)}{(3 \gamma -4) N}$, or 
   \item $1\leq \gamma <\frac{4}{3}, \; -\sqrt{6}<N<0, \; M<\frac{2 \left(N^2-3 \gamma \right)}{(3 \gamma -4) N}$
\end{enumerate}
It is an hyperbolic saddle. 
     \item[$B_{12}$:] $\left(1,\tan^{-1}\left[\sqrt{1-\frac{6 \gamma ^2}{\left(2 N+M (4-3 \gamma )\right)^2}},-\frac{\sqrt{6}
\gamma }{2 N+M (4-3 \gamma )}\right]+2 \pi  c_1,
\frac{4 N (2 M+N)-6 (2+M N) \gamma }{(2 N+M (4-3 \gamma ))^2},0,0\right), c_1\in \mathbb{Z}$ with eigenvalues 
${\left\{0,-\frac{3 N \gamma }{4 M+2 N-3 M \gamma },\frac{M p (4-3 \gamma )+2 N (p-3 \gamma )}{-2 N+M (-4+3 \gamma )},\right.}\\
{\frac{1}{4 N+M (8-6 \gamma )}(3 N (-2+\gamma )+3 M (-4+3 \gamma )+}\\
{\surd \left(-3 M^2 \left(-3+8 N^2-12 \gamma \right) (4-3 \gamma )^2+6 M^3 N (-4+3 \gamma )^3+\right.}\\
{\left.\left.\left.6 M N (-4+3 \gamma ) \left(-6+4 N^2+3 \gamma -6 \gamma ^2\right)-9 (-2+\gamma ) \left(N^2 (2-9 \gamma )+24 \gamma ^2\right)\right)\right)\right\},}\\
{\frac{1}{4 N+M (8-6 \gamma )}(3 N (-2+\gamma )+3 M (-4+3 \gamma )-}\\
{\surd \left(-3 M^2 \left(-3+8 N^2-12 \gamma \right) (4-3 \gamma )^2+6 M^3 N (-4+3 \gamma )^3+\right.}\\
{\left.\left.\left.6 M N (-4+3 \gamma ) \left(-6+4 N^2+3 \gamma -6 \gamma ^2\right)-9 (-2+\gamma ) \left(N^2 (2-9 \gamma )+24 \gamma ^2\right)\right)\right)\right\}}$.
 It exists for:
\begin{enumerate}
   \item $M\in \mathbb{R}, \; \gamma =\frac{4}{3}, \; N<-2$, or 
   \item $M\in \mathbb{R}, \; \gamma =\frac{4}{3}, \; N>2$, or 
   \item $\frac{4}{3}<\gamma \leq 2, \; N=-\sqrt{6}, \; M>\frac{\sqrt{6} (\gamma -2)}{3 \gamma
   -4}$, or 
   \item $\frac{4}{3}<\gamma \leq 2, \; 0<N\leq \sqrt{6}, \; M<\frac{2 \left(N^2-3 \gamma \right)}{(3 \gamma -4) N}$, or 
   \item $\frac{4}{3}<\gamma \leq 2, \; N<-\sqrt{6}, \; M\geq \frac{\sqrt{6} \gamma +2
   N}{3 \gamma -4}$, or 
   \item $\frac{4}{3}<\gamma \leq 2, \; -\sqrt{6}<N<0, \; M>\frac{2 \left(N^2-3 \gamma \right)}{(3 \gamma -4) N}$, or 
   \item $\frac{4}{3}<\gamma \leq 2, \; N>\sqrt{6}, \; M\leq -\frac{\sqrt{6}
   \gamma -2 N}{3 \gamma -4}$, or 
   \item $1\leq \gamma <\frac{4}{3}, \; N=-\sqrt{6}, \; M<\frac{\sqrt{6} (\gamma -2)}{3 \gamma -4}$, or 
   \item $1\leq \gamma <\frac{4}{3}, \; N>\sqrt{6}, \; M\geq -\frac{\sqrt{6} \gamma
   -2 N}{3 \gamma -4}$, or 
   \item $1\leq \gamma <\frac{4}{3}, \; N<-\sqrt{6}, \; M\leq \frac{\sqrt{6} \gamma +2 N}{3 \gamma -4}$, or 
   \item $1\leq \gamma <\frac{4}{3}, \; 0<N\leq \sqrt{6}, \; M>\frac{2 \left(N^2-3
   \gamma \right)}{(3 \gamma -4) N}$, or 
   \item $1\leq \gamma <\frac{4}{3}, \; -\sqrt{6}<N<0, \; M<\frac{2 \left(N^2-3 \gamma \right)}{(3 \gamma -4) N}$.
\end{enumerate}
The situation of physical interest is when it is non--hyperbolic with a 4D stable manifold for:
\begin{enumerate}
    \item $\gamma =\frac{4}{3}, \;  2<N\leq 2.06559, \;  p>4$, or 
    \item $\gamma =\frac{4}{3}, \;  -2.06559\leq N<-2, \;  p>4$, or 
    \item $\gamma=1, \;  N=-2.42061, \;  M=M_{1,1}, \;  p>\frac{6 N}{M+2 N}$, or
   \item $\gamma=1, \;  N=2.42061, \;  M=-2.13014, \;  p>\frac{6 N}{M+2 N}$, or 
   \item $\gamma=1, \;  N=2.44949, \;  \frac{6\, -2 N^2}{N}<M\leq M_{1,3}, \;  p>\frac{6 N}{M+2
   N}$, or 
   \item $\gamma=1, \;  N\leq -2.44949, \;  M_{1,1}\leq M<-\sqrt{N^2-6}-N, \;  p>\frac{6 N}{M+2 N}$, or 
   \item $\gamma=1, \;  N>2.44949, \;  \sqrt{N^2-6}-N<M\leq
   M_{1,1}, \;  p>\frac{6 N}{M+2 N}$, or 
   \item $\gamma=1, \;  2.42061<N<2.44949, \;  M_{1,2}\leq M\leq M_{1,3}, \; 
   p>\frac{6 N}{M+2 N}$, or 
   \item $\gamma=1, \;  -2.44949<N<-2.42061, \;  M_{1,1}\leq M\leq M_{1,2}, \;  p>\frac{6 N}{M+2 N}$, or 
   \item $\gamma  =1, \;  -2.42061<N<0, \;  M_{1,1}\leq M<\frac{6\, -2 N^2}{N}, \;  p>\frac{6 N}{M+2 N}$, or 
   \item $\gamma=1, \;  -2.44949<N\leq -2.42061, \; 
   M_{1,3}\leq M<\frac{6\, -2 N^2}{N}, \;  p>\frac{6 N}{M+2 N}$, or 
   \item $\gamma=1, \;  0<N<2.44949, \;  \frac{6\, -2 N^2}{N}<M\leq M_{1,1}, \; 
   p>\frac{6 N}{M+2 N}$, or 
   \item $\gamma =1.62562, \;  N>2.44949, \;  M_{2,1}\leq M<\frac{N}{3\gamma -4 }- \sqrt{\frac{6 \gamma ^2-12\gamma +N^2}{(3\gamma -4 )^2}}, \; 
   p>\frac{6 \gamma  N}{M(4-3\gamma)+2 N}$, or 
   \item $1<\gamma <\frac{9}{8}, \;  N=N_{3,2}, \;  M=M_{2,1}, \;  p>\frac{6 \gamma  N}{M(4-3\gamma) +2
   N}$, or 
   \item $1<\gamma <\frac{9}{8}, \;  N=N_{3,3}, \;  M=M_{2,2}, \;  p>\frac{6 \gamma  N}{M(4-3\gamma)+2 N}$, or 
   \item $\gamma =1.62562, \; 
   N<N_{3,1}, \;  \sqrt{\frac{6 \gamma ^2-12\gamma +N^2}{(3\gamma -4 )^2}}+\frac{N}{3\gamma -4 }<M\leq M_{2,1}, \;  p>\frac{6 \gamma  N}{-3\gamma  M+4
   M+2 N}$, or 
   \item $\gamma =1.62562, \;  -2<N<0, \;  \frac{2 N^2-6 \gamma }{3\gamma  N-4 N}<M\leq M_{2,1}, \;  p>\frac{6 \gamma  N}{M(4-3\gamma)+2 N}$, or
   \item $\gamma =1.62562, \;  N_{3,1}\leq N<-2.44949, \;  \sqrt{\frac{6 \gamma ^2-12\gamma +N^2}{(3\gamma -4 )^2}}+\frac{N}{3\gamma -4 }<M\leq M_{2,3}, \; 
   p>\frac{6 \gamma  N}{M(4-3\gamma)+2 N}$, or 
   \item $\gamma =1.62562, \;  0<N<2.44949, \;  M_{2,1}\leq M<\frac{2 N^2-6 \gamma }{N(3 \gamma -4)}, \;  p>\frac{6 \gamma  N}{-3
   \gamma  M+4 M+2 N}$, or 
   \item $1<\gamma <\frac{9}{8}, \;  N=-2.44949, \;  M_{2,1}\leq M<\frac{2 N^2-6 \gamma }{N(3 \gamma -4)}, \;  p>\frac{6 \gamma  N}{M(4-3\gamma) +2
   N}$, or 
   \item $\frac{4}{3}<\gamma <1.62562, \;  N>2.44949, \;  M_{2,1}\leq M<\frac{N}{3\gamma -4 }-\sqrt{\frac{6 \gamma ^2-12\gamma +N^2}{(3\gamma -4 )^2}}, \;  p>\frac{6 \gamma 
   N}{M(4-3\gamma)+2 N}$, or 
   \item $1.62562<\gamma <2,   \;  N>2.44949, \;  M_{2,1}\leq M<\frac{N}{3\gamma -4 }-\sqrt{\frac{6 \gamma ^2-12\gamma +N^2}{(3\gamma -4 )^2}}, \; 
   p>\frac{6 \gamma  N}{M(4-3\gamma)+2 N}$, or 
   \item $1<\gamma <\frac{9}{8}, \;  N<-2.44949, \;  M_{2,1}\leq M<\frac{N}{3\gamma -4 }-\sqrt{\frac{6 \gamma ^2-12\gamma +N^2}{(3
   \gamma -4)^2}}, \;  p>\frac{6 \gamma  N}{M(4-3\gamma)+2 N}$, or 
   \item $\frac{9}{8}<\gamma <\frac{4}{3}, \;  N<-2.44949, \;  M_{2,1}\leq M<\frac{N}{3\gamma -4 }-\sqrt{\frac{6 \gamma
   ^2-12\gamma +N^2}{(3\gamma -4 )^2}}, \;  p>\frac{6 \gamma  N}{M(4-3\gamma)+2 N}$, or
   \item $1<\gamma <\frac{9}{8}, \;  N=N_{3,2}, \;  M_{2,3}\leq
   M<\frac{2 N^2-6 \gamma }{N(3 \gamma -4)}, \;  p>\frac{6 \gamma  N}{M(4-3\gamma)+2 N}$, or 
   \item $\gamma =1.62562, \;  -2.44949<N\leq -2, \;  \frac{2 N^2-6 \gamma }{N(3 \gamma -4)}<M\leq
   M_{2,3}, \;  p>\frac{6 \gamma  N}{M(4-3\gamma)+2 N}$, or 
   \item $1<\gamma <\frac{9}{8}, \;  N>N_{3,4}, \;  \sqrt{\frac{6 \gamma ^2-12\gamma +N^2}{(3
   \gamma -4)^2}}+\frac{N}{3\gamma -4 }<M\leq M_{2,1}, \;  p>\frac{6 \gamma  N}{M(4-3\gamma)+2 N}$, or 
   \item $1<\gamma <\frac{9}{8}, \;  N=N_{3,3}, \; 
   \frac{2 N^2-6 \gamma }{N(3 \gamma -4)}<M\leq M_{2,1}, \;  p>\frac{6 \gamma  N}{M(4-3\gamma)+2 N}$, or 
   \item $1<\gamma <\frac{9}{8}, \;  N=2.44949, \;  \frac{2 N^2-6 \gamma }{3
   \gamma  N-4 N}<M\leq M_{2,3}, \;  p>\frac{6 \gamma  N}{M(4-3\gamma)+2 N}$, or 
   \item $\frac{9}{8}<\gamma <\frac{4}{3}, \;  N>N_{3,4}, \;  \sqrt{\frac{6
   \gamma ^2-12\gamma +N^2}{(3\gamma -4 )^2}}+\frac{N}{3\gamma -4 }<M\leq M_{2,1}, \;  p>\frac{6 \gamma  N}{M(4-3\gamma)+2 N}$, or 
   \item $\frac{4}{3}<\gamma <1.62562, \; 
   N<N_{3,1}, \;  \sqrt{\frac{6 \gamma ^2-12\gamma +N^2}{(3\gamma -4 )^2}}+\frac{N}{3\gamma -4 }<M\leq M_{2,1}, \;  p>\frac{6 \gamma  N}{-3 \gamma  M+4
   M+2 N}$, or 
   \item $1.62562<\gamma <2,   \;  N<N_{3,1}, \;  \sqrt{\frac{6 \gamma ^2-12\gamma +N^2}{(3\gamma -4 )^2}}+\frac{N}{3\gamma -4 }<M\leq
   M_{2,1}, \;  p>\frac{6 \gamma  N}{M(4-3\gamma)+2 N}$, or 
   \item $\frac{4}{3}<\gamma <1.62562, \;  N_{3,1}\leq N<-2.44949, \;  \sqrt{\frac{6 \gamma
   ^2-12\gamma +N^2}{(3\gamma -4 )^2}}+\frac{N}{3\gamma -4 }<M\leq M_{2,3}, \;  p>\frac{6 \gamma  N}{M(4-3\gamma)+2 N}$, or 
   \item $1.62562<\gamma <2,   \; 
   N_{3,1}\leq N<-2.44949, \;  \sqrt{\frac{6 \gamma ^2-12\gamma +N^2}{(3\gamma -4 )^2}}+\frac{N}{3\gamma -4 }<M\leq M_{2,3}, \;  p>\frac{6 \gamma  N}{-3
   \gamma  M+4 M+2 N}$, or 
   \item $\frac{4}{3}<\gamma <1.62562, \;  0<N<2.44949, \;  M_{2,1}\leq M<\frac{2 N^2-6 \gamma }{N(3 \gamma -4)}, \;  p>\frac{6 \gamma  N}{M(4-3\gamma) +2
   N}$, or 
   \item $1.62562<\gamma <2,   \;  0<N<2.44949, \;  M_{2,1}\leq M<\frac{2 N^2-6 \gamma }{N(3 \gamma -4)}, \;  p>\frac{6 \gamma  N}{M(4-3\gamma)+2 N}$, or
   \item $\frac{9}{8}<\gamma <\frac{4}{3}, \;  -2.44949<N<0, \;  M_{2,1}\leq M<\frac{2 N^2-6 \gamma }{N(3 \gamma -4)}, \;  p>\frac{6 \gamma  N}{M(4-3\gamma)+2 N}$, or 
   \item $1<\gamma
   <\frac{9}{8}, \;  N_{3,2}<N<0, \;  M_{2,1}\leq M<\frac{2 N^2-6 \gamma }{N(3 \gamma -4)}, \;  p>\frac{6 \gamma  N}{M(4-3\gamma)+2 N}$, or
   \item $1<\gamma <\frac{9}{8}, \;  -2.44949<N<N_{3,2}, \;  M_{2,1}\leq M\leq M_{2,2}, \;  p>\frac{6 \gamma  N}{M(4-3\gamma) +2
   N}$, or 
   \item $1<\gamma <\frac{9}{8}, \;  0<N<N_{3,3}, \;  \frac{2 N^2-6 \gamma }{N(3 \gamma -4)}<M\leq M_{2,1}, \;  p>\frac{6 \gamma  N}{-3 \gamma 
   M+4 M+2 N}$, or 
   \item $\frac{9}{8}<\gamma <\frac{4}{3}, \;  0<N<N_{3,3}, \;  \frac{2 N^2-6 \gamma }{N(3 \gamma -4)}<M\leq M_{2,1}, \;  p>\frac{6 \gamma 
   N}{M(4-3\gamma)+2 N}$, or 
   \item $1<\gamma <\frac{9}{8}, \;  -2.44949<N<N_{3,2}, \;  M_{2,3}\leq M<\frac{2 N^2-6 \gamma }{N(3 \gamma -4)}, \; 
   p>\frac{6 \gamma  N}{M(4-3\gamma)+2 N}$, or 
   \item $1<\gamma <\frac{9}{8}, \;  2.44949<N\leq N_{3,4}, \;  \sqrt{\frac{6 \gamma ^2-12\gamma +N^2}{(3\gamma -4 )^2}}+\frac{N}{3
   \gamma -4}<M\leq M_{2,3}, \;  p>\frac{6 \gamma  N}{M(4-3\gamma)+2 N}$, or 
   \item $\frac{9}{8}<\gamma <\frac{4}{3}, \;  2.44949<N\leq N_{3,4}, \; 
   \sqrt{\frac{6 \gamma ^2-12\gamma +N^2}{(3\gamma -4 )^2}}+\frac{N}{3\gamma -4 }<M\leq M_{2,3}, \;  p>\frac{6 \gamma  N}{M(4-3\gamma)+2 N}$, or 
   \item $\frac{4}{3}<\gamma
   <1.62562, \;  -2.44949<N\leq N_{3,2}, \;  \frac{2 N^2-6 \gamma }{N(3 \gamma -4)}<M\leq M_{2,3}, \;  p>\frac{6 \gamma  N}{M(4-3\gamma) +2
   N}$, or 
   \item $1.62562<\gamma <2,   \;  -2.44949<N\leq N_{3,2}, \;  \frac{2 N^2-6 \gamma }{N(3 \gamma -4)}<M\leq M_{2,3}, \;  p>\frac{6 \gamma 
   N}{M(4-3\gamma)+2 N}$, or 
   \item $1<\gamma <\frac{9}{8}, \;  N_{3,3}<N<2.44949, \;  M_{2,2}\leq M\leq M_{2,3}, \; 
   p>\frac{6 \gamma  N}{M(4-3\gamma)+2 N}$, or 
   \item $1<\gamma <\frac{9}{8}, \;  N_{3,3}<N<2.44949, \;  \frac{2 N^2-6 \gamma }{N(3 \gamma -4)}<M\leq
   M_{2,1}, \;  p>\frac{6 \gamma  N}{M(4-3\gamma)+2 N}$, or 
   \item $\frac{4}{3}<\gamma <1.62562, \;  N_{3,2}<N<0, \;  \frac{2 N^2-6 \gamma }{3 \gamma 
   N-4 N}<M\leq M_{2,1}, \;  p>\frac{6 \gamma  N}{M(4-3\gamma)+2 N}$, or 
   \item $1.62562<\gamma <2,   \;  N_{3,2}<N<0, \;  \frac{2 N^2-6 \gamma
   }{N(3 \gamma -4)}<M\leq M_{2,1}, \;  p>\frac{6 \gamma  N}{M(4-3\gamma)+2 N}$, or 
   \item $\frac{9}{8}<\gamma <\frac{4}{3}, \;  N_{3,3}\leq N<2.44949, \; 
   \frac{2 N^2-6 \gamma }{N(3 \gamma -4)}<M\leq M_{2,3}, \;  p>\frac{6 \gamma  N}{M(4-3\gamma)+2 N}$, or 
   \item $\gamma =\frac{9}{8}, \;  N>2.61227, \;  \frac{1}{5} \sqrt{64 N^2-378}-\frac{8}{5}
   N<M\leq M_{4,1}, \;  p>\frac{54 N}{5 M+16 N}$, or 
   \item $\gamma =\frac{9}{8}, \;  N<-2.44949, \;  M_{4,1}\leq M<-\frac{1}{5} \sqrt{64 N^2-378}-\frac{8}{5} N, \; 
   p>\frac{54 N}{5M+16 N}$, or 
   \item $\gamma =\frac{9}{8}, \;  2.44949<N\leq 2.61227, \;  \frac{1}{5} \sqrt{64 N^2-378}-\frac{8}{5} N<M\leq M_{4,3}, \;  p>\frac{54 N}{5. M+16 N}$, or 
   \item $\gamma
   =\frac{9}{8}, \;  -2.44949<N<0, \;  M_{4,1}\leq M<\frac{\left(54\, -16 N^2\right)}{5N}, \;  p>\frac{54 N}{5. M+16 N}$, or 
   \item $\gamma =\frac{9}{8}, \;  0<N<2.44949, \;  \frac{
   \left(54-16 N^2\right)}{5N}<M\leq M_{4,1}, \;  p>\frac{54 N}{5M+16 N}$, or 
   \item $\gamma=2, \;  0<N<2.44949, \;  \frac{\left(8N^2-27\right)}{8N}-\frac{1}{8}
   \sqrt{\frac{729 -48 N^2}{N^2}}\leq M<\frac{N^2-6}{N}, \;  p>\frac{6 N}{N-M}$, or 
   \item $\gamma=2, \;  -2.44949<N<0, \;  \frac{N^2-6}{N}<M\leq \frac{1}{8} \sqrt{\frac{729-48N^2}{N^2}}+\frac{\left(8 N^2-27\right)}{8N}, \;  p>\frac{6 N}{N-M}$. 
\end{enumerate}
Where $M_{1,1}$, $M_{1,2}$, $M_{1,3}$ are the first, the second and the third root of the polynomial
$P_1(M)=2 M^3 N+M^2 \left(8 N^2-15\right)+M \left(8 N^3-18 N\right)+21 N^2-72$, respectively. 

$M_{2,1}$, $M_{2,2}$ and $M_{2,3}$ are the first, the second and the third root of the polynomial 
$P_2(M)=-72 \gamma ^3+144 \gamma ^2+M^3 \left(54 \gamma ^3 N-216 \gamma ^2 N+288 \gamma  N-128 N\right)+M^2 \left(108 \gamma ^3-261
   \gamma ^2+120 \gamma -72 \gamma ^2 N^2+192 \gamma  N^2-128 N^2+48\right)$\\
   $+M \left(24 \gamma  N^3-32 N^3-36 \gamma ^3 N+66 \gamma ^2 N-60 \gamma  N+48 N\right)+27 \gamma ^2 N^2-60 \gamma  N^2+12 N^2$, respectively.
$N_{3,1}$, $N_{3,2}$, $N_{3,3}$ and $N_{3,4}$ are the first, the second, the third and the fourth root of the polynomial:   
   $P_3(N)=-1152 \gamma
   ^4+1440 \gamma ^3+1512 \gamma ^2+414 \gamma +64 N^4+\left(321 \gamma ^2-1176 \gamma +96\right) N^2+36$, respectively.
  Finally, $M_{4,1}$  is first root and  $M_{4,3}$ is the third root of the polynomial:  
   $P_4(M)=125 M^3 N+M^2 \left(800 N^2-1650\right)+M \left(1280 N^3-3270 N\right)+5460 N^2-20412$.
\end{enumerate}

\end{document}